\documentclass[
 reprint,
 amsmath,amssymb,
 aps,
]{revtex4-2}

\usepackage[dvipdfm]{graphicx} 
\usepackage{bm}

\usepackage{algorithm}
\usepackage{algorithmic}
\usepackage{braket}
\usepackage{flushend}
\usepackage{multirow}
\usepackage{siunitx}  
\usepackage{url}
\usepackage{qcircuit}

\usepackage[hang,small,bf]{caption}
\usepackage[subrefformat=parens]{subcaption}
\renewcommand{\algorithmicrequire}{\textbf{Input:}}
\renewcommand{\algorithmicensure}{\textbf{Output:}}

\begin{document}

\title{Hamiltonian simulation using quantum singular value transformation:\\
        complexity analysis and application to the linearized Vlasov-Poisson equation}

\author{Kiichiro Toyoizumi,${}^{\text{1}}$ Naoki Yamamoto,${}^{\text{1, 2}}$ and Kazuo Hoshino${}^{\text{1}}$}
  \affiliation{$^{\text{1}} {}$Department of Applied Physics and Physico-Informatics,
                Keio University, Hiyoshi 3-14-1, Kohoku-ku, Yokohama, 223-8522, Japan\\
              $^{\text{2}} {}$Quantum Computing Center, Keio University, Hiyoshi 3-14-1, Kohoku-ku, Yokohama, 223-8522, Japan}

\begin{abstract}
  Quantum computing can be used to speed up the simulation time
  (more precisely, the number of queries of the algorithm) for physical systems;
  one such promising approach is the Hamiltonian simulation (HS) algorithm. Recently, it was proven that 
  the quantum singular value transformation (QSVT) achieves the minimum simulation time for HS. 
  An important subroutine of the QSVT-based HS algorithm is the amplitude amplification operation, 
  which can be realized via the oblivious amplitude amplification or the fixed-point amplitude amplification 
  in the QSVT framework. In this work, we execute a detailed analysis of the error and 
  number of queries of the QSVT-based HS and show that the oblivious method is better than the fixed-point one 
  in the sense of simulation time. Based on this finding, 
  we apply the QSVT-based HS to the one-dimensional linearized Vlasov-Poisson equation and demonstrate that 
  the linear Landau damping can be successfully simulated.
\end{abstract}

\maketitle
  
\section{Introduction}
\subsection{Background}
Quantum computers are expected to outperform classical counterparts in some problems.
Several quantum algorithms have obtained speedups over classical ones, such as
the Grover search algorithm~\cite{grover1996fast}, Shor's algorithm
for integer factorization~\cite{shor1994algorithms}, and the HHL algorithm~\cite{harrow2009quantum, childs2017quantum}.
Quantum computing also gives us algorithms for solving physics problems.
In particular, the algorithms~\cite{feynman1982simulating, lloyd1996universal} realize exponential speedup
for the simulation of quantum systems.
This seems natural since quantum computing is based on quantum mechanics.
In recent years, some quantum algorithms for simulating classical physical systems have been developed, such as
the Navier-Stokes equation~\cite{gaitan2020finding,gaitan2021finding},
plasma equations~\cite{engel2019quantum,dodin2021quantum,novikau2022quantum,ameri2023quantum},
the Poisson equation~\cite{cao2013quantum,wang2020quantum},
and the wave equation~\cite{costa2019quantum,suau2021practical}.

One of the quantum algorithms for simulating physical systems is a Hamiltonian simulation (HS)
algorithm~\cite{berry2007efficient,childs2010limitations,berry2012black,childs2012hamiltonian,
berry2014exponential,berry2015hamiltonian,berry2015simulating,berry2016corrected,novo2017improved,childs2018toward}, which
implements $U=\exp(-iHt)$, where $H$ is a time-independent Hamiltonian and $t$ is an evolution time.
The optimal HS result was shown by Low and Chuang using quantum signal processing
(QSP)~\cite{low2017optimal,low2019hamiltonian}. This result has been generalized to
the quantum singular value transformation (QSVT) in Ref.~\cite{gilyen2019quantum}.
QSVT is a quantum algorithm for applying a polynomial transformation
$P^{(\mathrm{SV})}(A)$ to the singular values of a given matrix $A$, called the singular value transformation.
Notably, QSVT can formulate major quantum algorithms in a unified way, such as the Grover search, 
phase estimation~\cite{nielsen2010quantum}, matrix inversion, quantum walks~\cite{szegedy2004quantum},
and HS algorithm. Therefore, QSVT is called a grand unification of quantum algorithms in Ref.~\cite{martyn2021grand}.

The HS algorithm using QSVT has been proposed in Refs.~\cite{gilyen2019quantum,martyn2021grand}.
The algorithm includes an amplitude amplification algorithm that can be implemented by QSVT as a subroutine.
There are two QSVT-based amplitude amplification algorithms proposed for HS;
one is the oblivious amplitude amplification (OAA) algorithm~\cite{gilyen2019quantum}, and the other is 
the fixed-point amplitude amplification (FPAA) algorithm~\cite{martyn2021grand}.
However, there have been no discussion to compare those two schemes from neither theoretical nor numerical viewpoints.
It would be helpful to clarify
which one is preferable when the QSVT-based HS algorithm is applied to physical systems.

\subsection{Contribution of this paper}
We elaborate the QSVT-based HS using explicit quantum circuits and discuss
the approximation error and query complexity.
As a result, the number of queries for the OAA-based HS scales as $\mathcal{O}(t + \log(1/\varepsilon))$, whereas
the FPAA-based one scales as $\mathcal{O}(t\log(1/\varepsilon) + \log^2(1/\varepsilon))$, where
$t$ is an evolution time and $\varepsilon$ is an error tolerance.
To support this fact, we perform numerical experiments:
we plot the number of queries for a wide range of parameters $t$ and $\varepsilon$;
we curve-fit the data to identify the constant factors 
and coefficients of the number of queries hidden behind the asymptotic scalings.
Our findings indicate that the OAA-based method is both theoretically and numerically advantageous than
the FPAA-based method in the sense of the number of queries. 
Importantly, this advantage is consistent, independent of the type of the Hamiltonian.

To demonstrate the effectiveness of the QSVT-based HS combined with the above-described
detailed analysis, we apply the OAA-based HS to the simulation of the linearized Vlasov-Poisson system.
This system can be transformed into the same form of the unitary time evolution of quantum systems~\cite{engel2019quantum}.
In addition to the simulation, we discuss several issues unaddressed in 
previous studies on quantum algorithms for plasma simulation~\cite{engel2019quantum,dodin2021quantum,novikau2022quantum,ameri2023quantum}:
We discuss the computational complexity of extending evolution time using sequential
short HS circuits; we propose an algorithm to obtain a quantity related
to the distribution function; we provide explicit quantum circuits for the higher dimensional systems.
Furthermore, we show a potential advantage of applying HS to physical systems
compared to the classical Euler method.

\subsection{Comparison to prior work}
The QSVT-based HS algorithms have been introduced in Refs.~\cite{gilyen2019quantum,martyn2021grand}.
The authors of Ref.~\cite{gilyen2019quantum} have originally proposed the QSVT framework.
Within the framework, they developed a method for implementing the exponential function
and the OAA algorithm, which constructs the Chebyshev polynomial of the first kind.
Combining these methods, they have realized QSVT-based HS and shown that its asymptotic complexity
is consistent with the result of HS by Low and Chuang~\cite{low2017optimal,low2019hamiltonian}, known to be optimal.
The authors of Ref.~\cite{martyn2021grand} have 
reviewed that several major quantum algorithms can be described in a unified way
within the QSVT framework and suggested for HS the use of FPAA, which constructs
the approximate polynomial of the sign function.
These authors have independently proposed using OAA and FPAA for HS, respectively,
with rough analyses of the approximation error and query complexity.
However, a theoretical or numerical comparison of these methods remains absent.

To our knowledge, no investigation exists to compare OAA and FPAA in a non-QSVT framework.
This is probably because these algorithms have been originally developed for distinct purposes.
The OAA algorithm has been initially developed to simulate a sparse Hamiltonian evolution~\cite{berry2014exponential,berry2015simulating},
achieving amplitude amplification without the reflection operator about an unknown initial state.
It is also used to decompose single-qubit unitaries~\cite{paetznick2014repeat}
and compute matrix products for non-unitary matrices~\cite{daskin2017ancilla}.
On the other hand, the FPAA algorithm is an algorithm that ensures amplitude amplification regardless of an unknown amplitude.
Notably, several non-QSVT-based FPAA algorithms have been developed, such as the $\pi/3$-algorithm~\cite{grover2005quantum},
the measurement-based algorithm~\cite{tulsi2006new}, and the FPAA technique by Yoder et al.~\cite{yoder2014fixed}.

Several studies have addressed quantum algorithms for plasma simulations.
The authors of Ref.~\cite{dodin2021quantum} have conducted an extensive survey on applying 
quantum computers to plasma simulations.
The authors of Ref.~\cite{engel2019quantum} have introduced the quantum algorithm for
calculating the time evolution of the one-dimensional linearized Vlasov-Poisson system 
using the HS algorithm by Low and Chuang~\cite{low2017optimal,low2019hamiltonian}.
They have concluded that the algorithm achieves exponential speedup for a velocity grid size.
They have also discussed an estimation of the electric field with the quantum amplitude estimation
algorithm~\cite{Brassard2002quantum} and simulated its linear Landau damping.
While the state's amplitude comprises the electric field and the distribution function,
no method has been proposed to extract the latter's quantity from the state.
They have indicated that the above findings can be extended to systems with higher dimensions, 
yet without providing explicit circuits.
The authors of Ref.~\cite{ameri2023quantum} have examined the computational complexity for a system size, of a quantum algorithm
for the one-dimensional Vlasov-Poisson system with collisions.
They have adopted the Hermite representation, reducing the equations to a linear ODE problem---distinct
from our work and Ref.~\cite{engel2019quantum}.
The authors of Ref.~\cite{novikau2022quantum} have implemented the HS algorithm for one-dimensional cold plasma waves,
dividing the HS circuit into shorter circuits to avoid a large evolution time $t$, yet without discussing its cost.

\subsection{Organization of the paper}
The rest of this paper is organized as follows. In Sec.~\ref{section:preliminary}
we present a brief description of QSVT with application to the trigonometric functions;
then we show the transformation from the linearized Vlasov-Poisson system
to a form of the Schr\"{o}dinger equation. We discuss the error and query complexity of QSVT-based HS
in Sec.~\ref{section:HS_using_QSVT}.
The quantum algorithm for the linearized Vlasov-Poisson system is discussed in Sec.~\ref{section:VQS}.
We show the numerical results in Sec.~\ref{section:numerical_results}.
The paper is then concluded in Sec.~\ref{section:summary}.

\section{Preliminary} \label{section:preliminary}
\subsection{Quantum singular value transformation} \label{subsection:QSVT}
Quantum singular value transformation (QSVT)~\cite{gilyen2019quantum,martyn2021grand} is a quantum algorithm for applying
a polynomial transformation $P^{(\mathrm{SV})}(A)$ to the singular values of a given matrix $A$.
As mentioned above, QSVT has been applied to many problems, incluing HS.
In these problems, a degree-$d$ polynomial $P_{\varepsilon}$ is used to
$\varepsilon$-approximate the corresponding objective function $P$. 
How much quantum speedup is obtained depends on the degree $d$.
Recall that QSVT generalizes the result of QSP~\cite{low2016methodology, low2017hamiltonian,low2017optimal,low2019hamiltonian}.
We present a brief description of the derivation from QSP to QSVT in Appendix~\ref{appendix:from_QSP_to_QSVT}.

We introduce the block-encoding~\cite{gilyen2019quantum}, which represents a matrix $A$ as
the upper-left block of a unitary matrix $U$. Let $A$ be a matrix acting on $s$ qubits,
$U$ be a unitary matrix acting on $a+s$ qubits. Then, for $\alpha > 0$ and $\varepsilon > 0$,
$U$ is called an $(\alpha, a, \varepsilon)$-block-encoding of $A$, if 
\begin{equation}
  \left\|
    A - \alpha(\bra{0}_a\otimes I)U(\ket{0}_a\otimes I)
  \right\| \leq \varepsilon,
\end{equation}
where $\ket{0}_a = \ket{0}^{\otimes a}$. Note that, since $\|U\|=1$, we necessarily have $\|A\| \leq \alpha+\varepsilon$.
If $\varepsilon = 0$, then we can represent $A$ as the upper-left block of $U$:
\begin{equation}
  \label{eq:definition_of_block_encoding} 
  U = \begin{bmatrix}
    \frac{A}{\alpha} & \cdot \\
    \cdot            & \cdot
  \end{bmatrix},
\end{equation}
where the dot . denotes a matrix with arbitrary elements.

Recall now that the singular value decomposition of $A$; that is,
any matrix $A\in \mathbb{C}^{m\times n}$ can be decomposed as
\begin{equation}
  A = W\Sigma V,
\end{equation}
where $W\in\mathbb{C}^{m\times m}$ and $V\in\mathbb{C}^{n\times n}$ are unitary matrices;
$\Sigma$ is diagonal and contains the set of non-negative real numbers $\{\sigma_k\}$, called
the singular values of $A$. The matrix $A$ is also expressed as
\begin{equation}
  A = \sum_{k=1}^{r}\sigma_k\ket{w_k}\bra{v_k},
\end{equation}
where $\{\ket{w_k}\}$ and $\{\ket{v_k}\}$ are right and left singular vectors,
and $r=\mathrm{rank}(A)$.

The singular value transformation is defined from the singular value decomposition as follows:
for an odd polynomial $P\in\mathbb{C}$,
\begin{equation}
  P^{(\mathrm{SV})}(A) \equiv \sum_{k}P(\sigma_k)\ket{w_k}\bra{v_k},
\end{equation}
and for an even polynomial $P\in\mathbb{C}$,
\begin{equation}
  P^{(\mathrm{SV})}(A) \equiv \sum_{k}P(\sigma_k)\ket{v_k}\bra{v_k}.
\end{equation}
If $A$ is Hermitian and positive semidefinite, then $P^{(\mathrm{SV})}(A)$ is equal
to the eigenvalue transformation $P(A)$.

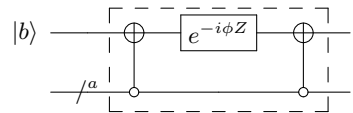
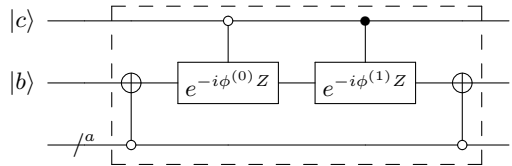
\begin{figure}[tbp]
  \begin{minipage}[c]{\linewidth}
    \[\Qcircuit @C=1.5em @R=1.5em {
      \lstick{\ket{b}} & \qw        & \targ      & \gate{e^{-i\phi Z}} & \targ      & \qw \\
                      &  {/}^a \qw & \ctrlo{-1} & \qw                 & \ctrlo{-1} & \qw \gategroup{1}{3}{2}{5}{1.2em}{--}
    }\]
    \subcaption{$\ket{b}\bra{b}\otimes e^{(-1)^b i\phi\Pi}$}
    \label{fig:circuit_of_exp_single_phi}
  \end{minipage} \\
  \begin{minipage}[c]{\linewidth}
    \[\Qcircuit @C=1.5em @R=1.5em {
      \lstick{\ket{c}} & \qw        & \qw        & \ctrlo{1}                 & \ctrl{1}                  & \qw        & \qw \\
      \lstick{\ket{b}} & \qw        & \targ      & \gate{e^{-i\phi^{(0)} Z}} & \gate{e^{-i\phi^{(1)} Z}} & \targ      & \qw \\
                      &  {/}^a \qw & \ctrlo{-1} & \qw                       & \qw                       & \ctrlo{-1} & \qw 
                      \gategroup{1}{3}{3}{6}{1.2em}{--}
    }\]
    \subcaption{$\ket{cb}\bra{cb}\otimes e^{(-1)^b i\phi^{(c)}\Pi}$}
    \label{fig:circuit_of_exp_multiple_phi}
  \end{minipage}
  \caption{Quantum circuits used to implement the unitary matrix $\exp(i\phi\Pi)$.
            In Fig.~\ref{fig:gate_used_for_QSVT}(a), a single phase $\phi$ is used,
            and the series of gates surrounded by dashed lines is denoted by $S_1(\phi)$,
            which is used in Fig.~\ref{fig:circuit_of_U_OAA} and~\ref{fig:circuit_of_U_FPAA}.
            In Fig.~\ref{fig:gate_used_for_QSVT}(b), two phases $\phi^{(0)}$ and $\phi^{(1)}$ are used,
            and the series of gates surrounded by dashed lines is denoted by $S_2(\phi^{(0)},\phi^{(1)})$,
            which is used in Fig.~\ref{fig:circuit_of_U_exp}.}
  \label{fig:gate_used_for_QSVT}
\end{figure}

Suppose that $U$ is an $(\alpha, a, 0)$-block-encoding of $A$ as in Eq.~\eqref{eq:definition_of_block_encoding}.
Then, a unitary matrix $U_{\Phi}$ called the alternating phase modulation sequence
in Ref.~\cite{gilyen2019quantum} is defined as follows: for odd $d$,
\begin{equation} \label{eq:alternating_phase_modulation_sequence_for_odd_polynomial}
  U_{\Phi} \equiv e^{i\phi_0\Pi} U e^{i\phi_1\Pi}\prod_{k=1}^{(d-1)/2}\left(
      U^{\dagger} e^{i\phi_{2k}\Pi}U e^{i\phi_{2k+1}\Pi}
      \right),
\end{equation}
and for even $d$,
\begin{equation} \label{eq:alternating_phase_modulation_sequence_for_even_polynomial}
  U_{\Phi} \equiv e^{i\phi_0\Pi}\prod_{k=1}^{d/2}\left(
      U^{\dagger} e^{i\phi_{2k-1}\Pi} U e^{i\phi_{2k}\Pi}
    \right),
\end{equation}
where $\Phi=\{\phi_0, \phi_1, \cdots, \phi_{d}\}\in\mathbb{R}^{d+1}$ is called the \textit{phases} and
$\Pi = 2\ket{0}_a\bra{0} - I$. The unitary matrix $\exp(i\phi\Pi)$ can be
implemented as in Fig.~\ref{fig:gate_used_for_QSVT}.
The phases are calculated efficiently from the degree-$d$ polynomial $P$ on a classical computer.
The details of the calculation can be found in Refs.~\cite{gilyen2019quantum, chao2020finding}.
In this work, we use the code provided in Ref.~\cite{chao2021finding} to calculate the phases.

Given a degree-$d$ polynomial $P\in\mathbb{C}$ and the corresponding phases $\Phi\in\mathbb{R}^{d+1}$, 
the unitary $U_{\Phi}$ can be represented as a $(1,a,0)$-block-encoding of $P^{(\mathrm{SV})}(A/\alpha)$:
\begin{equation}
  \begin{split}
    U_{\Phi} &= \begin{bmatrix}
      P^{(\mathrm{SV})}\left(
        \frac{A}{\alpha}
      \right) & \cdot \\
      \cdot            & \cdot
    \end{bmatrix},\\
    P^{(\mathrm{SV})}\left(
      \frac{A}{\alpha}
    \right) &= (\bra{0}_a\otimes I)U_{\Phi}(\ket{0}_a\otimes I).
  \end{split}
\end{equation}
Note that $P$ is not arbitrary and has some constraints. $P$ satisfies the following conditions
\cite{gilyen2019quantum,martyn2021grand}:
\begin{itemize}
  \item[(i)] $P$ has parity $d \bmod{2}$
  \item[(ii)] $\forall x\in[-1,1]: |P(x)|\leq 1$
  \item[(iii)] $\forall x\in(-\infty,-1]\cup [1,\infty): |P(x)|\geq 1$
  \item[(iv)] if $d$ is even, then $\forall x\in\mathbb{R}: P(ix)P^{*}(ix)\geq 1$.
\end{itemize}
These conditions are complicated, but taking the real part of $P$ relaxes them.
That is, $P_{\mathbb{\Re}}\equiv \mathrm{Re}(P)$ satisfies the following conditions:
\begin{itemize}
  \item[(v)] $P_{\mathbb{\Re}}$ has parity $d \bmod{2}$
  \item[(vi)] $\forall x\in[-1,1]: |P_{\mathbb{\Re}}(x)|\leq 1$,
\end{itemize}
and the corresponding phases can be calculated by~\cite{gilyen2019quantum,martyn2021grand}

\subsection{Applying QSVT to trigonometric functions} \label{subsection:applying_QSVT_to_trigonometric_functions}
Let $U$ be a $(1,a,0)$-block-encoding of $H$:
\begin{equation} \label{eq:block_encoding_of_H} 
  U = \begin{bmatrix}
    H     & \cdot \\
    \cdot & \cdot
  \end{bmatrix}, \quad H = (\bra{0}_a\otimes I)U(\ket{0}_a\otimes I),
\end{equation}
where $\|H\| \leq 1$ is a Hermitian matrix that is positive semidefinite.
We will discuss later the case $H$ is negative and normalized by $\alpha$.
The goal of HS is to construct a quantum circuit $U_{\mathrm{HS}}$
that is a block-encoding of $\exp(-iHt)$ using the unitary $U$, where $t$ is a evolution time.
Note that $U_{\mathrm{HS}}$ cannot be realized single $U_{\Phi}$ using QSVT,
because $\exp(-ixt)$ has no definite parity. To avoid this problem,
one can instead apply QSVT to two different functions: $\cos(xt)$ and $\sin(xt)$.

The functions $\cos(xt)$ and $\sin(xt)$ are given by the Jacobi-Anger expansion:
\begin{align}
  \cos(xt) &= J_0(t) + 2\sum_{k=1}^{\infty}(-1)^k J_{2k}(t)T_{2k}(x), \label{eq:jacobi_anger_expansion_cos} \\
  \sin(xt) &= 2\sum_{k=0}^{\infty}(-1)^k J_{2k+1}(t)T_{2k+1}(x),      \label{eq:jacobi_anger_expansion_sin}
\end{align}
where $J_m(t)$ is the $m$-th Bessel function of the first kind and $T_{k}(x)$ is the $k$-th Chebyshev polynomial
of the first kind. 
One can obtain $\varepsilon_{\mathrm{tri}}$-approximation to $\cos(xt)$ and $\sin(xt)$ by truncating 
Eqs.~\eqref{eq:jacobi_anger_expansion_cos} and~\eqref{eq:jacobi_anger_expansion_sin} at an index $R$:
\begin{align}
  \left|
    \cos(xt) - J_0(t) - 2\sum_{k=1}^{R}(-1)^k J_{2k}(t)T_{2k}(x)
  \right| &\leq \varepsilon_{\mathrm{tri}}, \label{eq:error_jacobi_anger_expansion_cos} \\
  \left|
    \sin(xt) - 2\sum_{k=0}^{R}(-1)^k J_{2k+1}(t)T_{2k+1}(x)
  \right| &\leq \varepsilon_{\mathrm{tri}}, \label{eq:error_jacobi_anger_expansion_sin}
\end{align}
where $0<\varepsilon_{\mathrm{tri}}<1/e$ and 
\begin{align} 
  R(t,\varepsilon_{\mathrm{tri}}) &= \left\lfloor
    \frac{1}{2}r\left(
      \frac{et}{2},\frac{5}{4}\varepsilon_{\mathrm{tri}}
    \right)
  \right\rfloor, \label{eq:degree_R} \\
  r(t,\varepsilon_{\mathrm{tri}}) &= \Theta\left(
    t + \frac{\log(\frac{1}{\varepsilon_{\mathrm{tri}}})}{\log\left(
      e+\log(\frac{1}{\varepsilon_{\mathrm{tri}}})/t
    \right)}
  \right) \notag \\
  &\leq \mathcal{O}\left(
    t + \log(1/\varepsilon_{\mathrm{tri}})
  \right). \label{eq:r}
\end{align}
For more details see Ref.~\cite{gilyen2019quantum}.

We denote the approximate polynomials of Eqs.~\eqref{eq:error_jacobi_anger_expansion_cos} and
\eqref{eq:error_jacobi_anger_expansion_sin} by $P_{\varepsilon_{\mathrm{tri}}}^{\cos}(x)$ and $P_{\varepsilon_{\mathrm{tri}}}^{\sin}(x)$.
Since cosine and sine are bounded in magnitude by $1$, these polynomials obey
$|P_{\varepsilon_{\mathrm{tri}}}^{\cos}(x)|, |P_{\varepsilon_{\mathrm{tri}}}^{\sin}(x)| \leq 1+\varepsilon_{\mathrm{tri}}$. Therefore,
the condition (vi) is violated. Here,
we introduce rescaled polynomials 
\begin{equation} \label{eq:kappa_approximate_polynomials}
  P_{\varepsilon_{\mathrm{tri}}, \kappa}^{\cos}(x) = \kappa P_{\varepsilon_{\mathrm{tri}}}^{\cos}(x), \quad
  P_{\varepsilon_{\mathrm{tri}}, \kappa}^{\sin}(x) = \kappa P_{\varepsilon_{\mathrm{tri}}}^{\sin}(x),
\end{equation}
where $\kappa=1/(1+\varepsilon_{\mathrm{tri}})$.
These polynomials satisfy the following inequality:
\begin{align}
  &\left|
    \frac{\kappa}{2}e^{-ixt}
      - \frac{P_{\varepsilon_{\mathrm{tri}}, \kappa}^{\cos}(x)-iP_{\varepsilon_{\mathrm{tri}}, \kappa}^{\sin}(x)}{2} 
  \right| \notag \\
  &  \quad \leq \frac{1}{2}\left|
    \kappa\cos (xt) - P_{\varepsilon_{\mathrm{tri}}, \kappa}^{\cos}(x)
  \right| + \frac{1}{2}\left|
    \kappa\sin (xt) - P_{\varepsilon_{\mathrm{tri}}, \kappa}^{\sin}(x)
  \right| \notag \\
  & \quad \leq \frac{\kappa\varepsilon_{\mathrm{tri}}+\kappa\varepsilon_{\mathrm{tri}}}{2} = \kappa\varepsilon_{\mathrm{tri}},
  \label{eq:rescaled_polynomials_inequality}
\end{align}
where in the first inequality we used the triangle inequality.

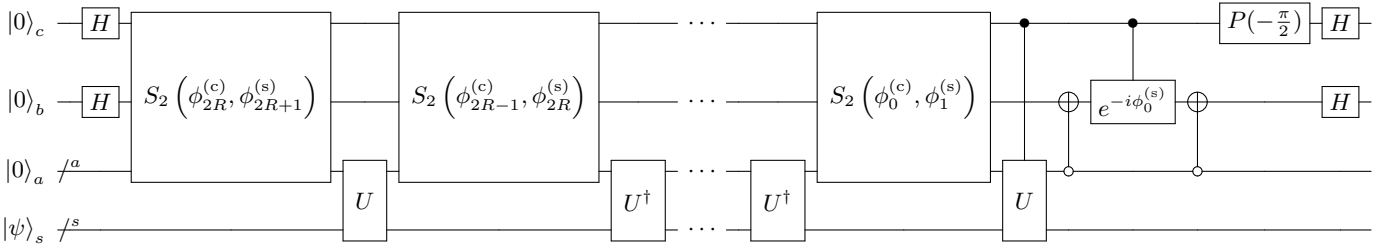
\begin{figure*}[tbp]
  \[\Qcircuit @C=0.5em @R=1.5em {
    \lstick{\ket{0}_c}    & \qw        & \gate{H} & \multigate{2}{S_2\left(\phi_{2R}^{(\mathrm{c})},\phi_{2R+1}^{(\mathrm{s})}\right)} & \qw              & \multigate{2}{S_2\left(\phi_{2R-1}^{(\mathrm{c})},\phi_{2R}^{(\mathrm{s})}\right)}
        & \qw                        & \qw & & \cdots & & & \qw & \qw                        & \multigate{2}{S_2\left(\phi_{0}^{(\mathrm{c})},\phi_{1}^{(\mathrm{s})}\right)} & \ctrl{2}         & \qw        & \ctrl{1}                    
        & \qw        & \gate{P(-\frac{\pi}{2})} & \gate{H} & \qw \\
    \lstick{\ket{0}_b}    & \qw        & \gate{H} & \ghost{S_2\left(\phi_{2R}^{(\mathrm{c})},\phi_{2R+1}^{(\mathrm{s})}\right)}        & \qw              & \ghost{S_2\left(\phi_{2R-1}^{(\mathrm{c})},\phi_{2R}^{(\mathrm{s})}\right)}
        & \qw                        & \qw & & \cdots & & & \qw & \qw                        & \ghost{S_2\left(\phi_{0}^{(\mathrm{c})},\phi_{1}^{(\mathrm{s})}\right)}        & \qw              & \targ      & \gate{e^{-i\phi_{0}^{(\mathrm{s})}}} 
        & \targ      & \qw                      & \gate{H} & \qw \\
    \lstick{\ket{0}_a}    &  {/}^a \qw & \qw      & \ghost{S_2\left(\phi_{2R}^{(\mathrm{c})},\phi_{2R+1}^{(\mathrm{s})}\right)}        & \multigate{1}{U} & \ghost{S_2\left(\phi_{2R-1}^{(\mathrm{c})},\phi_{2R}^{(\mathrm{s})}\right)} 
        & \multigate{1}{U^{\dagger}} & \qw & & \cdots & & & \qw & \multigate{1}{U^{\dagger}} & \ghost{S_2\left(\phi_{0}^{(\mathrm{c})},\phi_{1}^{(\mathrm{s})}\right)}        & \multigate{1}{U} & \ctrlo{-1} & \qw                         
        & \ctrlo{-1} & \qw                      & \qw      & \qw \\
    \lstick{\ket{\psi}_s} & {/}^s \qw  & \qw      & \qw                                                                    & \ghost{U}        &  \qw
        & \ghost{U^{\dagger}}        & \qw & & \cdots & & & \qw & \ghost{U^{\dagger}}        & \qw                                                 & \ghost{U}        & \qw        & \qw                         
        & \qw        & \qw                      & \qw      & \qw       
  }\]
  \caption{Quantum circuit $U_{\exp}$ that is a $(1,a+2,\kappa\varepsilon_{\mathrm{tri}})$-block-encoding
            of $\kappa \exp(-iHt)/2$. The gate $S_2$ is shown in Fig.~\ref{fig:gate_used_for_QSVT}(b),
            the unitary $U$ is given by Eq.~\eqref{eq:block_encoding_of_H}, and
            the phases $\Phi^{(\mathrm{c})}\in\mathbb{R}^{2R+1}$ and $\Phi^{(\mathrm{s})}\in\mathbb{R}^{2R+2}$
            are calculated from the $2R$-th polynomial $P_{\varepsilon_{\mathrm{tri}}, \kappa}^{\cos}(x)$
            and $(2R+1)$-th polynomial $P_{\varepsilon_{\mathrm{tri}}, \kappa}^{\sin}(x)$
            in Eq.~\eqref{eq:kappa_approximate_polynomials}, where $R$ is given by Eq.~\eqref{eq:degree_R}.}
  \label{fig:circuit_of_U_exp}
\end{figure*}

Suppose the phases $\Phi^{(\mathrm{c})}\in\mathbb{R}^{2R+1}$ and
$\Phi^{(\mathrm{s})}\in\mathbb{R}^{2R+2}$ are calculated from
the $2R$-th polynomial $P_{\varepsilon_{\mathrm{tri}}, \kappa}^{\cos}(x)$
and $(2R+1)$-th polynomial $P_{\varepsilon_{\mathrm{tri}}, \kappa}^{\sin}(x)$.
The quantum circuit $U_{\exp}$ using these phases is shown in Fig.~\ref{fig:circuit_of_U_exp}. 
This circuit constructs the $(1,a+2,\kappa\varepsilon_{\mathrm{tri}})$-block-encoding of $\kappa e^{-iHt}/2$:
\begin{equation} \label{eq:U_exp}
  \left\|
    \frac{\kappa}{2}e^{-iHt} - (\bra{0}_{abc}\otimes I)U_{\exp}(\ket{0}_{abc}\otimes I)
  \right\| \leq \kappa\varepsilon_{\mathrm{tri}},
\end{equation}
with $R$ uses of $U$ and $U^{\dagger}$, and one use of the controlled-$U$.
Therefore, the query complexity of $U_{\exp}$ is
\begin{align} \label{eq:query_complexity_of_U_exp}
  R+R+1 &= 2\left\lfloor
      \frac{1}{2}r\left(
        \frac{et}{2},\frac{5}{4}\varepsilon_{\mathrm{tri}}
      \right)
    \right\rfloor + 1 \\
    &\leq \mathcal{O}\left(
      t + \log(1/\varepsilon_{\mathrm{tri}})
    \right).
\end{align}
To obtain the block-encoding of $\exp(-iHt)$, the amplitude amplification must be used.
In Sec.~\ref{section:HS_using_QSVT}, we discuss two types of the QSVT-based amplitude amplification algorithms:
the oblivious amplitude amplification (OAA)~\cite{gilyen2019quantum} and 
fixed-point amplitude amplification (FPAA)~\cite{martyn2021grand}.

\subsection{The linearized Vlasov-Poisson system} \label{subsection:vlasov_poisson_system}
The time evolution of the distribution function $f(\bm{x}, \bm{v}, t)$ for electrons with stationary ions and
the electric field $\bm{E}=(E_x,E_y,E_z)$ governed by the Vlasov-Poisson system
is described by the following equations:
\begin{equation}
  \frac{\partial f}{\partial t} + \bm{v}\cdot \nabla f
    - \frac{e}{m}\bm{E}\cdot \frac{\partial f}{\partial \bm{v}} = 0, \label{eq:vlasov_equation}
\end{equation}
\begin{equation}
  \frac{\partial \bm{E}}{\partial t} = \frac{1}{\varepsilon_0}\int e\bm{v}f d\bm{v}, \label{eq:amperes_maxwell_equation}
\end{equation}
where $e$ is the absolute value of the electron charge, $m$ is the electron mass,
and $\varepsilon_0$ is the permittivity of the vacuum.
The variables $f$ and $\bm{E}$ are expanded into the equilibrium
terms (labeled by 0) and perturbations (labeled by 1) to linearize Eq.~\eqref{eq:vlasov_equation}:
\begin{equation}
  \begin{split}
    f(\bm{x},\bm{v}, t) &= f_0(\bm{v}) + f_1(\bm{x},\bm{v}, t), \\
    \bm{E} &= \bm{E}_1(\bm{x},t).
  \end{split}
\end{equation}
Note that we do not deal with the case when the nonzero electric field $\bm{E}_0$ increases the system's energy,
i.e., $\bm{E}_0=0$.

We assume a Maxwellian background distribution $f_0=f_{\mathrm{M}}$
and apply the same transformations as in Ref.~\cite{engel2019quantum}: a Fourier transformation of the variables
in space, change of variables, and discretization in velocity space with the following dimensionless variables:
\begin{equation} \label{eq:dimensionless_vaiables1}
  \begin{split}
    \hat{\bm{k}} = \lambda_{D_e} \bm{k}, \quad \hat{t} = \omega_{pe} t, \quad \hat{\bm{v}} = \frac{\bm{v}}{\lambda_{D_e}\omega_{pe}}, \\
    \hat{f} = \frac{(\lambda_{D_e}\omega_{pe})^3}{n_e}f, \quad \hat{\bm{E}} = \frac{e\lambda_{D_e}}{k_{\mathrm{B}}T_e}\bm{E},
  \end{split}
\end{equation}
where $\bm{k}=(k_x,k_y,k_z)$ is the wave vector for the Fourier transformation, 
$\lambda_{D_e}$ is the Debye length with ions neglected, $\omega_{pe}$ is the electron plasma frequency,
$n_e$ is the electron number density, $T_e$ is the electron temperature,
and $k_{\mathrm{B}}$ is the Boltzmann constant.
As a result, Eqs.~\eqref{eq:vlasov_equation} and~\eqref{eq:amperes_maxwell_equation} becomes
\begin{align}
  \begin{split}
    \frac{d F_{\bm{j}}}{d t} 
      &= - i(k_x v_{j_x}+k_y v_{j_y}+k_z v_{j_z})F_{\bm{j}} \\
      & \qquad - i\mu_{\bm{j}}(v_{j_x} E_x + v_{j_y} E_y + v_{j_z} E_z),
  \end{split} \label{eq:changed_3D_vlasov_equation} \\
  \frac{d E_p}{d t} 
      &= -i\sum_{\bm{j}} \mu_{\bm{j}} v_{j_{p}} F_{\bm{j}} \qquad (p=x,y,z), \label{eq:changed_3D_amperes_maxwell_equation}
\end{align}
where the subscripts 1 have been dropped,
\begin{align}
  \mu_{\bm{j}} &= \mu(v_{j_x}, v_{j_y}, v_{j_z}) = \sqrt{\Delta v f_{\mathrm{M}}(v_{j_x}, v_{j_y}, v_{j_z})}, \\
  \begin{split}
    F_j &= F(v_{j_x}, v_{j_y}, v_{j_z},t) \\
        &= i\sqrt{\frac{\Delta v}{f_{\mathrm{M}}(v_{j_x}, v_{j_y}, v_{j_z})}}f(v_{j_x}, v_{j_y}, v_{j_z}, t),
  \end{split} \\
  \sum_{\bm{j}}\cdot &= \sum_{j_x=0}^{N_{v_x}-1}\sum_{j_y=0}^{N_{v_y}-1}\sum_{j_z=0}^{N_{v_z}-1}\cdot,
\end{align}
where $\Delta v=\Delta v_x\Delta v_y\Delta v_z$ is the product of the mesh sizes,
$N_{v_x}=2^{n_{v_x}},N_{v_y}=2^{n_{v_y}}$ and $N_{v_z}=2^{n_{v_z}}$ are the grid sizes
in velocity space, and the velocity space grid is represented by index $\bm{j}=(j_x, j_y, j_z)$.

Equations~\eqref{eq:changed_3D_vlasov_equation} and~\eqref{eq:changed_3D_amperes_maxwell_equation} can be
rewritten in a form of the Schr\"{o}dinger equation:
\begin{equation} \label{eq:schrodinger_equation_form}
  \frac{d\ket{\psi(t)}}{dt} = -i H \ket{\psi(t)},
\end{equation}
where $H$ is a time-independent Hamiltonian and $\ket{\psi(t)}$ is a quantum state whose amplitudes
are the variables, which is written in bra-ket notation as
\begin{align} \label{eq:psi}
  \begin{split}
    \ket{\psi(t)} = \frac{1}{\eta}\Biggl(
      &\sum_{\bm{j}} F_{\bm{j}}\ket{0}_r\ket{\bm{j}}_v + E_x\ket{1}_r\ket{\bm{0}}_v\\
      &\qquad + E_y\ket{2}_r\ket{\bm{0}}_v + E_y\ket{3}_r\ket{\bm{0}}_v
    \Biggr),
  \end{split}
\end{align}
where $\ket{\bm{j}}_v=\ket{j_x}_{v_x}\ket{j_y}_{v_y}\ket{j_z}_{v_z}, \ket{\bm{0}}_v=\ket{0}_{v_x}\ket{0}_{v_y}\ket{0}_{v_z}$
and the normalization constant
\begin{equation}
  \eta = \sqrt{\sum_{\bm{j}}|F_{\bm{j}}|^{2} + |E_x|^2 + |E_y|^2+ |E_z|^2}.  
\end{equation}
$\ket{\psi(t)}$ has two registers labeled by $r$ and $v$. The $r$ register 
encodes the variable index: $\ket{0}_r \leftrightarrow F, \ket{1}_r \leftrightarrow E_x, \ket{2}_r \leftrightarrow E_y$ and
$\ket{3}_r \leftrightarrow E_z$. The $v$ register stores the velocity space dependence of $F$:
$\ket{0}_r\ket{\bm{j}}_v \leftrightarrow F_{\bm{j}}$. The corresponding Hamiltonian $H$,
which acts on these registers, is given by
\begin{equation} \label{eq:3D_hamiltonian1}
  \begin{split}
    H &= \sum_{\bm{j}}\Bigl[
      (k_x v_{j_x}+k_y v_{j_y}+k_z v_{j_z})\ket{0}_r\ket{\bm{j}}_v\bra{0}_r\bra{\bm{j}}_v \\
      & \qquad + \mu_{\bm{j}}v_{j_x}\left(
          \ket{0}_r\ket{\bm{j}}_v\bra{1}_r\bra{\bm{0}}_v + \ket{1}_r\ket{\bm{0}}_v\bra{0}_r\bra{\bm{j}}_v
        \right) \\
      & \qquad + \mu_{\bm{j}}v_{j_y}\left(
          \ket{0}_r\ket{\bm{j}}_v\bra{2}_r\bra{\bm{0}}_v + \ket{2}_r\ket{\bm{0}}_v\bra{0}_r\bra{\bm{j}}_v
        \right) \\
      & \qquad + \mu_{\bm{j}}v_{j_z}\left(
        \ket{0}_r\ket{\bm{j}}_v\bra{3}_r\bra{\bm{0}}_v + \ket{3}_r\ket{\bm{0}}_v\bra{0}_r\bra{\bm{j}}_v
      \right)
  \Bigr].
  \end{split}
\end{equation}
The solution of Eq.~\eqref{eq:schrodinger_equation_form} is given by
\begin{equation}
  \ket{\psi(t)} = e^{-iHt}\ket{\psi(t=0)}.
\end{equation}
Therefore, the time evolution of Eqs.~\eqref{eq:changed_3D_vlasov_equation} and~\eqref{eq:changed_3D_amperes_maxwell_equation}
can be computed by HS.

\section{QSVT-based Hamiltonian simulation} \label{section:HS_using_QSVT}
\subsection{Oblivious amplitude amplification} \label{subsection:oblivious_amplitude_amplification_by_QSVT}
\begin{figure*}[t]
  \[\Qcircuit @C=1em @R=1.5em {
    \lstick{\ket{0}_d} & \qw        & \multigate{3}{S_1\left(\phi_{3}^{(\mathrm{OAA})}\right)} & \qw              & \multigate{3}{S_1\left(\phi_{2}^{(\mathrm{OAA})}\right)}
        & \qw                               & \multigate{3}{S_1\left(\phi_{1}^{(\mathrm{OAA})}\right)} & \qw                     & \multigate{3}{S_1\left(\phi_{0}^{(\mathrm{OAA})}\right)} & \gate{X} & \gate{Z} & \gate {X} & \rstick{\ket{0}_d} \qw \\
    \lstick{\ket{0}_c} & \qw        & \ghost{S_1\left(\phi_{3}^{(\mathrm{OAA})}\right)}        & \multigate{3}{U_{\exp}} & \ghost{S_1\left(\phi_{2}^{(\mathrm{OAA})}\right)}
        & \multigate{3}{U_{\exp}^{\dagger}} & \ghost{S_1\left(\phi_{1}^{(\mathrm{OAA})}\right)}        & \multigate{3}{U_{\exp}} & \ghost{S_1\left(\phi_{0}^{(\mathrm{OAA})}\right)}        & \qw      & \qw      & \qw       & \qw \\
    \lstick{\ket{0}_b} & \qw        & \ghost{S_1\left(\phi_{3}^{(\mathrm{OAA})}\right)}        & \ghost{U_{\exp}}              & \ghost{S_1\left(\phi_{2}^{(\mathrm{OAA})}\right)}
        & \ghost{U_{\exp}^{\dagger}}        & \ghost{S_1\left(\phi_{1}^{(\mathrm{OAA})}\right)}        & \ghost{U_{\exp}}        & \ghost{S_1\left(\phi_{0}^{(\mathrm{OAA})}\right)}        & \qw      & \qw      & \qw       & \qw \\
    \lstick{\ket{0}_a} &  {/}^a \qw & \ghost{S_1\left(\phi_{3}^{(\mathrm{OAA})}\right)}        & \ghost{U_{\exp}} & \ghost{S_1\left(\phi_{2}^{(\mathrm{OAA})}\right)} 
        & \ghost{U_{\exp}^{\dagger}}        & \ghost{S_1\left(\phi_{1}^{(\mathrm{OAA})}\right)}        & \ghost{U_{\exp}}        & \ghost{S_1\left(\phi_{0}^{(\mathrm{OAA})}\right)}        & \qw      & \qw      & \qw       & \qw \\
                       & {/}^s \qw  & \qw                                          & \ghost{U_{\exp}}        &  \qw
        & \ghost{U_{\exp}^{\dagger}}        & \qw                                                      & \ghost{U_{\exp}}        & \qw                                                      & \qw      & \qw      & \qw       & \qw  
  }\]
  \caption{Quantum circuit $U_{\mathrm{OAA}}$ that is a $(1,a+2,\varepsilon)$-block-encoding of $\exp(-iHt)$.
            The unitary $U_{\exp}$ is shown in Fig.~\ref{fig:circuit_of_U_exp},
            the gate $S_1$ is shown in Fig.~\ref{fig:gate_used_for_QSVT}(a), and
            the phases $\Phi^{(\mathrm{OAA})}\in\mathbb{R}^{4}$ is given by Eq.~\eqref{eq:OAA_phases}.}
  \label{fig:circuit_of_U_OAA}
\end{figure*}
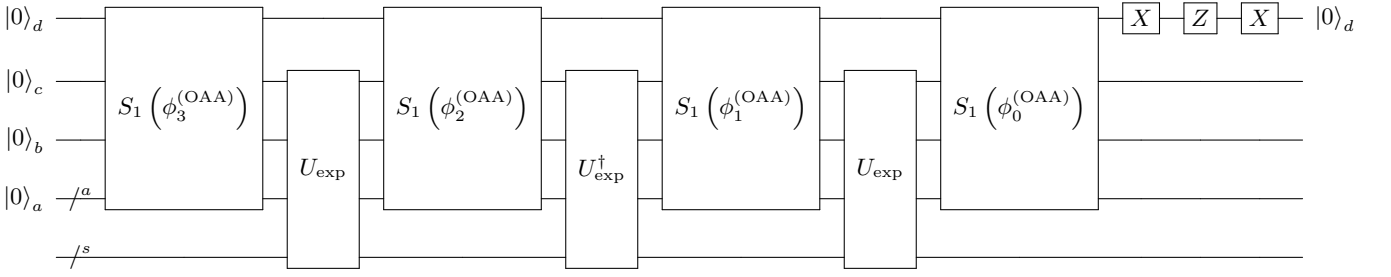

\begin{figure}[tbp]
  \begin{algorithm}[H]
    \caption{The OAA-based Hamiltonian simulation} \label{algorithm:HS_using_OAA}
    \begin{algorithmic}[1]
      \renewcommand{\algorithmicrequire}{\textbf{Input:}}
      \renewcommand{\algorithmicensure}{\textbf{Output:}}
      \REQUIRE a $(1, a, 0)$-block-encoding of a Hamiltonian matrix $H$,
              an evolution time $t$, and an error tolerance $\varepsilon$.
      \ENSURE a $(1, a+2, \varepsilon)$-block-encoding of $e^{-iHt}$
      \renewcommand{\algorithmicrequire}{\textbf{Runtime:}}
      \renewcommand{\algorithmicensure}{\textbf{Procedure:}}
      \REQUIRE $Q_{\mathrm{HS}}^{(\mathrm{OAA})}$ queries to the block-encoding of $H$,
               where $Q_{\mathrm{HS}}^{(\mathrm{OAA})}$ is given by Eq.~\eqref{eq:query_complexity_of_HS_using_OAA}
      \ENSURE
      \STATE Calculate the phases $\Phi^{(\mathrm{c})}\in\mathbb{R}^{2R+1}$ and $\Phi^{(\mathrm{s})}\in\mathbb{R}^{2R+2}$
            on a classical computer from the $2R$-th polynomial $P_{\varepsilon_{\mathrm{tri}}, \kappa}^{\cos}(x)$ and
            $(2R+1)$-th polynomial $P_{\varepsilon_{\mathrm{tri}}, \kappa}^{\sin}(x)$ in Eq.~\eqref{eq:kappa_approximate_polynomials},
            where $\kappa=1/(1+\varepsilon_{\mathrm{tri}})$, $\varepsilon_{\mathrm{tri}}=\varepsilon/9$ and
            $R$ is given by Eq.~\eqref{eq:degree_R}. 
      \STATE Construct the circuit $U_{\mathrm{exp}}$ in Fig.~\ref{fig:circuit_of_U_exp}
             using the phases $\Phi^{(\mathrm{c})}$ and $\Phi^{(\mathrm{s})}$, which is
             a $(1,a+2,\kappa\varepsilon_{\mathrm{tri}})$-block-encoding of $\kappa e^{-iHt}/2$.
      \STATE Run the circuit $U_{\mathrm{OAA}}$ in Fig.~\ref{fig:circuit_of_U_OAA}
             using the phases $\Phi^{(\mathrm{OAA})}\in\mathbb{R}^{4}$ in Eq.~\eqref{eq:OAA_phases}.
    \end{algorithmic}
  \end{algorithm}
\end{figure}

Oblivious amplitude amplification (OAA) using QSVT has been proposed in Ref.~\cite{gilyen2019quantum}.
In this section, we show the circuit of OAA and discuss the error and number of queries of the OAA-based HS.
In the QSVT-based OAA, the $d$-th Chebyshev polynomial of the first kind, defined by $T_{d}(x)=\cos(d\arccos(x))$,
is used as an objective function. For odd $d$, the corresponding phases $\Phi\in\mathbb{R}^{d+1}$ is given by 
\begin{equation}
  \begin{cases}
    \phi_{0} = -\frac{d\pi}{2} \\
    \phi_{k} = \frac{\pi}{2} & (k=1,2,\ldots,d).
  \end{cases}
\end{equation}

From Eq.~\eqref{eq:rescaled_polynomials_inequality}, the following inequality holds:
\begin{align}
  &\left|
    \frac{e^{-ixt}}{2}
      - \frac{P_{\varepsilon_{\mathrm{tri}}, \kappa}^{\cos}(x)-iP_{\varepsilon_{\mathrm{tri}}, \kappa}^{\sin}(x)}{2} 
  \right| \notag \\
  &  \quad \leq \left|
    \frac{\kappa\varepsilon_{\mathrm{tri}}e^{-ixt}}{2}
  \right| + \left|
    \frac{\kappa e^{-ixt}}{2}-\frac{P_{\varepsilon_{\mathrm{tri}}, \kappa}^{\cos}(x)-iP_{\varepsilon_{\mathrm{tri}}, \kappa}^{\sin}(x)}{2} 
  \right| \notag \\
  & \quad \leq \frac{\kappa\varepsilon_{\mathrm{tri}}}{2}+\kappa\varepsilon_{\mathrm{tri}} \notag \\
  & \quad = \frac{3\varepsilon_{\mathrm{tri}}}{2(1+\varepsilon_{\mathrm{tri}})}<\frac{3}{2}\varepsilon_{\mathrm{tri}},
  \label{eq:rescaled_polynomials_inequality_2}
\end{align}
where in the first inequality we used the triangle inequality.
Letting $U$ be $U_{\exp}$ in Eq.~\eqref{eq:alternating_phase_modulation_sequence_for_odd_polynomial} and
using the following phases:
\begin{equation} \label{eq:OAA_phases}
  \begin{cases}
    \phi_{0}^{(\mathrm{OAA})} = -\frac{3\pi}{2} \\
    \phi_{k}^{(\mathrm{OAA})} = \frac{\pi}{2} & (k=1,2,3),
  \end{cases}
\end{equation}
then one can get the block-encoding of
\begin{equation}
  T_{3}\left(\frac{e^{-iHt}}{2}\right)=T_{3}\left(\cos\left(\frac{\pi}{3}\right)\right)e^{-iHt}=-e^{-iHt},
\end{equation}
where in the first equality we used that the singular value of a unitary matrix is 1.
OAA multiplies the error by a factor of $2d$~\cite{gilyen2019quantum}.
Therefore, the quantum circuit $U_{\mathrm{OAA}}$ in Fig.~\ref{fig:circuit_of_U_OAA} using the phases $\Phi^{(\mathrm{OAA})}$
in Eq.~\eqref{eq:OAA_phases} constructs the $(1,a+2,9\varepsilon_{\mathrm{tri}})$-block-encoding of $\exp(-iHt)$:
\begin{align} \label{eq:U_OAA}
  \left\|
    e^{-iHt} - (\bra{0}_{abc}\otimes I)U_{\mathrm{OAA}}(\ket{0}_{abc}\otimes I)
  \right\| \leq 9\varepsilon_{\mathrm{tri}},
\end{align}
with $2$ uses of $U_{\exp}$ and $1$ use of $U_{\exp}^{\dagger}$.

Given an error tolerance $\varepsilon$, the functions $\cos(xt)$ and $\sin(xt)$ should be
$\frac{\varepsilon}{9}$-approximated. Therefore, the number of queries of the OAA-based HS is given by
\begin{equation} \label{eq:query_complexity_of_HS_using_OAA}
  Q_{\mathrm{HS}}^{(\mathrm{OAA})}
    = 3\left(
      2R\left(t,\frac{\varepsilon}{9}\right)+1
    \right) \leq \mathcal{O}\left(
      t + \log(1/\varepsilon)
    \right).
\end{equation}
We summarize the OAA-based HS in Algorithm~\ref{algorithm:HS_using_OAA}.

\subsection{Fixed-point amplitude amplification} \label{subsection:fixed_point_amplitude_amplification_by_QSVT}
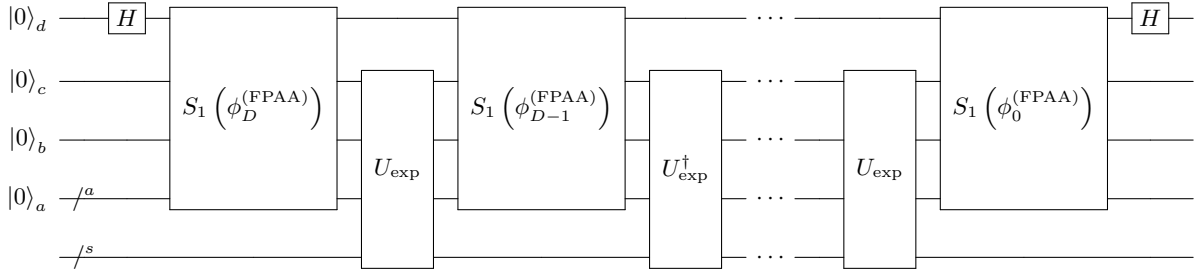
\begin{figure*}[t]
  \[\Qcircuit @C=1em @R=1.5em {
    \lstick{\ket{0}_d} & \qw       & \gate{H}  & \multigate{3}{S_1\left(\phi_{D}^{(\mathrm{FPAA})}\right)} & \qw                     & \multigate{3}{S_1\left(\phi_{D-1}^{(\mathrm{FPAA})}\right)}
        & \qw                        & \qw & \cdots & & \qw & \qw              & \multigate{3}{S_1\left(\phi_{0}^{(\mathrm{FPAA})}\right)}            & \gate{H} & \qw \\
    \lstick{\ket{0}_c} & \qw       & \qw       & \ghost{S_1\left(\phi_{D}^{(\mathrm{FPAA})}\right)}        & \multigate{3}{U_{\exp}} & \ghost{S_1\left(\phi_{D-1}^{(\mathrm{FPAA})}\right)}
        & \multigate{3}{U_{\exp}^{\dagger}} & \qw & \cdots & & \qw & \multigate{3}{U_{\exp}} & \ghost{S_1\left(\phi_{0}^{(\mathrm{FPAA})}\right)}     & \qw      & \qw \\
    \lstick{\ket{0}_b} & \qw       & \qw       & \ghost{S_1\left(\phi_{D}^{(\mathrm{FPAA})}\right)}        & \ghost{U_{\exp}}        & \ghost{S_1\left(\phi_{D-1}^{(\mathrm{FPAA})}\right)}
        & \ghost{U_{\exp}^{\dagger}}        & \qw & \cdots & & \qw & \ghost{U_{\exp}}        & \ghost{S_1\left(\phi_{0}^{(\mathrm{FPAA})}\right)}     & \qw      & \qw \\
    \lstick{\ket{0}_a} & {/}^a \qw & \qw       & \ghost{S_1\left(\phi_{D}^{(\mathrm{FPAA})}\right)}        & \ghost{U_{\exp}} & \ghost{S_1\left(\phi_{D-1}^{(\mathrm{FPAA})}\right)} 
        & \ghost{U_{\exp}^{\dagger}}        & \qw & \cdots & & \qw & \ghost{U_{\exp}}        & \ghost{S_1\left(\phi_{0}^{(\mathrm{FPAA})}\right)}     & \qw      & \qw \\
                      & {/}^s \qw & \qw       & \qw                                                       & \ghost{U_{\exp}}        &  \qw
        & \ghost{U_{\exp}^{\dagger}}        & \qw & \cdots & & \qw & \ghost{U_{\exp}}        & \qw                                                    & \qw      & \qw       
  }\]
  \caption{Quantum circuit $U_{\mathrm{FPAA}}$ that is a $(1,a+3,\varepsilon)$-block-encoding of $\exp(-iHt)$.
            The unitary $U_{\exp}$ is shown in Fig.~\ref{fig:circuit_of_U_exp},
            the gate $S_1$ is shown in Fig.~\ref{fig:gate_used_for_QSVT}(a), and
            the phases $\Phi^{(\mathrm{FPAA})}\in\mathbb{R}^{D+1}$ is calculated from 
            the $D$-th polynomial $P_{\varepsilon_{\mathrm{sign}}, \Delta}^{\mathrm{sign}}$ that
            satisfies Eq.~\eqref{eq:P_sign_2}.}
  \label{fig:circuit_of_U_FPAA}
\end{figure*}

\begin{figure}[tbp]
  \begin{algorithm}[H]
    \caption{The FPAA-based Hamiltonian simulation} \label{algorithm:HS_using_FPAA}
    \begin{algorithmic}[1]
      \renewcommand{\algorithmicrequire}{\textbf{Input:}}
      \renewcommand{\algorithmicensure}{\textbf{Output:}}
      \REQUIRE a $(1, a, 0)$-block-encoding of a Hamiltonian matrix $H$,
              an evolution time $t$, and an error tolerance $\varepsilon$.
      \ENSURE a $(1, a+3, \varepsilon)$-block-encoding of $e^{-iHt}$
      \renewcommand{\algorithmicrequire}{\textbf{Runtime:}}
      \renewcommand{\algorithmicensure}{\textbf{Procedure:}}
      \REQUIRE $Q_{\mathrm{HS}}^{(\mathrm{FPAA})}$ queries to the block-encoding of $H$,
               where $Q_{\mathrm{HS}}^{(\mathrm{FPAA})}$ is given by Eq.~\eqref{eq:query_of_HS_using_FPAA}
      \ENSURE
      \STATE Calculate $\varepsilon_{\mathrm{tri}}$ and $\varepsilon_{\mathrm{sign}}$ that minimize
             $Q_{\mathrm{HS}}^{(\mathrm{FPAA})}$, where
             $R$ and $D$ are given by Eqs.~\eqref{eq:degree_R} and~\eqref{eq:degree_D}.
      \STATE Calculate the phases $\Phi^{(\mathrm{c})}\in\mathbb{R}^{2R+1}$ and $\Phi^{(\mathrm{s})}\in\mathbb{R}^{2R+2}$
            on a classical computer from the $2R$-th polynomial $P_{\varepsilon_{\mathrm{tri}}, \kappa}^{\cos}(x)$ and
            $(2R+1)$-th polynomial $P_{\varepsilon_{\mathrm{tri}}, \kappa}^{\sin}(x)$ in Eq.~\eqref{eq:kappa_approximate_polynomials},
            where $\kappa=1/(1+\varepsilon_{\mathrm{tri}})$.
      \STATE Construct the circuit $U_{\mathrm{exp}}$ in Fig.~\ref{fig:circuit_of_U_exp}
             using the phases $\Phi^{(\mathrm{c})}$ and $\Phi^{(\mathrm{s})}$, which is
             a $(1,a+2,\kappa\varepsilon_{\mathrm{tri}})$-block-encoding of $\kappa e^{-iHt}/2$.
      \STATE Calculate the phases $\Phi^{(\mathrm{FPAA})}\in\mathbb{R}^{D+1}$ on a classical computer
             from the $D$-th polynomial $P_{\varepsilon_{\mathrm{sign}}, \kappa}^{\mathrm{sign}}$ that
             satisfies Eq.~\eqref{eq:P_sign_2}.
      \STATE Run the circuit $U_{\mathrm{FPAA}}$ in Fig.~\ref{fig:circuit_of_U_FPAA}
             using the phases $\Phi^{(\mathrm{FPAA})}$.
    \end{algorithmic}
  \end{algorithm}
\end{figure}

Fixed-point amplitude amplification (FPAA) using QSVT has been proposed in Refs.~\cite{gilyen2019quantum,martyn2021grand}.
In this section, the error and number of queries of the FPAA-based HS are investigated in detail.
In QSVT-based FPAA, the sign function
\begin{equation}
  \mathrm{sign}(x) = \begin{cases}
    -1 & x<0 \\
    0  & x=0 \\
    1  & x>0
  \end{cases}
\end{equation}
is chosen as an objective function.
The sign function can be estimated by a polynomial approximation to
an error function $\mathrm{erf}(kx)$ for large enough $k$~\cite{martyn2021grand}.
Let $D$ be odd, $\Delta>0$, and $\varepsilon_{\mathrm{sign}}\in(0,\sqrt{2/e\pi}]$.
The phases $\Phi^{(\mathrm{FPAA})}\in\mathbb{R}^{D+1}$ can be calculated from the $D$-th polynomial
$P_{\varepsilon_{\mathrm{sign}}, \Delta}^{\mathrm{sign}}$~\cite{gilyen2019quantum,martyn2021grand}
(the explicit form of $P_{\varepsilon_{\mathrm{sign}}, \Delta}^{\mathrm{sign}}$ is seen in Ref.~\cite{low2017quantum}):
\begin{align}
  \begin{split} \label{eq:P_sign_2}
    & \left|
      \mathrm{sign}(x) - P_{\varepsilon_{\mathrm{sign}}, \Delta}^{\mathrm{sign}}(x)
    \right| \leq \varepsilon_{\mathrm{sign}} \\
    & \qquad \mathrm{for}\ x\in\left[-1,-\frac{\Delta}{2}\right]\cup\left[\frac{\Delta}{2},1\right].
  \end{split}
\end{align}
The degree $D$ was given asymptotically in Ref.~\cite{gilyen2019quantum,martyn2021grand}.
We give it explicitly using the result of Ref.~\cite{low2017quantum,mitarai2022quantum}. 
If $k=\frac{\sqrt{2}}{\Delta}\log^{\frac{1}{2}}(8/[\pi\varepsilon_{\mathrm{sign}}^{2}])$ and 
\begin{align} \label{eq:degree_D}
  D(k, \varepsilon_{\mathrm{sign}}) 
  &= 2\left\lceil 
    \frac{16k}{\sqrt{\pi}\varepsilon_{\mathrm{sign}}}\exp\left[
      -\frac{1}{2}W\left(
        \frac{512}{\pi\varepsilon_{\mathrm{sign}}^{2}e^2}
      \right)
    \right]
  \right\rceil + 1 \notag \\
  &=\mathcal{O}\left(
    \frac{1}{\Delta}\log(1/\varepsilon_{\mathrm{sign}})
  \right)
\end{align}
where $W$ is the Lambert W function,
then $P_{\varepsilon_{\mathrm{sign}}, \Delta}^{\mathrm{sign}}$ is $\varepsilon_{\mathrm{sign}}$-approximation
to the sign function in the region $\left[-1,-\frac{\Delta}{2}\right]\cup\left[\frac{\Delta}{2},1\right]$.
We require that $\Delta/2\leq\kappa/2$ because we desire that $\kappa/2$ be mapped to a value
greater than $1-\varepsilon_{\mathrm{sign}}$, and then $1/\Delta\geq1/\kappa$. If $1/\Delta$ increases, then 
$D$ increases because of $D \propto k \propto 1/\Delta$. Therefore, we should choose $\Delta = \kappa$.

We discuss the upper bound of the error of the FPAA-based HS.
Let us denote $A=\kappa\exp(-iHt)/2$ and $\tilde{A}=(\bra{0}_{abc}\otimes I)U_{\exp}(\ket{0}_{abc}\otimes I)$.
Equation~\eqref{eq:U_exp} can be rewritten as
\begin{equation} \label{eq:U_exp_2}
  \|A - \tilde{A}\| \leq \kappa\varepsilon_{\mathrm{tri}},
\end{equation}
and the following inequality holds:
\begin{align}
  \|A + \tilde{A}\| &\leq \|A\| + \|A\| + \|\tilde{A} - A\| \notag \\
                    &\leq \frac{\kappa}{2} + \frac{\kappa}{2} + \kappa\varepsilon_{\mathrm{tri}} \notag \\
                    &= \kappa(1+\varepsilon_{\mathrm{tri}}) = 1,
\end{align}
where in the first inequality we used the triangle inequality.
Therefore, the matrices $A$ and $\tilde{A}$ satisfy the following inequality:
\begin{align}
  \|A - \tilde{A}\| + \left\|\frac{A + \tilde{A}}{2}\right\|^{2}
  &\leq \kappa\varepsilon_{\mathrm{tri}} + \frac{1}{4}  \notag \\
  &= \frac{5}{4} - \frac{1}{1+\varepsilon_{\mathrm{tri}}} \notag \\
  &< \frac{5}{4} - \frac{e}{1+e} < 1,
\end{align}
where in the second last inequality we used $\varepsilon_{\mathrm{tri}}<1/e$.
According to Lemma 23 in Ref.~\cite{gilyen2019quantum}, we have that
\begin{align}
  & \left\|
    P_{\varepsilon_{\mathrm{sign}}, \Delta}^{\mathrm{sign}}(A)
      - P_{\varepsilon_{\mathrm{sign}}, \Delta}^{\mathrm{sign}}(\tilde{A})
  \right\| \notag \\
  &\qquad \leq D\sqrt{\frac{2}{1-\|\frac{A + \tilde{A}}{2}\|^{2}}}\|A-\tilde{A}\| \notag \\
  &\qquad \leq \sqrt{\frac{8}{3}}D\frac{\varepsilon_{\mathrm{tri}}}{1+\varepsilon_{\mathrm{tri}}}<\sqrt{3}D\varepsilon_{\mathrm{tri}}.
\end{align}
Therefore, we have that
\begin{align}
  & \quad \left\|
    \mathrm{sign}(A) - P_{\varepsilon_{\mathrm{sign}}, \Delta}^{\mathrm{sign}}(\tilde{A})
  \right\| \notag \\
  & \leq \left\|
    \mathrm{sign}(A) - P_{\varepsilon_{\mathrm{sign}}, \Delta}^{\mathrm{sign}}(A)
  \right\| + \left\|
    P_{\varepsilon_{\mathrm{sign}}, \Delta}^{\mathrm{sign}}(A)
      - P_{\varepsilon_{\mathrm{sign}}, \Delta}^{\mathrm{sign}}(\tilde{A})
  \right\| \notag \\
  & \leq \varepsilon_{\mathrm{sign}}
    + \sqrt{3}D\varepsilon_{\mathrm{tri}} \equiv \varepsilon'. \label{eq:FPAA_delta}
\end{align}
From the above inequality, we have that $\|P_{\varepsilon_{\mathrm{sign}}, \Delta}^{\mathrm{sign}}(\tilde{A})\|\leq 1+\varepsilon'$,
which violates the condition (vi). Therefore, we must consider the rescaled polynomial
$\frac{1}{1+\varepsilon'}P_{\varepsilon_{\mathrm{sign}}, \Delta}^{\mathrm{sign}}(\tilde{A})$ such that
$\frac{1}{1+\varepsilon'}\|P_{\varepsilon_{\mathrm{sign}}, \Delta}^{\mathrm{sign}}(\tilde{A})\|\leq 1$.
Now, we obtain the following inequality:
\begin{align}
  &\left\|
    \mathrm{sign}(A) - \frac{1}{1+\varepsilon'}P_{\varepsilon_{\mathrm{sign}}, \Delta}^{\mathrm{sign}}(\tilde{A})
  \right\| \notag \\
  &\leq \frac{1}{1+\varepsilon'}\left(
    \|\mathrm{sign}(A)- P_{\varepsilon_{\mathrm{sign}}, \Delta}^{\mathrm{sign}}(\tilde{A})\| + \|\varepsilon'\mathrm{sign}(A)\|
  \right) \notag \\
  &\leq\frac{\varepsilon'+\varepsilon'}{1+\varepsilon'}< 2\varepsilon',
\end{align}
and the error tolerance is defined as
\begin{equation} \label{eq:error_tolerance_of_HS_using_FPAA}
  \varepsilon \equiv 2\varepsilon' =2\left(
    \varepsilon_{\mathrm{sign}}+ \sqrt{3}D\varepsilon_{\mathrm{tri}}
  \right).
\end{equation}
The quantum circuit $U_{\mathrm{FPAA}}$ using the phases $\Phi^{(\mathrm{FPAA})}\in\mathbb{R}^{D+1}$
in Fig.~\ref{fig:circuit_of_U_FPAA} constructs the $(1,a+3,\varepsilon)$-block-encoding of $\exp(-iHt)$:
\begin{equation} \label{eq:U_FPAA}
  \left\|
    e^{-iHt} - (\bra{0}_{abcd}\otimes I)U_{\mathrm{FPAA}}(\ket{0}_{abcd}\otimes I)
  \right\| \leq \varepsilon,
\end{equation}
with $(D+1)/2$ uses of $U_{\exp}$ and $(D-1)/2$ uses of $U_{\exp}^{\dagger}$.
Therefore, the number of queries of the FPAA-based HS is given by $D(2R+1)$. 
It varies depending on $\varepsilon_{\mathrm{tri}}$ and $\varepsilon_{\mathrm{sign}}$
satisfying Eq.~\eqref{eq:error_tolerance_of_HS_using_FPAA}. The number of queries of the FPAA-based HS is defined as
\begin{equation} \label{eq:query_of_HS_using_FPAA}
  Q_{\mathrm{HS}}^{(\mathrm{FPAA})}=\min_{\varepsilon_{\mathrm{tri}},\varepsilon_{\mathrm{sign}}}
                      D(\kappa, \varepsilon_{\mathrm{sign}})(2R(t, \varepsilon_{\mathrm{tri}})+1).
\end{equation}
This is asymptotically given by
\begin{align} \label{eq:query_complexity_of_HS_using_FPAA}
  Q_{\mathrm{HS}}^{(\mathrm{FPAA})}
    &=\mathcal{O}\left(
      \log(1/\varepsilon_{\mathrm{sign}})\left[
        t+\log(1/\varepsilon_{\mathrm{tri}})
      \right]
    \right) \notag \\
    &=\mathcal{O}\left(
      \log(1/\varepsilon)t + \log^2(1/\varepsilon)
    \right).
\end{align}
We summarize the FPAA-based HS in Algorithm~\ref{algorithm:HS_using_FPAA}; 
also we summarize the comparison between OAA-based and FPAA-based HS
in Table~\ref{tb:comparison_between_OAA_and_FPAA}.

\begin{table}[tbp]
  \centering
  \caption{The comparison between OAA-based and FPAA-based HS.}
    \begin{tabular}{ccc} \hline
               & OAA-based HS & FPAA-based HS \\ \hline \hline
      \begin{tabular}{c}
        Polynomial used
      \end{tabular} & $T_3(x)$ & \begin{tabular}{c}
                              The approximate \\ polynomial of $\mathrm{sign}(x)$ 
                            \end{tabular}\\ \hline
      \begin{tabular}{c}
        The number \\of queries
      \end{tabular} & $3\left(
                        2R\left(t,\frac{\varepsilon}{9}\right)+1
                      \right)$ & $D(\kappa, \varepsilon_{\mathrm{sign}})(2R(t, \varepsilon_{\mathrm{tri}})+1)$ \\ \hline
      \begin{tabular}{c}
        Asymptotic \\query complexity
      \end{tabular} & $\mathcal{O}\left(
                        t + \log(1/\varepsilon)
                      \right)$ & $\mathcal{O}\left(
                                          \log(1/\varepsilon)t + \log^2(1/\varepsilon)
                                        \right)$\\ \hline        
    \end{tabular}
  \label{tb:comparison_between_OAA_and_FPAA}
\end{table}

\subsection{Hamiltonian simulation for general Hermitian matrix and extension of evolution time} \label{subsection:HS_for_general_H_and_extension_of_time}
\begin{figure}[tbp]
  \[\Qcircuit @C=1.5em @R=1.5em {
    \lstick{\ket{0}_{a'}} & \gate{H} & \ctrlo{1}        & \gate{H} & \qw \\
                          & \qw      & \multigate{1}{U} & \qw      & \qw \\
                          & \qw      & \ghost{U}        & \qw      & \qw
  }\]
  \caption{Construction of the unitary $U'$ that is a $(1,a+1,0)$-block-encoding of $(H/\alpha + I)/2$
            from the unitary $U$, defined as in Eq.~\eqref{eq:block_encoding_of_H_alpha}.}
  \label{fig:circuit_of_U_dash}
\end{figure}
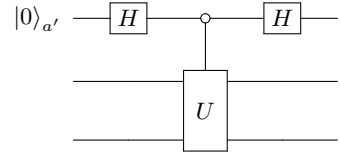

We describe the way to implement HS for a general Hermitian matrix $H$
that is not positive semidefinite and normalized by $\alpha>0$,
i.e., $\|H/\alpha\|\leq 1$. Suppose that $U$ is an $(\alpha, a, 0)$-block-encoding of $H$:
\begin{equation} \label{eq:block_encoding_of_H_alpha} 
  U = \begin{bmatrix}
    \frac{H}{\alpha}     & \cdot \\
    \cdot & \cdot
  \end{bmatrix}, \quad \frac{H}{\alpha} = (\bra{0}_a\otimes I)U(\ket{0}_a\otimes I).
\end{equation}
The authors of Ref.~\cite{martyn2021grand} have proposed the unitary $U'$ as in Fig.~\ref{fig:circuit_of_U_dash},
which is a $(1,a+1,0)$-block-encoding of the positive semidefinite Hermitian matrix $(H/\alpha + I)/2$. Instead of $U$, this unitary is used
in the circuits $U_{\exp}$, and $U_{\mathrm{OAA}}$ or $U_{\mathrm{FPAA}}$, denoted by $U_{\mathrm{HS}}$.
Then, $U_{\mathrm{HS}}$ becomes a $(1,a+3, \varepsilon)$-block-encoding of $\exp(-i\frac{H/\alpha + I}{2}t)$:
\begin{equation} \label{eq:modified_U_HS}
  \left\|
    e^{-i\frac{H/\alpha + I}{2}t} - (\bra{0}_{aa'bc}\otimes I)U_{\mathrm{HS}}(\ket{0}_{aa'bc}\otimes I)
  \right\| \leq \varepsilon.
\end{equation}
If the evolution time $t$ is modified to $2\alpha t$, then $U_{\mathrm{HS}}$ becomes 
a $(1, a+3, \varepsilon)$-block-encoding of $e^{-iHt}$ up to a global phase.
Therefore, the query complexity of QSVT-based HS can be generally represented as
\begin{equation} \label{eq:query_complexity_of_HS_using_QSVT}
  \mathcal{O}\left(
    2\alpha t + \log(1/\varepsilon)
  \right).
\end{equation}
Note that the factor $\log(1/\varepsilon)$ is multiplied by
the above complexity for the FPAA-based HS, but it is ignored for simplicity.

Larger $t$ requires calculating a higher number of the phases, which can be challenging. 
One can instead use $N_t$ sequential $U_{\mathrm{HS}}$ for the smaller time step $\Delta t = t/N_t$
and extend the evolution time as follows:
\begin{equation}
  e^{-iHt} \approx (\bra{0}_{aa'bc}\otimes I)U_{\mathrm{HS}}^{N_t}(\bra{0}_{aa'bc}\otimes I).
\end{equation}
This split is also found in Ref.~\cite{novikau2022quantum}. We discuss the query complexity of the extension of 
the evolution time of HS.
According to Lemma 53 in Ref.~\cite{gilyen2019quantum}, the error of the product of two block-encoded matrices
does not exceed the sum of each error. Let $U_{\mathrm{HS}}$ be a $(1,a+3,\delta)$-block-encoding of $\exp(-iH\Delta t)$,
and then we have that
\begin{equation}
  \left\|
    e^{-iHt} - (\bra{0}_{aa'bc}\otimes I)U_{\mathrm{HS}}^{N_t}(\ket{0}_{aa'bc}\otimes I)
  \right\| \leq N_t \delta \equiv \varepsilon,
\end{equation}
where $\varepsilon$ is an error tolerance. The query complexity in Eq.~\eqref{eq:query_complexity_of_HS_using_QSVT}
can be rewritten as follows:
\begin{align} \label{eq:query_complexity_of_iterate_HS_using_QSVT}
  &N_t \mathcal{O}\left(
    2\alpha \Delta t + \log\left(\frac{1}{\delta}\right)
  \right) \notag \\
  &=\mathcal{O}\left(
    2\alpha t + N_t\log\left(\frac{N_t}{\varepsilon}\right)
  \right).
\end{align}

\section{Quantum algorithm for the linearized Vlasov-Poisson system} \label{section:VQS}
Quantum algorithms for the linearized Vlasov-Poisson system have three steps:
(1) initialization, (2) HS, and (3) Extracting data.
Initialization prepares a quantum state that represents initial physical conditions;
HS implements the time evolution of Eq.~\eqref{eq:schrodinger_equation_form};
Extracting physically meaningful data from a final state.

To encode data and construct the Hamiltonian $H$, we use a rotation gate called 
a variable rotation introduced in Ref.~\cite{engel2019quantum}, defined as
\begin{equation}
  R(x) \equiv \begin{cases}
    e^{-iY\arccos x} & (x \in \mathbb{R}) \\
    e^{-iX\arccos \mathrm{Im}(x)}e^{iZ\frac{\pi}{2}} & (x \in \mathbb{C}\setminus\mathbb{R}) ,
  \end{cases}
\end{equation}
where $x$ is a rotation angle such that $|x|\leq 1$. This gate acts as $R(x)\ket{0}=x\ket{0}+\sqrt{1-|x|^2}\ket{1}$.
We assume that the rotation angles can be calculated efficiently on temporary registers, 
and applying the rotations controlled on the angle qubits can be implemented efficiently.
Then the cost to implement the variable rotations is 
$\mathcal{O}\left(\mathrm{poly}(n_v)\right)=\mathcal{O}\left(\mathrm{poly}(\log N_v)\right)$ because
the input register has $n_v$ qubits.

\subsection{The one-dimensional linearized Vlasov-Poisson system}
Efficiently preparing a quantum state that represents the physical data is a very difficult problem.
This problem is related to amplitude encoding~\cite{prakash2014quantum}. 
Assuming quantum random access memory (QRAM)~\cite{giovannetti2008architectures},
amplitude encoding is also implemented efficiently.
Under this assumption, we desire that a quantum circuit for initialization $U_{\mathrm{ini}}^{\mathrm{ideal}}$
can prepare the following initial state:
\begin{equation}
  \begin{split} 
    &U_{\mathrm{ini}}^{\mathrm{ideal}}\ket{0}_r\ket{0}_v \\
    &\quad =\frac{1}{\eta}\left[
        \sum_{j_x=0}^{Nv-1}F_{j_x}(t=0)\ket{0}_{r}\ket{j_x}_{v_x} 
        + E(t=0)\ket{1}_{r}\ket{0}_{v_x}
      \right], \label{eq:ideal_initial_state}
  \end{split}
\end{equation}
where $\eta = \sqrt{\sum_{j_x}|F_{j_x}(t=0)|^{2} + |E(t=0)|^2}$. We assume
the gate complexity of $U_{\mathrm{ini}}^{\mathrm{ideal}}$ is $\mathcal{O}(\mathrm{poly}(\log N_{v_x}))$.

To implement HS for the linearized Vlasov-Poisson system, it is necessary to construct the corresponding unitary $U$
that is an $(\alpha, a, 0)$-block-encoding of $H$. The unitary for the one-dimensional system has been proposed
in Ref.~\cite{engel2019quantum}, which consists of two unitaries $U_{\mathrm{row}}$ and $U_{\mathrm{col}}$.
These unitaries are called state preparation unitaries in Ref.~\cite{gilyen2019quantum} and
satisfy $U=U_{\mathrm{row}}^{\dagger}U_{\mathrm{col}}$. The gate complexity of $U$ is
given by $\mathcal{O}(\mathrm{poly}(\log N_{v_x}))$. According to Ref.~\cite{engel2019quantum},
$\alpha$ satisfies the following inequality:
\begin{equation}
  \frac{4\Lambda}{5}\leq \alpha \leq \Lambda,
\end{equation}
where
\begin{equation}
  \begin{split}
    \Lambda &= |k_x|v_{x,\max} + \sqrt{\Delta v_x N_{v_x} v_{x,\max} G_{\max}}, \\
    v_{x,\max} &= \max_{j_x}|v_{j_x}|, \\
    G_{\max} &= \max_{j_x}|v_{j_x} f_{\mathrm{M}}(v_{v_{j_x}})|.
  \end{split}
\end{equation}
Since $\Delta v_x N_{v_x} = 2v_{x,\max} + \Delta v_x$, $\Lambda$ does not increase with increasing $N_{v_x}$,
i.e. $\alpha=\Theta(1)$. Therefore, the query complexity of QSVT-based HS is given by 
$\mathcal{O}(t + \log(1/\varepsilon))$.
The gate complexity of QSVT-based HS is the above equation multiplied by $\mathrm{poly}(\log N_{v_x})$.

One of the ways to obtain data from the final state is the quantum amplitude estimation (QAE) algorithm
\cite{Brassard2002quantum}, which can produce an estimate $\tilde{p}$ of the probability $p$ with
an error bounded by
\begin{equation} \label{eq:QAE_inequality}
  |\tilde{p}-p| \leq 2\pi \frac{\sqrt{p(1-p)}}{M} + \frac{\pi^2}{M^2},
\end{equation}
where $M$ is the number of iterations.
The authors of Ref.~\cite{engel2019quantum} have used QAE to obtain
an estimate of the magnitude $|E|$. They have also proposed
the algorithm for obtaining the real and imaginary parts of $E$ and showed its cost does not change asymptotically. 
We discuss the computational complexity of calculating the time evolution of $E$, and the algorithm for
obtaining quantities related to the distribution function $f$.

The state after HS for the evolution time $t$ is applied to the initial state
in Eq.~\eqref{eq:ideal_initial_state} becomes
\begin{equation}
  \ket{\psi(t)}=\frac{1}{\eta}\ket{0}_a\left[
      \sum_{j_x=0}^{N_{v_x}-1}F_{j_x}(t)\ket{0}_{r}\ket{j_x}_{v_x}+E(t)\ket{1}_{r}\ket{0}_{v_x}
    \right], \label{eq:ideal_psi_t}
\end{equation}
where label $a$ is represented as all ancilla qubits including qubits labeled $a, a', b, c,$ and $d$.
Let $p=|E(t)|^2/\eta^2$ and its estimate be $\tilde{p}$.
We introduce $a=|E(t)|^2$ and $\tilde{a}=\eta^2 \tilde{p}$ and then they satisfy the following inequality
from Eq.~\eqref{eq:QAE_inequality}:
\begin{align} \label{eq:QAE_for_ideal_E_1}
  |\tilde{a}-a| &\leq 2\pi \frac{\sqrt{a(\eta^2-a)}}{M} + \frac{\pi^2\eta^2}{M^2} \notag \\
                &\leq \frac{2\pi\eta|E(t)|}{M} + \frac{\pi^2\eta^2}{M^2},
\end{align}
where in the last inequality we used $\eta^2-a\leq \eta^2$.
Assuming we know the upper bound of $|E(t)|$ is $E_u$. Let $0<\delta<1$ and 
$M=\left\lceil \frac{(2E_{u}+1)\pi\eta}{\delta}\right\rceil$. The above inequality becomes
\begin{align} \label{eq:QAE_for_ideal_E_2}
  |\tilde{a}-a| &\leq \frac{2|E(t)|\delta}{2E_u+1} + \frac{\delta^2}{(2E_u+1)^2} \notag \\
                &\leq \frac{2|E(t)|\delta}{2E_u+1} + \frac{\delta}{2E_u+1}\leq \delta,
\end{align}
where in the last inequality we used $2|E(t)|+1\leq 2E_u+1$.
The value of $E(t)$ can also be obtained asymptotically with the same complexity at an additional cost 
using the algorithm proposed in Ref.~\cite{engel2019quantum}.

Since $M=\mathcal{O}(1/\delta)$, the gate complexity of the whole algorithm for obtaining the estimate of $E(t)$
including the initialization and HS steps is given by:
\begin{equation} \label{eq:gate_complexity_of_obtaining_E}
  \mathcal{O}\left(
    \frac{\mathrm{poly}(\log N_v)}{\delta}\left(
      t + \log(1/\varepsilon)
    \right)
  \right).
\end{equation}

Now, we discuss the cost of calculating the time evolution of $E$. For simplicity,
we assume the phases $\Phi^{(\mathrm{c})}$ and $\Phi^{(\mathrm{s})}$ for a large $t$, and given error tolerance
$\varepsilon$ can be calculated. Let $N_t$ be the number of time steps,
$t_{\max}$ be the maximum of the evolution time, and $\Delta t = t_{\max}/N_t$ be the time step.
The gate complexity of the algorithm for the evolution time $t_l = l\Delta t$ is given by
Eq.~\eqref{eq:gate_complexity_of_obtaining_E} with $t$ replaced by $t_l$, denoted by $Q_l$.
Since
\begin{equation}
  \sum_{l=1}^{N_t}t_l = \frac{\Delta t}{2}N_t(N_t+1)=\frac{t_{\max}(N_t+1)}{2},
\end{equation}
the gate complexity of calculating the time evolution of $E$ is
\begin{equation} \label{eq:gate_complexity_of_calculating_the_time_evolution_of_E}
  \sum_{l=1}^{N_t}Q_{l} = \mathcal{O}\left(
    \frac{\mathrm{poly}(\log N_v)N_t}{\delta}\left(
      t_{\max} + \log(1/\varepsilon)
    \right)
  \right).
\end{equation}
If we consider $t_{\max}$ to be a constant, then the cost of the quantum algorithm is
asymptotically the same for the number of time steps $N_t$ as that of a classical algorithm,
which scales linearly with $N_t$.

We show the way to obtain the deviation from the Maxwell distribution
\begin{align}
  D_{\mathrm{M}}(t)
  &\equiv \sum_{j_x=0}^{N_{v_x} -1} \left|
    f(v_{j_x}, t) - f_{\mathrm{M}}(v_{j_x})
  \right|^2 \Delta v_x \notag \\
  &= \sum_{j_x=0}^{N_{v_x} -1} \left|
    f_1(v_{j_x}, t)
  \right|^2 \Delta v_x.
\end{align}
Here, we write the subscript $1$ explicitly.
This quantity can be used to know how well the fluid approximation is applied to the system.
We add single ancilla qubit and variable rotation gates with angles
$\sqrt{f_{\mathrm{M}}(v_{j_x})}$, and the state in Eq.~\eqref{eq:ideal_psi_t}
becomes as follows:
\begin{equation} \label{eq:f_1_state}
  \frac{1}{\eta}\sum_{j_x=0}^{N_{v_x} - 1}i f_{1}(v_{j_x}, t)\sqrt{\Delta v_x}\ket{0}_a\ket{0}_r\ket{j_x}_{v_x}\ket{0} + \ket{\bot},
\end{equation}
where  $\ket{\bot}$ is the state of no interest.
QAE is applied to the above state, we obtain the estimate
\begin{equation}
  \tilde{p} \approx \frac{1}{\eta^2}\sum_{j_x=0}^{N_{v_x} - 1}|f_1(v_{j_x}, t)|^2\Delta v_x
            = \frac{1}{\eta^2} D_{\mathrm{M}}(t).
\end{equation}
Similar to the discussion of QAE to obtain the estimate of $E(t)$,
the number of iterations to obtain the estimate of $D_{\mathrm{M}}$ is also given by $M=\mathcal{O}(1/\delta)$.
Therefore, the gate complexity of the algorithm for calculating the deviation $D_{\mathrm{M}}$ is given
by Eq.~\eqref{eq:gate_complexity_of_obtaining_E}.

\subsection{The higher dimensional Vlasov-Poisson systems}
\begin{figure*}[tbp]
  \[\Qcircuit @C=1.0em @R=2.0em {
    \lstick{a_0} & \qw               & \gate{R(b_{\bm{j}}^{*})}    & \qw        & \qw      & \qw              & \qw        & \qw      & \qw              & \gate{R(p_{j_x}^{*})} & \gate{R(q_{j_x}^{*})} & \qw       & \qw                          & \qw                              & \qw             & \qw      & \qw  \\
    \lstick{a_1} & \qw               & \qw                         & \targ      & \ctrl{5} & \ctrl{7}         & \qw        & \qw      & \qw              & \ctrl{-1} \qwx[7]     & \qw                   & \targ     & \ctrl{3}                     & \ctrl{4}                         & \ctrlo{2}       & \ctrl{1} & \qw   \\
    \lstick{a_2} & \qw               & \qw                         & \qw        & \qw      & \qw              & \targ      & \ctrl{5} & \ctrl{6}         & \qw                   & \ctrl{-2} \qwx[7]     & \ctrl{-1} & \qw                          & \qw                              & \qw             & \gate{H} & \qw  \\
    \lstick{a_3} & \qw               & \qw                         & \qw        & \qw      & \qw              & \qw        & \qw      & \qw              & \qw                   & \qw                   & \qw       & \qw                          & \qw                              & \gate{R(c^{*})} & \qw      & \qw    \\
    \lstick{a_4} & \qw               & \qw                         & \qw        & \qw      & \qw              & \qw        & \qw      & \qw              & \qw                   & \qw                   & \qw       & \gate{R(b_{\bm{j}})} \qwx[4] & \qw                              & \qw             & \qw      & \qw \\
    \lstick{a_5} & \qw               & \qw                         & \qw        & \qw      & \qw              & \qw        & \qw      & \qw              & \qw                   & \qw                   & \qw       & \qw                          & \gate{R(d_{\bm{j}}^{*})} \qwx[3] & \qw             & \qw      & \qw    \\
    \lstick{r_0} & \qw               & \ctrlo{-6} \qwx[1]          & \ctrl{-5}  & \targ    & \qw              & \ctrlo{-4} & \qw      & \qw              & \qw                   & \qw                   & \qw       & \qw                          & \qw                              & \qw             & \qw      & \qw      \\
    \lstick{r_1} & \qw               & \ctrlo{1}                   & \ctrlo{-1} & \qw      & \qw              & \ctrl{-1}  & \targ    & \qw              & \qw                   & \qw                   & \qw       & \qw                          & \qw                              & \qw             & \qw      & \qw   \\
    \lstick{v_x} & {/}^{n_{v_x}} \qw & \multigate{1}{\ket{\bm{j}}} & \ctrlo{-1} & \qw      & \multigate{1}{H} & \ctrlo{-1} & \qw      & \multigate{1}{H} & \gate{\ket{j_x}}      & \qw                   & \qw       & \multigate{1}{\ket{\bm{j}}}  & \multigate{1}{\ket{\bm{j}}}      & \qw             & \qw      & \qw  \\
    \lstick{v_y} & {/}^{n_{v_y}} \qw & \ghost{\ket{\bm{j}}}        & \ctrlo{-1} & \qw      & \ghost{H}        & \ctrlo{-1} & \qw      & \ghost{H}        & \qw                   & \gate{\ket{j_y}}      & \qw       & \ghost{\ket{\bm{j}}}         & \ghost{\ket{\bm{j}}}             & \qw             & \qw      & \qw    \\
  }\]
  \caption{Quantum circuit of the unitary $U_{\mathrm{row}}$ for the two-dimensional linearized Vlasov-Poisson system.}
  \label{fig:circuit_of_2D_U_row}
  \[\Qcircuit @C=1.0em @R=2.0em {
    \lstick{a_0} & \qw               & \qw                         & \qw        & \qw      & \qw              & \qw        & \qw      & \qw              & \qw                       & \qw                       & \qw       & \gate{R(b_{\bm{j}}^{*})}     & \qw                          & \qw         & \qw       & \qw  \\
    \lstick{a_1} & \qw               & \qw                         & \qw        & \qw      & \qw              & \qw        & \qw      & \qw              & \qw                       & \qw                       & \qw       & \qw                          & \qw                          & \gate{R(c)} & \qw       & \qw   \\
    \lstick{a_2} & \qw               & \qw                         & \qw        & \qw      & \qw              & \targ      & \ctrl{5} & \ctrl{6}         & \qw                       & \ctrl{2}                  & \ctrl{1}  & \qw                          & \qw                          & \qw         & \gate{H}  & \qw  \\
    \lstick{a_3} & \qw               & \qw                         & \targ      & \ctrl{3} & \ctrl{5}         & \qw        & \qw      & \qw              & \ctrl{1}                  & \qw                       & \targ     & \ctrl{-3} \qwx[5]            & \ctrl{2}                     & \ctrlo{-2}  & \ctrl{-1} & \qw    \\
    \lstick{a_4} & \qw               & \gate{R(b_{\bm{j}})}        & \qw        & \qw      & \qw              & \qw        & \qw      & \qw              & \gate{R(p_{j_x})} \qwx[4] & \gate{R(q_{j_x})} \qwx[5] & \qw       & \qw                          & \qw                          & \qw         & \qw       & \qw \\
    \lstick{a_5} & \qw               & \qw                         & \qw        & \qw      & \qw              & \qw        & \qw      & \qw              & \qw                       & \qw                       & \qw       & \qw                          & \gate{R(d_{\bm{j}})} \qwx[3] & \qw         & \qw       & \qw    \\
    \lstick{r_0} & \qw               & \ctrlo{-2} \qwx[1]          & \ctrl{-3}  & \targ    & \qw              & \ctrlo{-4} & \qw      & \qw              & \qw                       & \qw                       & \qw       & \qw                          & \qw                          & \qw         & \qw       & \qw      \\
    \lstick{r_1} & \qw               & \ctrlo{1}                   & \ctrlo{-1} & \qw      & \qw              & \ctrl{-1}  & \targ    & \qw              & \qw                       & \qw                       & \qw       & \qw                          & \qw                          & \qw         & \qw       & \qw   \\
    \lstick{v_x} & {/}^{n_{v_x}} \qw & \multigate{1}{\ket{\bm{j}}} & \ctrlo{-1} & \qw      & \multigate{1}{H} & \ctrlo{-1} & \qw      & \multigate{1}{H} & \gate{\ket{j_x}}          & \qw                       & \qw       & \multigate{1}{\ket{\bm{j}}}  & \multigate{1}{\ket{\bm{j}}}  & \qw         & \qw       & \qw  \\
    \lstick{v_y} & {/}^{n_{v_y}} \qw & \ghost{\ket{\bm{j}}}        & \ctrlo{-1} & \qw      & \ghost{H}        & \ctrlo{-1} & \qw      & \ghost{H}        & \qw                       & \gate{\ket{j_y}}          & \qw       & \ghost{\ket{\bm{j}}}         & \ghost{\ket{\bm{j}}}         & \qw         & \qw       & \qw    \\
  }\]
  \caption{Quantum circuit of the unitary $U_{\mathrm{col}}$ for the two-dimensional linearized Vlasov-Poisson system.}
  \label{fig:circuit_of_2D_U_col}
\end{figure*}
\begin{figure*}[tbp]
  \[\Qcircuit @C=0.75em @R=1.5em {
    \lstick{a_0} & \qw               & \gate{R(b_{\bm{j}}^{*})}    & \qw        & \qw      & \qw              & \qw        & \qw      & \qw              & \qw        & \qw      & \qw      & \qw              & \gate{R(p_{j_x}^{*})} & \gate{R(q_{j_y}^{*})} & \gate{R(r_{j_z}^{*})} & \qw       & \qw       & \qw                         & \qw                              & \qw             & \qw      & \qw      & \qw           \\
    \lstick{a_1} & \qw               & \qw                         & \targ      & \ctrl{6} & \ctrl{8}         & \qw        & \qw      & \qw              & \qw        & \qw      & \qw      & \qw              & \ctrl{-1} \qwx[8]     & \qw                   & \qw                   & \targ     & \targ     & \ctrl{4}                    & \ctrl{5}                         & \ctrlo{3}       & \ctrl{1} & \ctrl{2} & \qw    \\
    \lstick{a_2} & \qw               & \qw                         & \qw        & \qw      & \qw              & \targ      & \ctrl{6} & \ctrl{7}         & \qw        & \qw      & \qw      & \qw              & \qw                   & \ctrl{-2} \qwx[8]     & \qw                   & \ctrl{-1} & \qw       & \qw                         & \qw                              & \qw             & \gate{H} & \qw      & \qw           \\
    \lstick{a_3} & \qw               & \qw                         & \qw        & \qw      & \qw              & \qw        & \qw      & \qw              & \targ      & \ctrl{4} & \ctrl{5} & \ctrl{6}         & \qw                   & \qw                   & \ctrl{-3} \qwx[8]     & \qw       & \ctrl{-2} & \qw                         & \qw                              & \qw             & \qw      & \gate{H} & \qw      \\
    \lstick{a_4} & \qw               & \qw                         & \qw        & \qw      & \qw              & \qw        & \qw      & \qw              & \qw        & \qw      & \qw      & \qw              & \qw                   & \qw                   & \qw                   & \qw       & \qw       & \qw                         & \qw                              & \gate{R(c^{*})} & \qw      & \qw      & \qw            \\
    \lstick{a_5} & \qw               & \qw                         & \qw        & \qw      & \qw              & \qw        & \qw      & \qw              & \qw        & \qw      & \qw      & \qw              & \qw                   & \qw                   & \qw                   & \qw       & \qw       & \gate{R(b_{\bm{j}})} \qwx[4]       & \qw                              & \qw             & \qw      & \qw      & \qw           \\
    \lstick{a_6} & \qw               & \qw                         & \qw        & \qw      & \qw              & \qw        & \qw      & \qw              & \qw        & \qw      & \qw      & \qw              & \qw                   & \qw                   & \qw                   & \qw       & \qw       & \qw                         & \gate{R(d_{\bm{j}}^{*})} \qwx[3] & \qw             & \qw      & \qw      & \qw             \\
    \lstick{r_0} & \qw               & \ctrlo{-7} \qwx[1]          & \ctrl{-6}  & \targ    & \qw              & \ctrlo{-5} & \qw      & \qw              & \ctrl{-4}  & \targ    & \qw      & \qw              & \qw                   & \qw                   & \qw                   & \qw       & \qw       & \qw                         & \qw                              & \qw             & \qw      & \qw      & \qw          \\
    \lstick{r_1} & \qw               & \ctrlo{1}                   & \ctrlo{-1} & \qw      & \qw              & \ctrl{-1}  & \targ    & \qw              & \ctrl{-1}  & \qw      & \targ    & \qw              & \qw                   & \qw                   & \qw                   & \qw       & \qw       & \qw                         & \qw                              & \qw             & \qw      & \qw      & \qw          \\
    \lstick{v_x} & {/}^{n_{v_x}} \qw & \multigate{2}{\ket{\bm{j}}} & \ctrlo{-1} & \qw      & \multigate{2}{H} & \ctrlo{-1} & \qw      & \multigate{2}{H} & \ctrlo{-1} & \qw      & \qw      & \multigate{2}{H} & \gate{\ket{j_x}}      & \qw                   & \qw                   & \qw       & \qw       & \multigate{2}{\ket{\bm{j}}} & \multigate{2}{\ket{\bm{j}}}      & \qw             & \qw      & \qw      & \qw          \\
    \lstick{v_y} & {/}^{n_{v_y}} \qw & \ghost{\ket{\bm{j}}}        & \ctrlo{-1} & \qw      & \ghost{H}        & \ctrlo{-1} & \qw      & \ghost{H}        & \ctrlo{-1} & \qw      & \qw      & \ghost{H}        & \qw                   & \gate{\ket{j_y}}      & \qw                   & \qw       & \qw       & \ghost{\ket{\bm{j}}}        & \ghost{\ket{\bm{j}}}             & \qw             & \qw      & \qw      & \qw        \\
    \lstick{v_z} & {/}^{n_{v_z}} \qw & \ghost{\ket{\bm{j}}}        & \ctrlo{-1} & \qw      & \ghost{H}        & \ctrlo{-1} & \qw      & \ghost{H}        & \ctrlo{-1} & \qw      & \qw      & \ghost{H}        & \qw                   & \qw                   & \gate{\ket{j_z}}      & \qw       & \qw       & \ghost{\ket{\bm{j}}}        & \ghost{\ket{\bm{j}}}             & \qw             & \qw      & \qw      & \qw           \\
  }\]
  \caption{Quantum circuit of the unitary $U_{\mathrm{row}}$ for the three-dimensional linearized Vlasov-Poisson system.}
  \label{fig:circuit_of_3D_U_row}
  \[\Qcircuit @C=0.75em @R=1.5em {
    \lstick{a_0} & \qw               & \qw                         & \qw        & \qw      & \qw              & \qw        & \qw      & \qw              & \qw        & \qw      & \qw      & \qw              & \qw                       & \qw                       & \qw                       & \qw       & \qw       & \gate{R(b_{\bm{j}}^{*})}    & \qw                          & \qw         & \qw       & \qw       & \qw           \\
    \lstick{a_1} & \qw               & \qw                         & \qw        & \qw      & \qw              & \qw        & \qw      & \qw              & \qw        & \qw      & \qw      & \qw              & \qw                       & \qw                       & \qw                       & \qw       & \qw       & \qw                         & \qw                          & \gate{R(c)} & \qw       & \qw       & \qw    \\
    \lstick{a_2} & \qw               & \qw                         & \qw        & \qw      & \qw              & \targ      & \ctrl{6} & \ctrl{7}         & \qw        & \qw      & \qw      & \qw              & \qw                       & \ctrl{3}                  & \qw                       & \ctrl{2}  & \qw       & \qw                         & \qw                          & \qw         & \gate{H}  & \qw       & \qw           \\
    \lstick{a_3} & \qw               & \qw                         & \qw        & \qw      & \qw              & \qw        & \qw      & \qw              & \targ      & \ctrl{4} & \ctrl{5} & \ctrl{6}         & \qw                       & \qw                       & \ctrl{2}                  & \qw       & \ctrl{1}  & \qw                         & \qw                          & \qw         & \qw       & \gate{H}  & \qw      \\
    \lstick{a_4} & \qw               & \qw                         & \targ      & \ctrl{3} & \ctrl{5}         & \qw        & \qw      & \qw              & \qw        & \qw      & \qw      & \qw              & \ctrl{1}                  & \qw                       & \qw                       & \targ     & \targ     & \ctrl{-4} \qwx[5]           & \ctrl{2}                     & \ctrlo{-3}  & \ctrl{-2} & \ctrl{-1} & \qw            \\
    \lstick{a_5} & \qw               & \gate{R(b_{\bm{j}})}        & \qw        & \qw      & \qw              & \qw        & \qw      & \qw              & \qw        & \qw      & \qw      & \qw              & \gate{R(p_{j_x})} \qwx[4] & \gate{R(q_{j_y})} \qwx[5] & \gate{R(r_{j_z})} \qwx[6] & \qw       & \qw       & \qw                         & \qw                          & \qw         & \qw       & \qw       & \qw           \\
    \lstick{a_6} & \qw               & \qw                         & \qw        & \qw      & \qw              & \qw        & \qw      & \qw              & \qw        & \qw      & \qw      & \qw              & \qw                       & \qw                       & \qw                       & \qw       & \qw       & \qw                         & \gate{R(d_{\bm{j}})} \qwx[3] & \qw         & \qw       & \qw       & \qw             \\
    \lstick{r_0} & \qw               & \ctrlo{-2} \qwx[1]          & \ctrl{-3}  & \targ    & \qw              & \ctrlo{-5} & \qw      & \qw              & \ctrl{-4}  & \targ    & \qw      & \qw              & \qw                       & \qw                       & \qw                       & \qw       & \qw       & \qw                         & \qw                          & \qw         & \qw       & \qw       & \qw          \\
    \lstick{r_1} & \qw               & \ctrlo{1}                   & \ctrlo{-1} & \qw      & \qw              & \ctrl{-1}  & \targ    & \qw              & \ctrl{-1}  & \qw      & \targ    & \qw              & \qw                       & \qw                       & \qw                       & \qw       & \qw       & \qw                         & \qw                          & \qw         & \qw       & \qw       & \qw          \\
    \lstick{v_x} & {/}^{n_{v_x}} \qw & \multigate{2}{\ket{\bm{j}}} & \ctrlo{-1} & \qw      & \multigate{2}{H} & \ctrlo{-1} & \qw      & \multigate{2}{H} & \ctrlo{-1} & \qw      & \qw      & \multigate{2}{H} & \gate{\ket{j_x}}          & \qw                       & \qw                       & \qw       & \qw       & \multigate{2}{\ket{\bm{j}}} & \multigate{2}{\ket{\bm{j}}}  & \qw         & \qw       & \qw       & \qw          \\
    \lstick{v_y} & {/}^{n_{v_y}} \qw & \ghost{\ket{\bm{j}}}        & \ctrlo{-1} & \qw      & \ghost{H}        & \ctrlo{-1} & \qw      & \ghost{H}        & \ctrlo{-1} & \qw      & \qw      & \ghost{H}        & \qw                       & \gate{\ket{j_y}}          & \qw                       & \qw       & \qw       & \ghost{\ket{\bm{j}}}        & \ghost{\ket{\bm{j}}}         & \qw         & \qw       & \qw       & \qw        \\
    \lstick{v_z} & {/}^{n_{v_z}} \qw & \ghost{\ket{\bm{j}}}        & \ctrlo{-1} & \qw      & \ghost{H}        & \ctrlo{-1} & \qw      & \ghost{H}        & \ctrlo{-1} & \qw      & \qw      & \ghost{H}        & \qw                       & \qw                       & \gate{\ket{j_z}}          & \qw       & \qw       & \ghost{\ket{\bm{j}}}        & \ghost{\ket{\bm{j}}}         & \qw         & \qw       & \qw       & \qw           \\
  }\]
  \caption{Quantum circuit of the unitary $U_{\mathrm{col}}$ for the three-dimensional linearized Vlasov-Poisson system.}
  \label{fig:circuit_of_3D_U_col}
\end{figure*}
The discussion of initialization and extracting data steps for the one-dimensional linearized Vlasov-Poisson system
can easily be extended to the higher system. In this section, we focus on the construction of unitaries
$U=U_{\mathrm{row}}^{\dagger}U_{\mathrm{col}}$ that are $(\alpha, a, 0)$-block-encoding of $H$ for the higher systems.

\begin{table*}[t]
  \centering
  \caption{Comparison of Hamiltonian simulation for the linearized Vlasov-Poisson system.
            The system register is defined as register labeled by $r$ and $v$.}
  \begin{tabular}{ccccc} \hline
    Dimension & \begin{tabular}{c}
      The total grid size \\ $N_v$
    \end{tabular}                       & \begin{tabular}{c}
      The number of \\system register
    \end{tabular}                                                           & \begin{tabular}{c}
      The number of \\ancilla qubits of $U$
    \end{tabular}                                                               & Gate complexity \rule[0mm]{0mm}{3mm} \\ \hline \hline
    1         & $N_{v_x}$               & $n_{v_x} + 1$                     & 4 &   \\ 
    2         & $N_{v_x}N_{v_y}$        & $n_{v_x} + n_{v_y} + 2$           & 6 & $\mathcal{O}\left(
                                                                                    \mathrm{poly}(\log N_v)\left(
                                                                                      t + \log(1/\varepsilon)
                                                                                    \right)
                                                                                  \right)$                  \rule[0mm]{0mm}{5mm} \\
    3         & $N_{v_x}N_{v_y}N_{v_z}$ & $n_{v_x} + n_{v_y} + n_{v_z} + 2$ & 7 &                     \rule[0mm]{0mm}{5mm} \\ \hline                   
  \end{tabular}
  \label{tb:HS_for_the_linearized_Vlasov_Poisson_system}
\end{table*}

We show the circuits $U_{\mathrm{row}}$ and $U_{\mathrm{col}}$ for the two-dimensional system
in Figs.~\ref{fig:circuit_of_2D_U_row} and~\ref{fig:circuit_of_2D_U_col}. The unitary $U$ is
an $(\alpha, 6, 0)$-block-encoding of the Hamiltonian $H$:
\begin{align}
    &\frac{H}{\alpha} = \sum_{\bm{j}}\biggl[
      c^2b_{\bm{j}}^{2}\ket{0}_r\ket{\bm{j}}_v\bra{0}_r\bra{\bm{j}}_v \notag \\
      &+\frac{\sqrt{1-|c|^2}d_{\bm{j}}}{\sqrt{2N_v}} \biggl(p_{j_x}\bigl(
          \ket{0}_r\ket{\bm{j}}_v\bra{1}_r\bra{\bm{0}}_v + \ket{1}_r\ket{\bm{0}}_v\bra{0}_r\bra{\bm{j}}_v
        \bigr)  \notag \\
      &+ q_{j_y}\bigl(
          \ket{0}_r\ket{\bm{j}}_v\bra{2}_r\bra{\bm{0}}_v + \ket{2}_r\ket{\bm{0}}_v\bra{0}_r\bra{\bm{j}}_v
        \bigr) \biggr)
    \biggr] + \hat{D}_2, \label{eq:2D_hamiltonian2}
\end{align}
where $\hat{D}_2$ is the unused subspace. Comparing Eq.~\eqref{eq:3D_hamiltonian1} for the two-dimensional system
with Eq.~\eqref{eq:2D_hamiltonian2}, we obtain
\begin{equation}
  \begin{split} \label{eq:2D_coefficient_comparison}
    c^2b_{\bm{j}}^{2} &= \frac{k_x v_{j_x}+k_y v_{j_y}}{\alpha}, \\
    \sqrt{1-|c|^2}\frac{p_{j_x}d_{\bm{j}}}{\sqrt{2N_v}} &= \frac{\mu_{\bm{j}}v_{j_x}}{\alpha},\\
    \sqrt{1-|c|^2}\frac{q_{j_y}d_{\bm{j}}}{\sqrt{2N_v}} &= \frac{\mu_{\bm{j}}v_{j_y}}{\alpha},
  \end{split}
\end{equation}
where $N_v=N_{v_x}N_{v_y}$. The angles $b_j,d_j,p_{j_x},q_{j_y}$ are chosen as follows:
\begin{equation}
  \begin{split} \label{eq:2D_angles}
    b_{\bm{j}} = &\sqrt{\frac{k_x v_{j_x}+k_y v_{j_y}}{K_{\max}}},\quad d_{\bm{j}} = \sqrt{\frac{f_{\mathrm{M}}(v_{\bm{j}})}{g_{\max}}},\\
    p_{j_x} &= \frac{v_{j_x}}{V_{\max}},\quad q_{j_y} = \frac{v_{j_y}}{V_{\max}},
  \end{split}
\end{equation}
where
\begin{equation}
  \begin{split}
    K_{\max} &= \max_{\bm{j}}|k_x v_{j_x}+k_y v_{j_y}|, \\
    g_{\max} &= \max_{\bm{j}}f_{\mathrm{M}}(v_{j_x}, v_{j_y}), \\
    V_{\max} &= \max_{j_x}|v_{j_x}|\max_{j_y}|v_{j_y}|=v_{x, \max}v_{y, \max}.
  \end{split}
\end{equation}
Note that we assume $v_{x, \max}\geq1, v_{y, \max}\geq1$ to make $|p_{j_x}|\leq 1, |q_{j_y}|\leq 1$ hold.
From Eqs.~\eqref{eq:2D_coefficient_comparison} and~\eqref{eq:2D_angles}, $c$ and $\alpha$ are given by
\begin{align}
  c^2 &= \frac{\Gamma}{2}\left(
    \sqrt{1+\frac{4}{\Gamma}} - 1
  \right), \label{eq:2D_c_squared} \\
  \alpha &= \frac{K_{\max}}{c^2}, \label{eq:2D_alpha}
\end{align}
where
\begin{equation}
  \Gamma = \frac{K_{\max}^{2}}{2\Delta v N_v V_{\max}^{2} g_{\max}},
\end{equation}
and $\Delta v=\Delta v_x\Delta v_y$. As for the one-dimensional system, $\alpha$ satisfies the following inequality:
\begin{equation} \label{eq:inequality_of_2D_alpha}
  \frac{4\Lambda}{5} \leq \alpha \leq \Lambda,
\end{equation}
where 
\begin{equation} \label{eq:2D_Lambda}
  \Lambda = K_{\max} + \sqrt{2\Delta v N_v V_{\max}^{2} g_{\max}}.
\end{equation}
Since $\Delta v N_v=(2v_{x, \max}+\Delta v_x)(2v_{y, \max}+\Delta v_y)$, if $N_v$ increases,
then $\Lambda$ does not increase., i.e. $\alpha=\Theta(1)$
Therefore, the query complexity also does not increase with increasing $N_v$.
In addition, the gate complexity scales logarithmically with $N_{v}$ because
the input register of the variable rotations
in Fig.~\ref{fig:circuit_of_2D_U_row} and~\ref{fig:circuit_of_2D_U_col} has $n_{v_x}+n_{v_y}$ qubits.
These results are the same as for the one-dimensional system.

We show a unitary $U=U_{\mathrm{row}}^{\dagger}U_{\mathrm{col}}$ for the three-dimensional Vlasov-Poisson system
and similar results to the ones for the lower dimensional systems. The circuits $U_{\mathrm{row}}$ and $U_{\mathrm{col}}$
are shown in Figs.~\ref{fig:circuit_of_3D_U_row} and~\ref{fig:circuit_of_3D_U_col}. The unitary $U$ is
an $(\alpha, 7, 0)$-block-encoding of the Hamiltonian $H$:
\begin{align}
  &\frac{H}{\alpha} = \sum_{\bm{j}}\biggl[
    c^2b_{\bm{j}}^{2}\ket{0}_r\ket{\bm{j}}_v\bra{0}_r\bra{\bm{j}}_v \notag \\
    &+ \frac{\sqrt{1-|c|^2}d_{\bm{j}}}{2\sqrt{N_v}}\biggl(p_{j_x}\left(
        \ket{0}_r\ket{\bm{j}}_v\bra{1}_r\bra{\bm{0}}_v + \ket{1}_r\ket{\bm{0}}_v\bra{0}_r\bra{\bm{j}}_v
      \right) \notag \\
    &+ q_{j_y}\left(
        \ket{0}_r\ket{\bm{j}}_v\bra{2}_r\bra{\bm{0}}_v + \ket{2}_r\ket{\bm{0}}_v\bra{0}_r\bra{\bm{j}}_v
      \right) \notag \\
    &+ r_{j_z}\left(
      \ket{0}_r\ket{\bm{j}}_v\bra{3}_r\bra{\bm{0}}_v + \ket{3}_r\ket{\bm{0}}_v\bra{0}_r\bra{\bm{j}}_v
    \right) \biggr)
  \biggr] + \hat{D}_3, \label{eq:3D_hamiltonian2}
\end{align}
where $N_v=N_{v_x}N_{v_y}N_{v_z}$ and $\hat{D}_3$ is the unused subspace.
The corresponding angles and $\alpha$ are as follows:
\begin{equation}
  \begin{split}
    b_j = \sqrt{\frac{k_x v_{j_x}+k_y v_{j_y}+k_z v_{j_z}}{K_{\max}}},\\
    d_j = \sqrt{\frac{f_{\mathrm{M}}(v_{\bm{j}})}{g_{\max}}},\quad p_{j_x} = \frac{v_{j_x}}{V_{\max}},\\
    q_{j_y} = \frac{v_{j_y}}{V_{\max}},\quad r_{j_z} = \frac{v_{j_z}}{V_{\max}},
  \end{split}
\end{equation}
\begin{align}
  c^2 &= \frac{\Gamma}{2}\left(
    \sqrt{1+\frac{4}{\Gamma}} - 1
  \right), \label{eq:3D_c_squared} \\
  \alpha &= \frac{K_{\max}}{c^2},
\end{align}
where
\begin{equation}
  \begin{split}
    K_{\max} &= \max_{\bm{j}}|k_x v_{j_x}+k_y v_{j_y}+k_z v_{j_z}|, \\
    g_{\max} &= \max_{\bm{j}}f_{\mathrm{M}}(v_{j_x}, v_{j_y}, v_{j_z}), \\
    V_{\max} &= v_{x, \max}v_{y, \max}v_{z, \max}, \\
    \Gamma &= \frac{K_{\max}^{2}}{4\Delta v N_v V_{\max}^{2} g_{\max}}.
  \end{split}
\end{equation}
As for the lower dimensional system, $\alpha$ satisfies the following inequality:
\begin{equation}
  \frac{4\Lambda}{5} \leq \alpha \leq \Lambda,
\end{equation}
where 
\begin{equation}
  \Lambda = K_{\max} + \sqrt{4\Delta v N_v V_{\max}^{2} g_{\max}}.
\end{equation}
Since $\Delta v N_v=(2v_{x, \max}+\Delta v_x)(2v_{y, \max}+\Delta v_y)(2v_{z, \max}+\Delta v_z)$, if $N_v$ increases,
then $\Lambda$ does not increase.
Therefore, the same result is obtained for the query complexity and the gate complexity.
We summarize the computational resources of HS
for the linearized Vlasov-Poisson system in Table~\ref{tb:HS_for_the_linearized_Vlasov_Poisson_system}.

\section{Numerical Results} \label{section:numerical_results}
\subsection{QSVT-based Hamiltonian simulation}
\begin{figure}[tbp]
  \begin{minipage}[c]{\linewidth}
    \includegraphics[width=\linewidth]{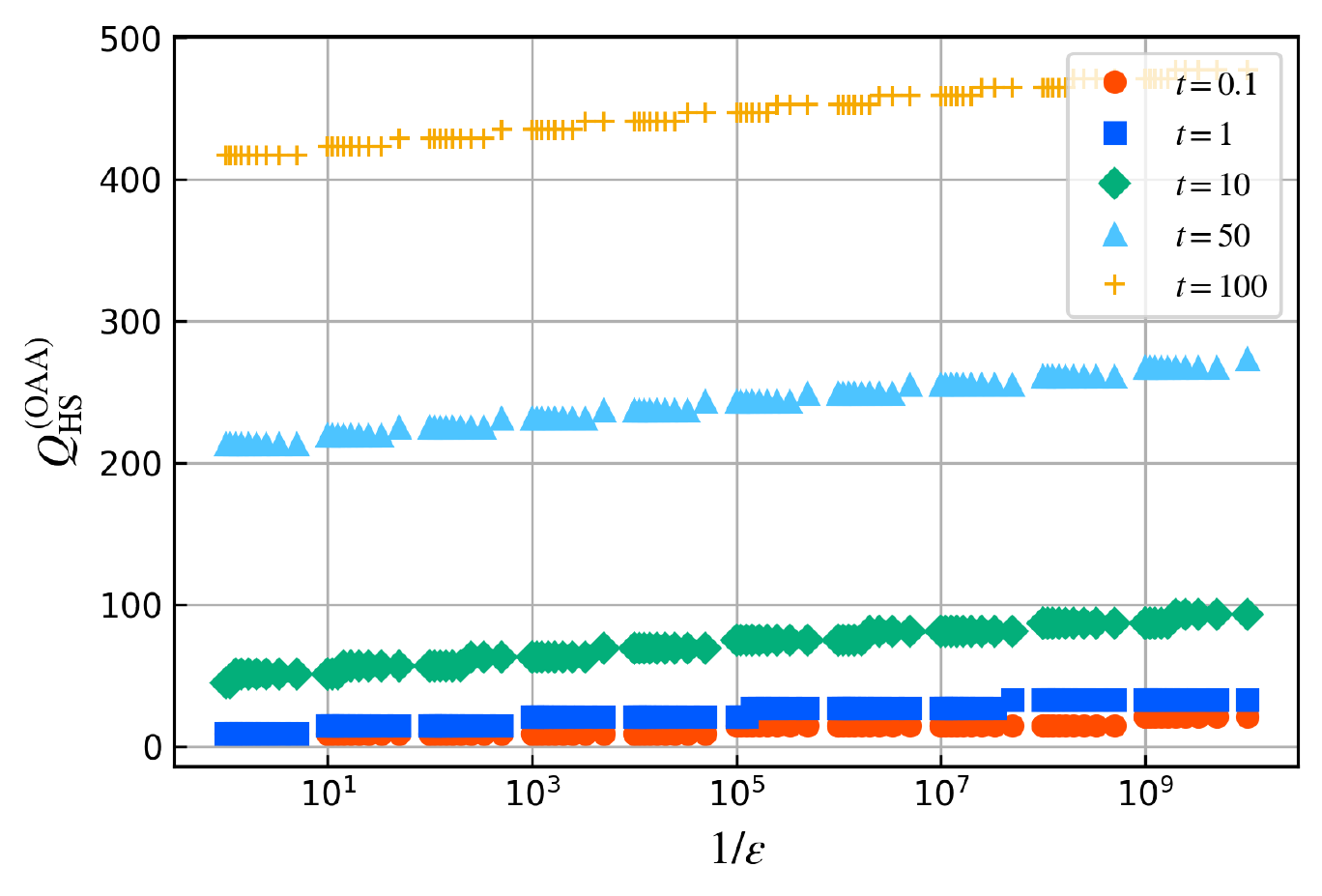}
    \subcaption{OAA-based HS.}
    \label{fig:number_of_queries_vs_error_for_OAA}
  \end{minipage} \\
  \begin{minipage}[c]{\linewidth}
    \includegraphics[width=\linewidth]{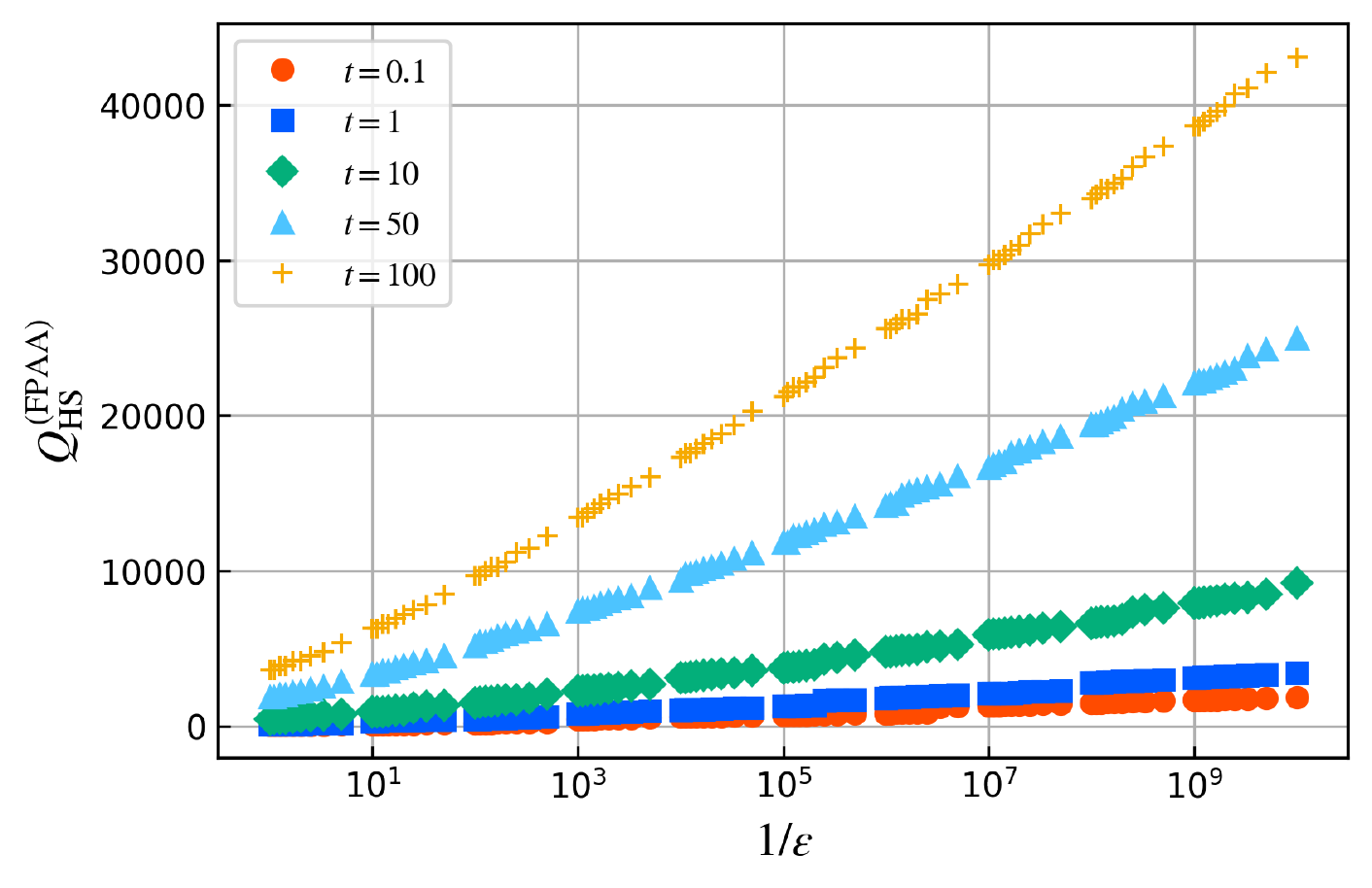}
    \subcaption{FPAA-based HS.}
    \label{fig:number_of_queries_vs_error_for_FPAA}
  \end{minipage}
  \caption{The number of queries vs. $1/\varepsilon$.}
  \label{fig:number_of_queries_vs_error}
\end{figure}

\begin{figure}[tbp]
  \begin{minipage}[c]{\linewidth}
    \includegraphics[width=\linewidth]{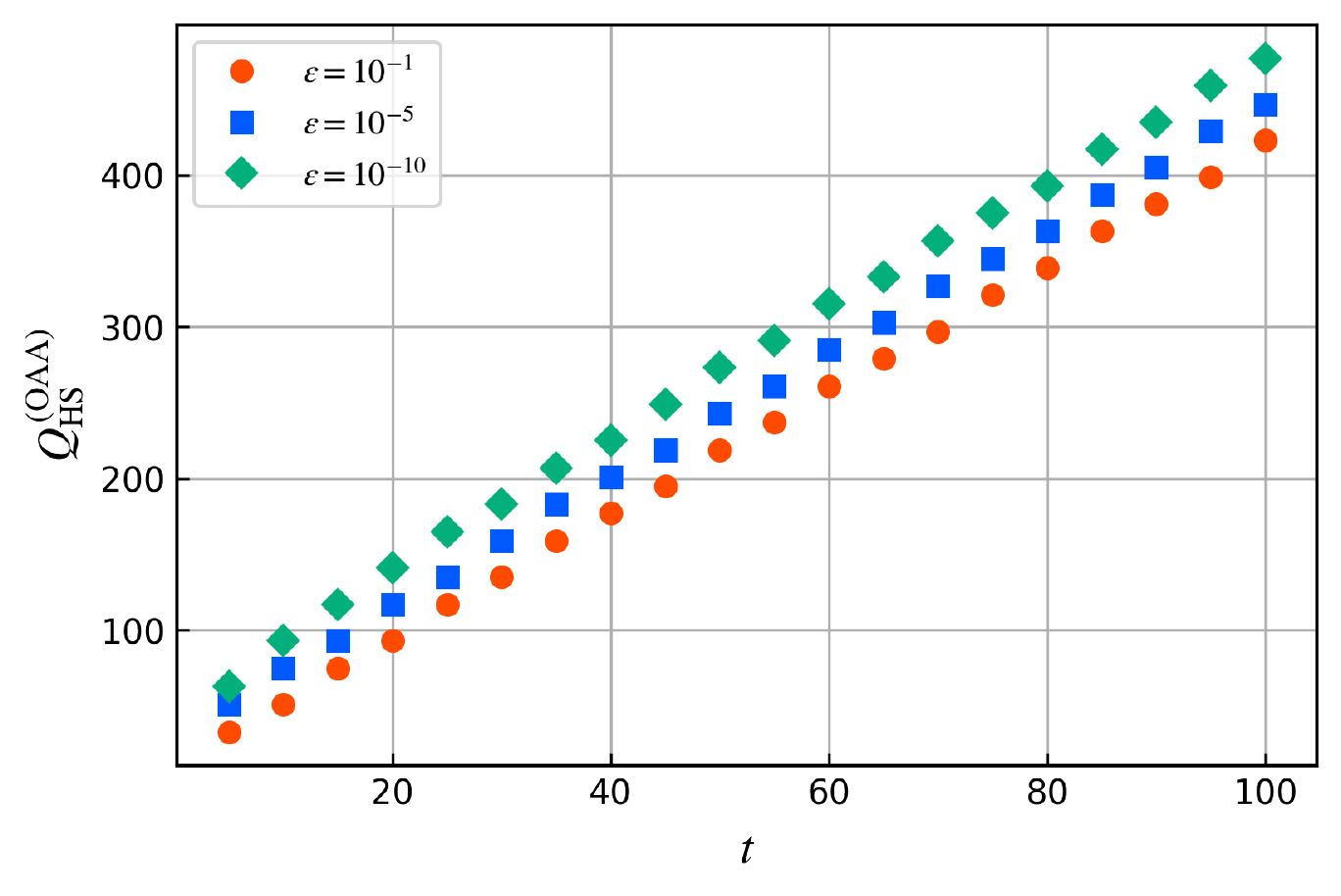}
    \subcaption{OAA-based HS.}
    \label{fig:number_of_queries_vs_time_for_OAA}
  \end{minipage} \\
  \begin{minipage}[c]{\linewidth}
    \includegraphics[width=\linewidth]{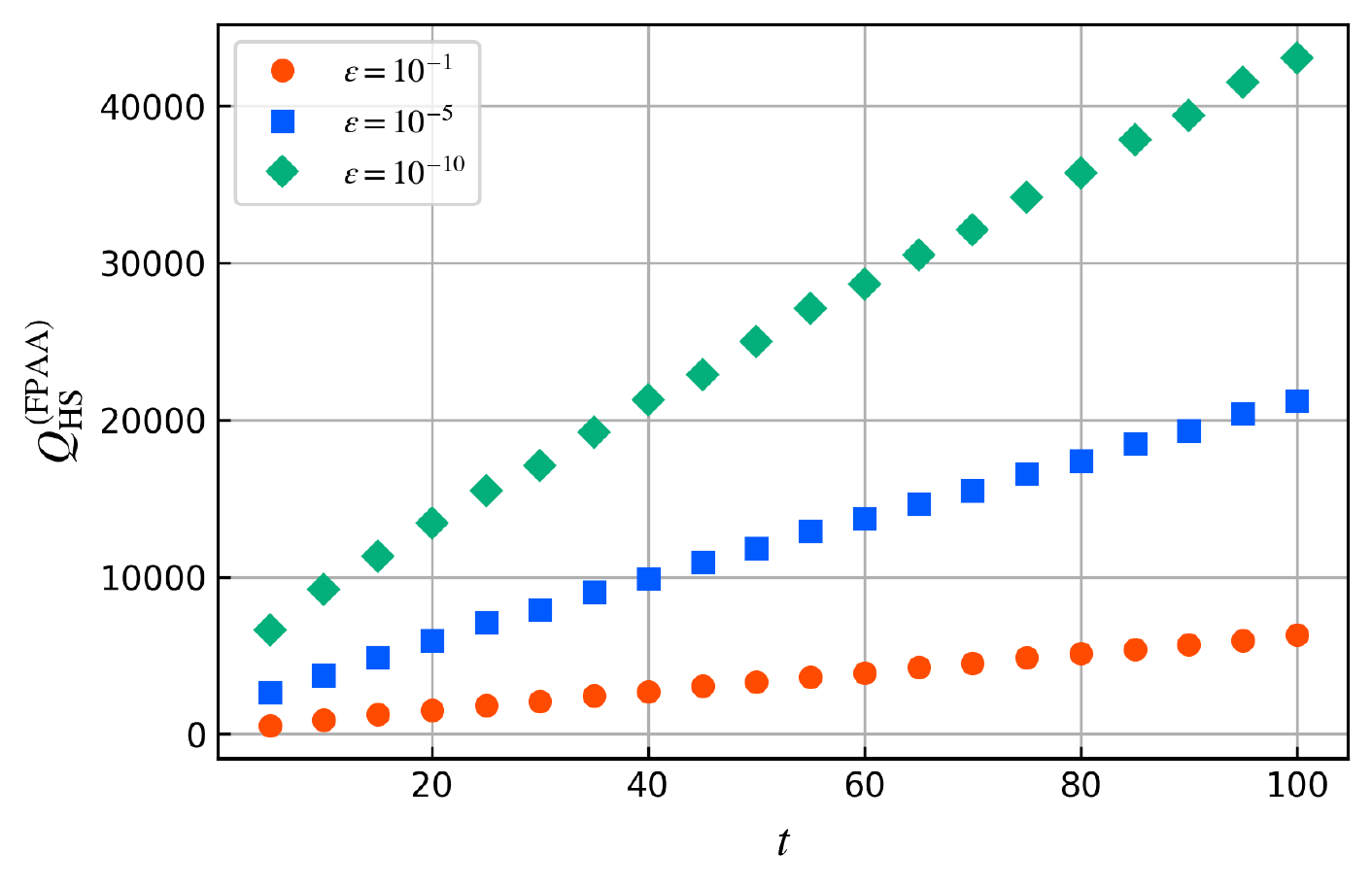}
    \subcaption{FPAA-based HS.}
    \label{fig:number_of_queries_vs_time_for_FPAA}
  \end{minipage}
  \caption{The number of queries vs. $t$.}
  \label{fig:number_of_queries_vs_time}
\end{figure}

We compare the number of queries of the OAA-based and FPAA-based HS algorithms.
The number of queries for some given error tolerances $\varepsilon$ and evolution times $t$
calculated from Eqs.~\eqref{eq:query_complexity_of_HS_using_OAA} and~\eqref{eq:query_of_HS_using_FPAA}
is shown in Figs.~\ref{fig:number_of_queries_vs_error} and \ref{fig:number_of_queries_vs_time}.
Notably, the number of queries of the OAA-based HS is significantly smaller than that of
the FPAA-based one for all parameters.
These figures show that the number of queries
$Q_{\mathrm{HS}}^{(\mathrm{OAA})}$ and $Q_{\mathrm{HS}}^{(\mathrm{FPAA})}$
scale linearly for $t$ and linearly and quadratically for $\log(1/\varepsilon)$, respectively.
These results are consistent with the asymptotic scaling of
Eqs.~\eqref{eq:query_complexity_of_HS_using_OAA} and~\eqref{eq:query_complexity_of_HS_using_FPAA}.


\begin{table}[tbp]
  \centering
  \caption{The constant factors and coefficients of Eqs.~\eqref{eq:query_complexity_of_HS_using_OAA},~\eqref{eq:query_of_HS_using_FPAA}.}
    \begin{tabular}{ccccccc} \hline
      Method                & Range of parameters  & $\alpha_0$ & $\alpha_1$ & $\alpha_2$ & $\alpha_3$ & $\alpha_4$ \\ \hline \hline
      \multirow{2}{*}{OAA}  & $0.1\leq t \leq 10$               & \multirow{2}{*}{2.73} & \multirow{2}{*}{4.88} & \multirow{2}{*}{1.78} &  &  \\ 
                            & $10^{-5}\leq\varepsilon\leq 0.9$  &  &  &  \\ \hline
      \multirow{2}{*}{OAA}  & $1\leq t \leq 100$                & \multirow{2}{*}{2.77} & \multirow{2}{*}{4.18} & \multirow{2}{*}{2.45} &  &  \\
                            & $10^{-10}\leq\varepsilon\leq 0.9$ &  &  &  \\ \hline
      \multirow{2}{*}{FPAA} & $0.1\leq t \leq 10$               & \multirow{2}{*}{142} & \multirow{2}{*}{28.3} & \multirow{2}{*}{11.9} & \multirow{2}{*}{20.9} & \multirow{2}{*}{7.26} \\ 
                            & $10^{-5}\leq\varepsilon\leq 0.9$  &  &  &  \\ \hline
      \multirow{2}{*}{FPAA} & $1\leq t \leq 100$                & \multirow{2}{*}{490}  & \multirow{2}{*}{21.7} & \multirow{2}{*}{-15.6} & \multirow{2}{*}{15.4} & \multirow{2}{*}{10.9} \\
                            & $10^{-10}\leq\varepsilon\leq 0.9$ &  &  &  \\ \hline                  
    \end{tabular}
  \label{tb:constant_factors_and_coefficients_of_number_of_queries}
\end{table}
To identify the constant factors and coefficients of the number of queries hidden behind the asymptotic scaling,
we fit the curve for OAA with
\begin{equation}
  Q_{\mathrm{HS}}^{(\mathrm{OAA})} = \alpha_{0} + \alpha_1 t + \alpha_2\log(1/\varepsilon),
\end{equation}
and the curve for FPAA with
\begin{align}
  Q_{\mathrm{HS}}^{(\mathrm{FPAA})} = \alpha_{0}
    &+ \alpha_1 t + \alpha_2\log(1/\varepsilon) \notag \\
    &+ \alpha_3t\log(1/\varepsilon) + \alpha_4\log^{2}(1/\varepsilon).
\end{align}
The results are presented
in Table~\ref{tb:constant_factors_and_coefficients_of_number_of_queries}. 
Both values of FPAA are greater than those of OAA.
Thus, OAA proves more effective for HS than FPAA in terms of the number of queries.

We emphasize that the advantage of OAA over FPAA in HS is a general result.
The reasons for this can be explained as follows.
The number of queries is calculated from Eqs.~\eqref{eq:query_complexity_of_HS_using_OAA}
and \eqref{eq:query_of_HS_using_FPAA}. These equations are derived
under the general assumption that the Hamiltonian is positive semidefinite and
its norm is less than 1; that is, Eqs.~\eqref{eq:query_complexity_of_HS_using_OAA}
and \eqref{eq:query_of_HS_using_FPAA} hold without respect to
the type of the Hamiltonian. Therefore, from the theoretical and numerical results of
the number of queries, OAA is more advantageous than FPAA in Hamiltonian simulations of general systems.

To specify which degree of approximation of trigonometric functions or sign function
dominates the number of queries of FPAA-based HS, we fit the curves for $R$ and $D$ in Eq.~\eqref{eq:query_of_HS_using_FPAA} with
\begin{align}
  R & = \alpha_0 + \alpha_1 t + \alpha_2\log(1/\varepsilon),\\
  D & = \alpha_0 + \alpha_2\log(1/\varepsilon),
\end{align}
for $0.1\leq t \leq 10, 10^{-5}\leq\varepsilon\leq 0.9$. Then we obtain the following results:
for $R$,
\begin{equation}
  \alpha_0 = 0.853,\quad \alpha_1=0.913,\quad \alpha_2=0.293,
\end{equation}
for $D$,
\begin{equation}
  \alpha_0=21.8,\quad \alpha_2=10.1.
\end{equation}
The constant factors and coefficients of $D$ are larger than those of $R$.
Thus, the large number of queries of FPAA-based HS is caused by requiring a high degree
of approximation of the sign function.

\begin{figure}[tbp]
  \centering
  \includegraphics[width=\linewidth]{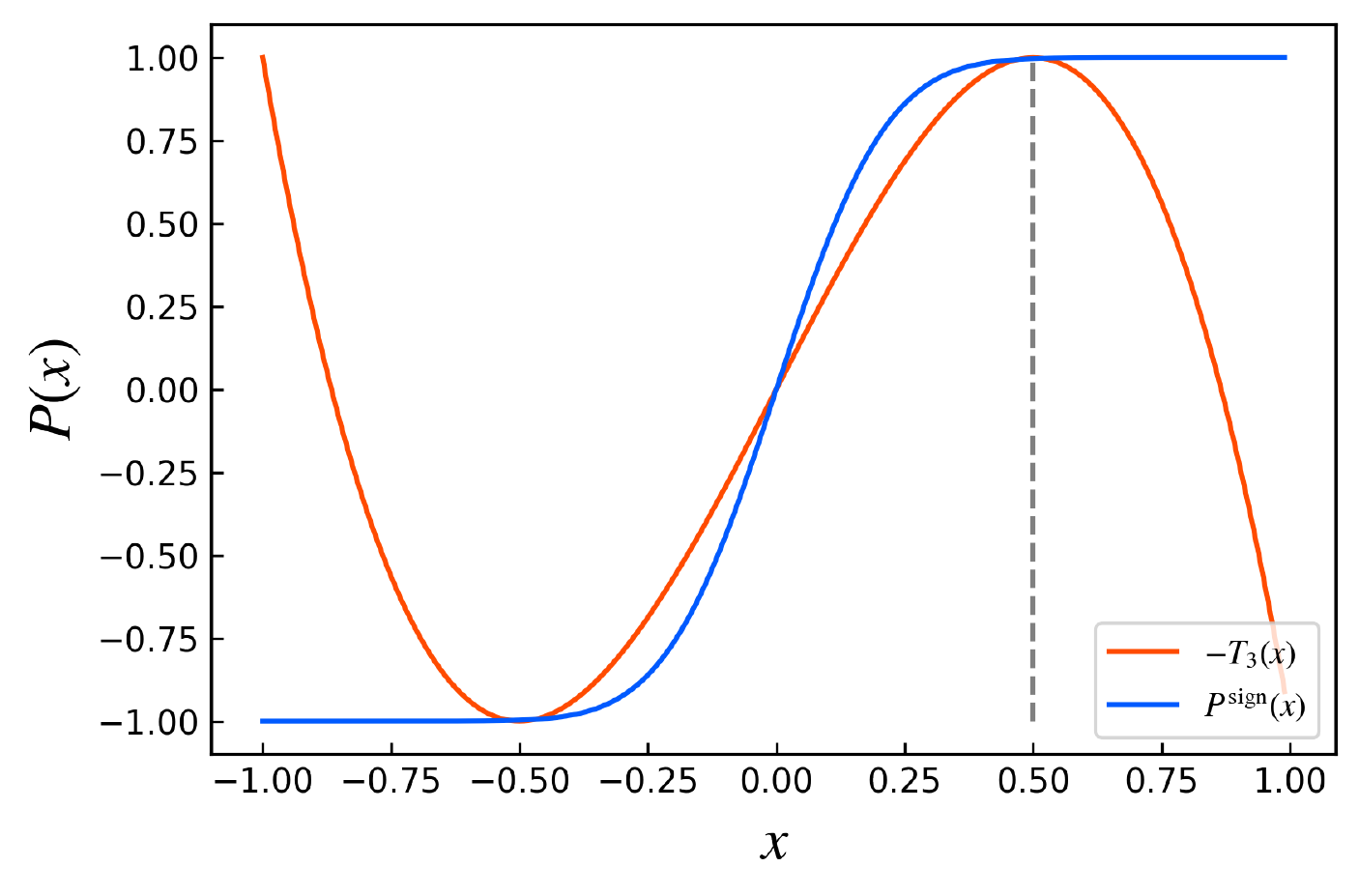}
  \caption{Illustration of the difference between amplitude amplifications in OAA-based and FPAA-based HS.}
  \label{fig:cheb_polynomial_vs_sign_function}
\end{figure}

We now explain the intuitive reason why it is not appropriate to 
use a sign function for amplitude amplification in QSVT-based HS.
As seen in Sec.~\ref{subsection:applying_QSVT_to_trigonometric_functions},
the specific amplitude value $\kappa/2\approx1/2$ must be amplified
in QSVT-based HS.
For OAA-based HS, the 3rd Chebyshev polynomial is precisely $|T_3(x)|=1$ at $x=1/2$, whereas
for the FPAA-based one, the sign function is $\mathrm{sign}(x)=1$ for $1/2 \leq x\leq 1$.
The OAA-based method amplifies the value exclusively at $x=1/2$. In contrast, the FPAA-based method
aims to amplify the values for $1/2\leq x\leq1$, leading to extra, unneeded effort for amplification
in this range, as depicted in Fig.~\ref{fig:cheb_polynomial_vs_sign_function}.
This results in a high degree of approximation of the sign function and many queries for FPAA-based HS.

\subsection{Application to the linearized Vlasov-Poisson system}
\begin{figure}[tbp]
  \centering
  \includegraphics[width=\linewidth]{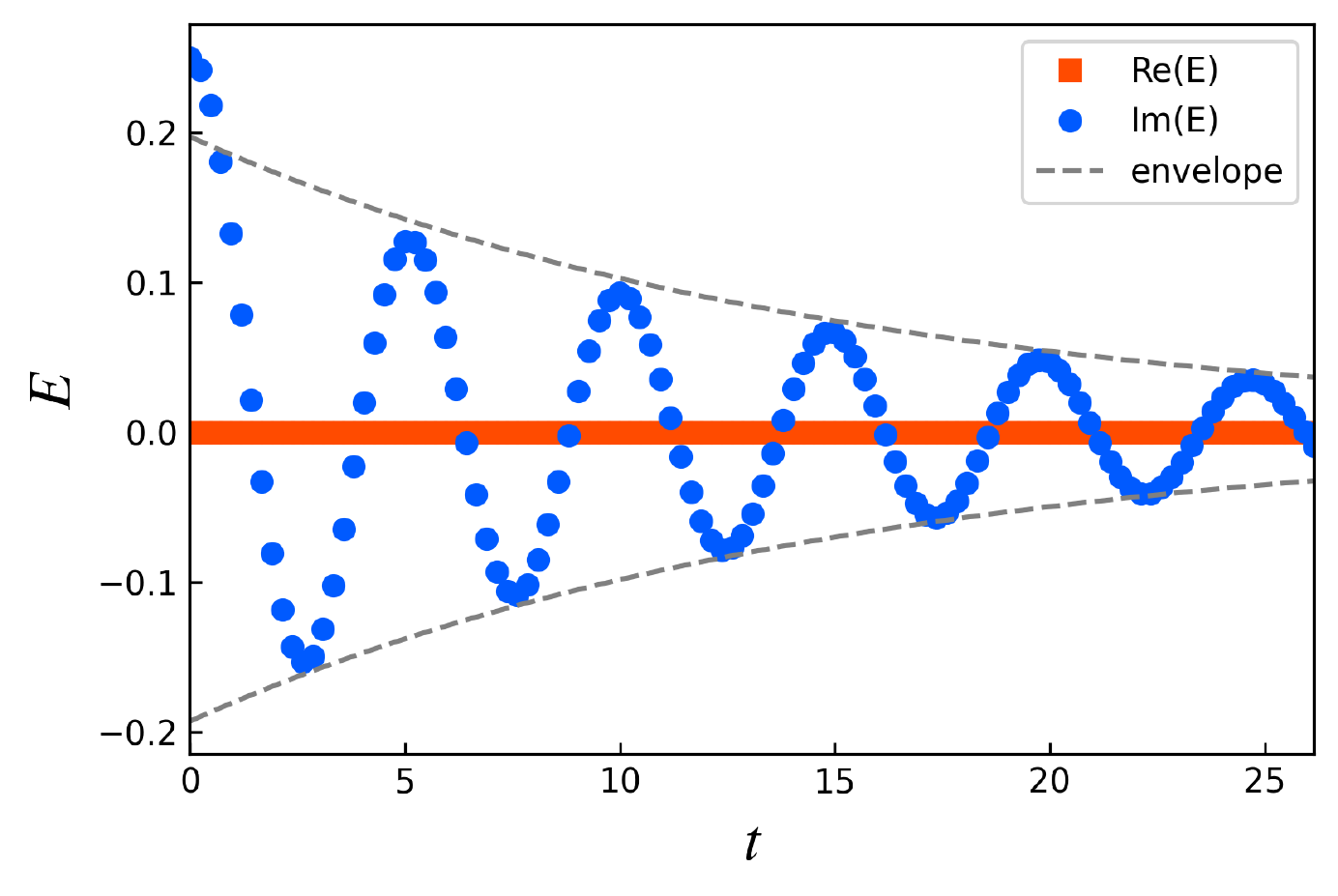}
  \caption{Time evolution of the electric field $E$ when using the quantum algorithm with $\Delta t = 0.238$ ($k=0.4$).
            The envelope is a fitted exponential $\pm \exp(\gamma t)$.}
  \label{fig:time_evolution_of_E_k_0.4}
\end{figure}
\begin{figure}[tbp]
  \centering
  \includegraphics[width=\linewidth]{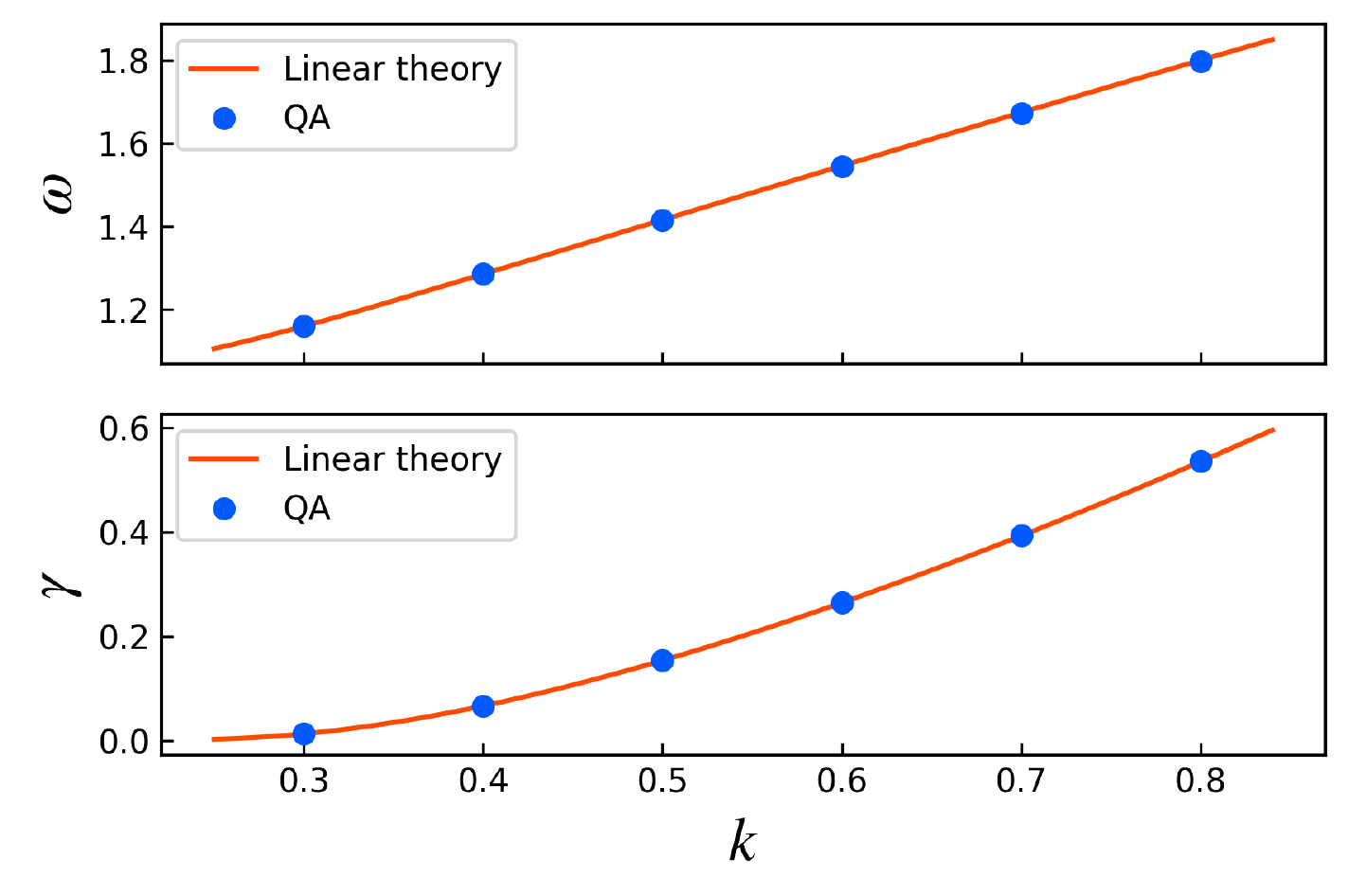}
  \caption{Comparison of the frequencies and damping rates obtained from the results of the quantum algorithm (QA)
            for various wavenumbers with those obtained from the linear Landau theory.}
  \label{fig:omega_and_gamma_vs_wavenumber}
\end{figure}
Here, the OAA-based HS is applied to the simulation of the one-dimensional linearized Vlasov-Poisson system.
The simulation is implemented on a classical emulator of a quantum computer using Qiskit~\cite{ANIS2021Qiskit},
especially \textit{statevector simulator} as the backend. This backend gives us access to the whole output space
at all moments, and we do not implement QAE directly for saving the number of qubits.
We compare the simulation results of the quantum algorithm using HS with those of a classical algorithm which have been obtained
by directly solving the Vlasov equation and the Poisson equation using the Euler method for the time.
These simulations are performed using the following parameters for a given wavenumber $k$:
\begin{align}
  N_{v} &= 32, & f_{\mathrm{M}}(v_j) &= \frac{1}{\sqrt{2\pi}}e^{-\frac{1}{2}v_{j}^{2}}, \notag \\
  v_{\max} &= 4.5, & f_{1}(v_j, t&=0) = 0.1 f_{\mathrm{M}}(v_j),
\end{align}
\begin{equation}
  E(t=0) = \frac{i}{k}\sum_{j}f_{1}(v_j, t=0)\Delta v. \notag
\end{equation}
We construct the unitary $U'$ in Fig.~\ref{fig:circuit_of_U_dash}
from the unitary $U=U_{\mathrm{row}}^{\dagger}U_{\mathrm{cal}}$ in Ref.~\cite{engel2019quantum}, 
which is the $(\alpha, 4, 0)$-block-encoding of $H$. Using the unitary $U'$,
we construct the circuit $U_{\mathrm{OAA}}$ for $t=2$ and choose an error tolerance $\varepsilon = 10^{-3}$. Then,
$U_{\mathrm{OAA}}$ is a $(1,7,10^{-3})$-block-encoding of $e^{-iH\Delta t}$, where
$\Delta t = 1/\alpha$. We implement HS for the evolution time $l\Delta t$ using $l$ sequential $U_{\mathrm{HS}}$
because it is difficult to compute the phases $\Phi^{(\mathrm{c})}$ and $\Phi^{(\mathrm{s})}$
for a large $t$ as mentioned in Sec.~\ref{subsection:HS_for_general_H_and_extension_of_time}.

Figure~\ref{fig:time_evolution_of_E_k_0.4} shows the time evolution of the electric field $E$ for $k=0.4$
using the quantum algorithm. In this case, the normalization of the Hamiltonian becomes $1/\alpha = 0.238$.
After a brief initial stage, the imaginary component of $E$ is damped and oscillating.
We fit the curve with the function $Ae^{-i\gamma (t-t_0)}\cos(\omega(t-t_0)-\rho)+E_0$ to obtain parameters of interest,
i.e., the frequency $\omega$ and damping rate $\gamma$:
\begin{equation}
  \omega = 1.28508, \quad \gamma = 0.06623,
\end{equation}
where $t_0 = 5.23$. One can find precise values of $\omega = 1.28506$ and
$\gamma = 0.06613$ from the linear Landau theory~\cite{chen1984introduction}.
Figure~\ref{fig:omega_and_gamma_vs_wavenumber} shows the comparison of the frequencies and damping rates
obtained from the results of the quantum algorithm with the linear theory for various wavenumbers.
The parameters obtained by fitting the curves agree well with the linear theory.
These results indicate that our quantum algorithm accurately reproduces the linear Landau damping.
Hereafter, the case $k=0.4$ is discussed in both the quantum and classical algorithms.

\begin{figure}[tbp]
  \centering
  \includegraphics[width=\linewidth]{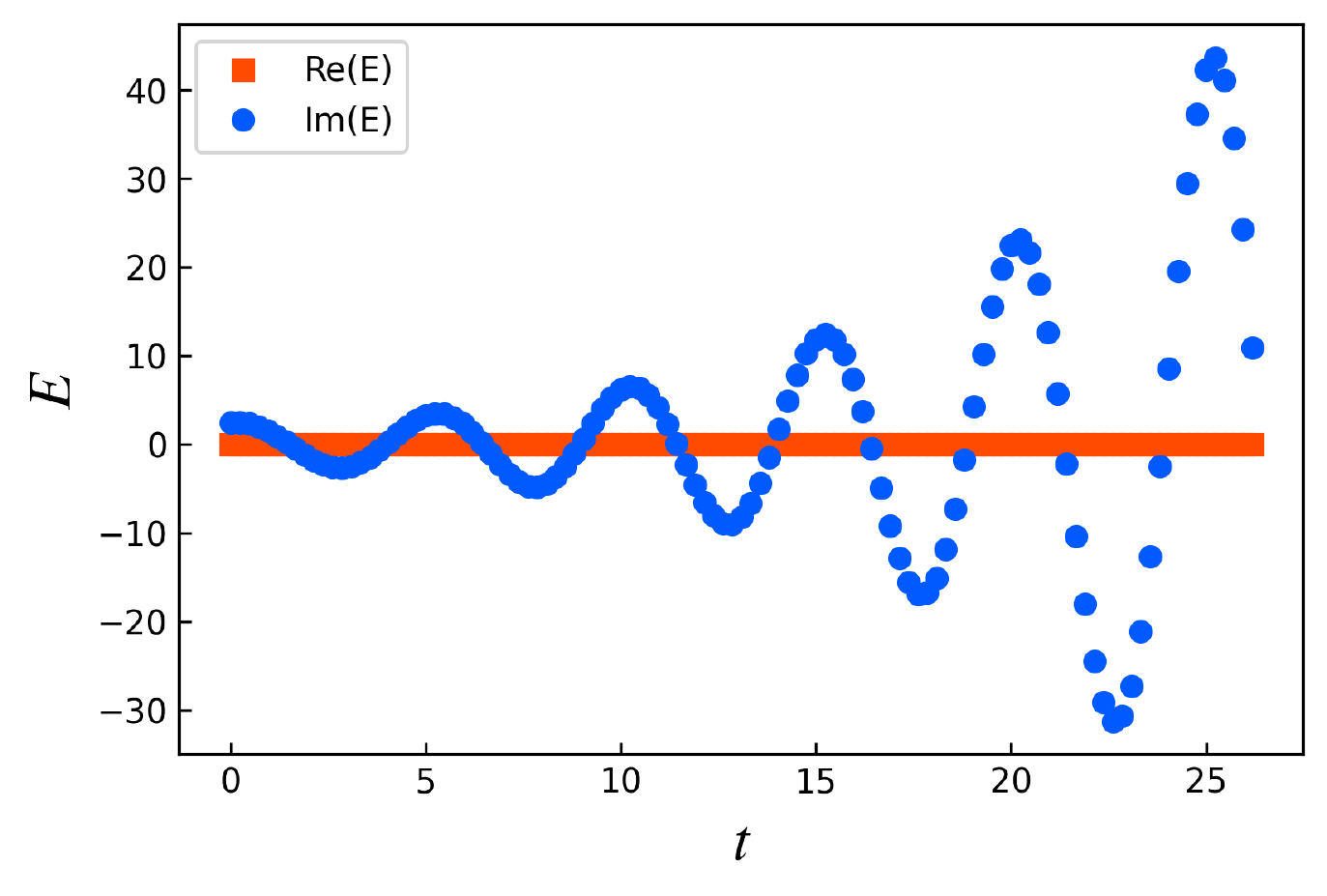}
  \caption{Time evolution of the electric field $E$ when using the classical algorithm using the Euler method
            with $\Delta t = 0.238$ ($k=0.4$).}
  \label{fig:time_evolution_of_E_k_0.4_dt_0.24}
\end{figure}

Figure~\ref{fig:time_evolution_of_E_k_0.4_dt_0.24} shows the time evolution of $E$ with the same time step
$\Delta t = 0.238$ in the classical algorithm using the Euler method.
Unlike the quantum algorithm, the imaginary component of $E$ diverges numerically because of the long time step. 
Table~\ref{tb:comparison_of_the_omega_and_gamma} shows the relative errors of $\omega$ and $\gamma$
for the quantum and classical algorithms with different time steps. 
The classical algorithm requires a smaller time step $\Delta t$ to obtain $\omega$ and $\gamma$
with the same order of accuracy as in the quantum algorithm.

\begin{table}[tbp]
  \centering
  \caption{Comparison of the frequencies and damping rates with different time steps
            obtained from the quantum algorithm (QA) and classical algorithm (CA) for $k=0.4$.}
    \begin{tabular}{lS[table-format=3.2e2]S[table-format=3.2e2]S[table-format=3.2e2]} \hline
                          & {Time step}     & \multicolumn{2}{c}{Relative error [\%]} \rule[0mm]{0mm}{2mm} \\ 
                          & {$\Delta t$}    & \multicolumn{1}{r}{frequency $\omega$} & \multicolumn{1}{r}{damping rate $\gamma$} \\ \hline \hline
      QA                  & 2.38E-1         & 1.67E-3                                & 1.60E-1    \rule[0mm]{0mm}{2mm}       \\ \hline
      \multirow{4}{*}{CA} & 1.00E-2         & 6.25E-2                                & 1.24E+1    \rule[0mm]{0mm}{2mm}  \\
                          & 1.00E-3         & 8.77E-3                                & -1.12      \rule[0mm]{0mm}{2mm} \\
                          & 5.00E-4         & 5.52E-3                                & -4.92E-1   \rule[0mm]{0mm}{2mm}  \\
                          & 1.00E-4         & 2.91E-3                                & 6.51E-3    \rule[0mm]{0mm}{2mm} \\ \hline                   
    \end{tabular}
  \label{tb:comparison_of_the_omega_and_gamma}
\end{table}

\begin{figure}[tbp]
  \centering
  \includegraphics[width=\linewidth]{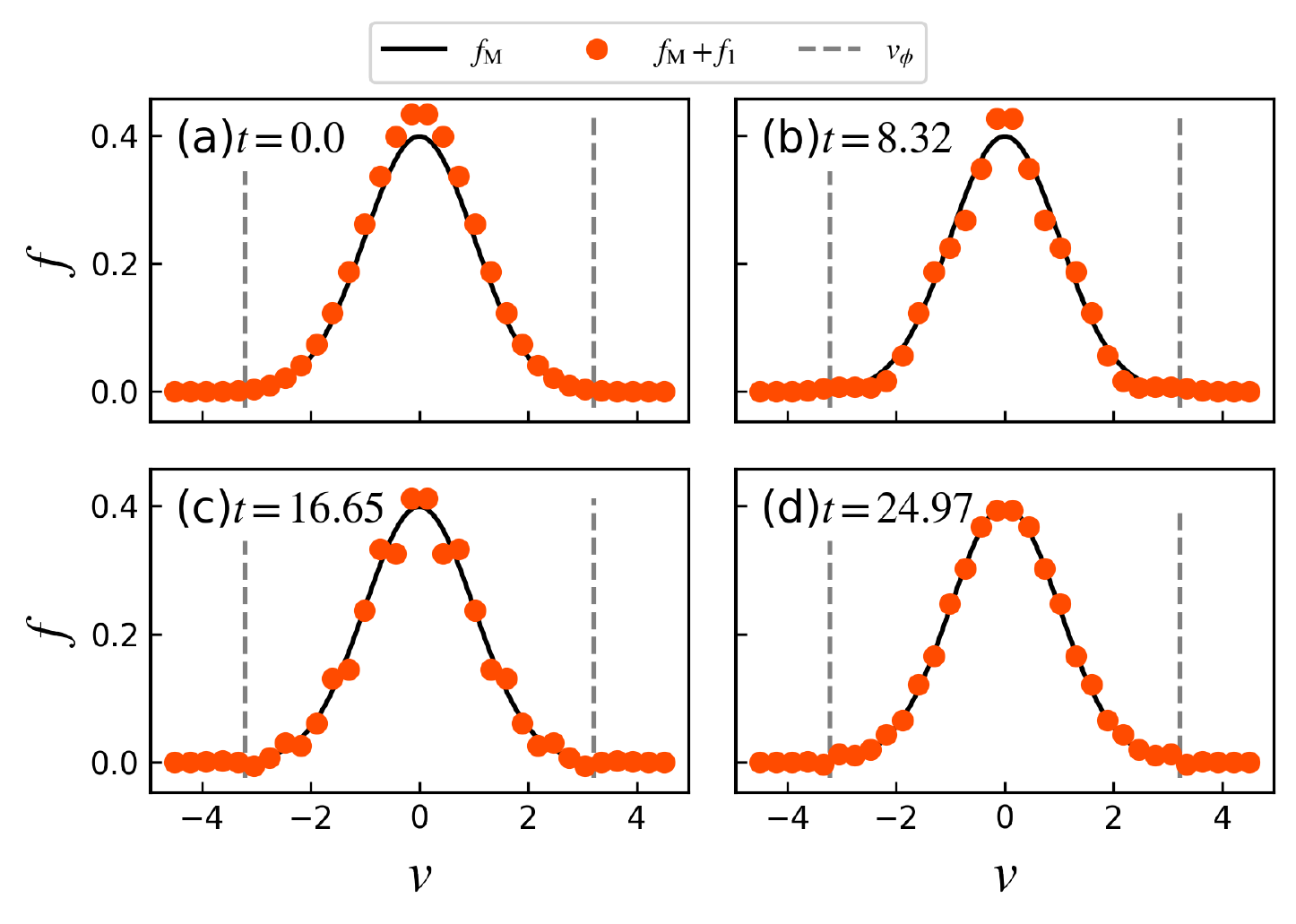}
  \caption{Time evolution of the distribution function $f=f_{\mathrm{M}}+f_1$ ($k=0.4$).
            The solid black curve shows the Maxwellian $f=f_{\mathrm{M}}$ distribution function.
            The gray dashed lines show the theoretical phase velocity $v_{\phi} = 3.21$.}
  \label{fig:time_evolution_of_distribution_function}
\end{figure}

In the linearized Vlasov-Poisson system, the energy transfer between the particles and the electric field occurs,
and the distribution functions have a wavy structure.
Figure~\ref{fig:time_evolution_of_distribution_function} shows the velocity profiles of the
distribution functions at different times. The distribution functions
at $t=8.32, 16.65$, and $24.97$ have a wavy structure. At $t=16.65$ and $24.97$, the structure appears mainly around the phase velocity
$v_{\phi}=\omega/k=3.213$. These results are consistent with the linear Landau theory~\cite{chen1984introduction}.

We validate the results of the distribution functions by comparing them with those of the classical algorithm.
The error between different distribution functions is defined as
\begin{equation} \label{eq:error_between_different_distribution_functions}
  \delta(f, g) = \sum_{j}\left|
    f_{j} - g_{j}
  \right|^2\Delta v,
\end{equation}
where $f$ and $g$ are different distribution functions. We denote $f_{\mathrm{QA}}$ and $f_{\mathrm{CA}}$ to
the distribution functions obtained from the quantum algorithm (QA) and the classical algorithm (CA).
The distribution function $f_{\mathrm{CA}}$ is corresponding to the case when $\Delta t = 1\times 10^{-4}$ and $k=0.4$.
We consider the result of this case as accurate because the time step $\Delta t$ is sufficiently small.
Table~\ref{tb:errors_between_distribution_functions_in_q_and_c} shows the errors
between these distribution functions are sufficiently small.
For reference, we show the error $\delta=2.02\times 10^{-5}$ between a Maxwellian distribution $f_{\mathrm{M}}(v)$ and
a drift-Maxwellian distribution $f_{\mathrm{M}}(v-10^{-4})$ for $N_v=32$. These results indicate
that our quantum algorithm can reproduce precisely the structure of the distribution function.

\begin{table}[tbp]
  \centering
  \caption{Errors between the distribution functions $f_{\mathrm{QA}}$ and $f_{\mathrm{CA}}$,
            defined as in Eq.~\eqref{eq:error_between_different_distribution_functions}: 
            $f_{\mathrm{QA}}$ is obtained from the quantum algorithm with $\Delta t = 0.238$ ($k=0.4$);
            $f_{\mathrm{CA}}$ is obtained from the classical algorithm using the Euler method
            with $\Delta t = 1\times 10^{-4}$ ($k=0.4$).}
  \begin{tabular}{SS} \hline
    $t$   & {$\delta$ [$\times 10^{-5}$]} \\ \hline
    0     & 0.000  \\
    8.32  & 0.560 \\
    16.65 & 0.925 \\
    24.97 & 1.06  \\ \hline                   
  \end{tabular}
  \label{tb:errors_between_distribution_functions_in_q_and_c}
\end{table}

\section{Summary} \label{section:summary}
In this study, we have shown how to apply the quantum singular value transformation (QSVT) to 
the Hamiltonian simulation (HS) algorithm and discussed the error and query complexity of HS using 
oblivious amplitude amplification (OAA) and fixed-point amplitude amplification (FPAA) within the QSVT framework. 
As a result, the number of queries for the OAA-based HS scales as $\mathcal{O}(t + \log(1/\varepsilon))$, whereas
the FPAA-based one scales as $\mathcal{O}(t\log(1/\varepsilon) + \log^2(1/\varepsilon))$, where 
$t$ is an evolution time and $\varepsilon$ is an error tolerance. In addition, we
have numerically compared the number of
queries of these HS algorithms, showing that the
number of queries of the OAA-based HS
is smaller than that of the FPAA-based one, regardless of parameters $t$ and $\varepsilon$.
Fitting the curve of the plotted data, we computed the constant factors and coefficients hidden
behind the asymptotic scaling. Then, we found that the values of OAA-based HS 
are smaller than those of FPAA-based HS. We also identified that the large number of queries of FPAA-based HS
is due to the high degree required to approximate the sign function.
Therefore, the OAA method is more appropriate for HS than the FPAA one.

Based on the above findings, applying OAA-based HS to 
the one-dimensional linearized Vlasov-Poisson system,
we simulated the case of electrostatic Landau damping for various wavenumbers
on a classical emulator of a quantum computer using Qiskit~\cite{ANIS2021Qiskit}.
The frequencies $\omega$ and damping rates $\gamma$ obtained by curve fitting the time evolutions
of the electric field $E$ are in agreement with the linear Landau theory~\cite{chen1984introduction}.
Moreover, the velocity profiles of the distribution function $f$ that the quantum algorithm produces 
match the classical ones for the same velocity grid size $N_v$.
These results show that the quantum algorithm can reproduce precisely
the linear Landau damping with the structure of the distribution function.

We have compared the results of the quantum algorithm using HS with those of the classical algorithm using
the Euler method for time. The classical algorithm with a large time step $\Delta t=0.238$ causes
numerical divergence. On the other hand, the quantum algorithm remains stable for the same $\Delta t$.
This stability is because the state at the next time can be
analytically determined by $U=\exp(-iHt)$, which is 
one of the features of the HS algorithm. 
The classical algorithm requires a smaller time step $\Delta t$
to obtain $\omega$ and $\gamma$ with the same order of accuracy as in the quantum algorithm.
These results show that the HS algorithm has advantages in time step over the classical algorithm
using the Euler method.

We have discussed the gate complexity of the algorithm for calculating the time evolution of the electric field $E$.
The complexity scales logarithmically with the total grid size in velocity space $N_v$ and
linearly with the number of time steps $N_t$. We have proposed the algorithm for obtaining the deviation
from the Maxwell distribution. The gate complexity of this algorithm also scales logarithmically with $N_v$.
The circuits of the unitaries which are block-encodings of the Hamiltonian for the higher dimensional systems
have been developed. The gate complexities of HS using the circuits can be represented in the same form
as the one-dimensional system and scales logarithmically with $N_v$.
This result indicates the quantum algorithm for the linearized Vlasov-Poisson system has exponential speedups
over classical algorithms.

\section*{ACKNOWLEDGMENTS}
This work is supported by MEXT Quantum Leap Flagship Program Grant Number JPMXS0118067285 and JPMXS0120319794,
JSPS KAKENHI Grant Number 20H05966, and JST Grant Number JPMJPF2221.

\appendix
\renewcommand{\theequation}{A\arabic{equation} }
\setcounter{equation}{0}
\section{From QSP to QSVT} \label{appendix:from_QSP_to_QSVT} 
Quantum singular value transformation (QSVT) is based on the results of
quantum signal processing (QSP)~\cite{gilyen2019quantum}.
QSP is performed using a series of two gates $W$ and $S$ defined as
\begin{equation}
  W(x) \equiv e^{i\arccos(x)X} =
    \begin{bmatrix}
      x & i\sqrt{1-x^2} \\
      i\sqrt{1-x^2} & x \\
    \end{bmatrix},
\end{equation}
for $x\in[-1,1]$ and
\begin{equation}
  S(\phi) \equiv e^{i\phi Z} =
    \begin{bmatrix}
      e^{\phi} & 0 \\
      0 & e^{-\phi} \\
    \end{bmatrix}.
\end{equation}
These gates construct the following gate sequence
\begin{equation} \label{eq:QSP_sequence_Wx_convention}
  W_{\Phi'} \equiv e^{i\phi'_0 Z}\prod_{k=1}^{d}W(x)e^{i\phi'_k Z},
\end{equation}
where $\Phi'=(\phi'_0,\phi'_1, \ldots, \phi'_d)\in \mathbb{R}^{d+1}$. This convention is called the Wx convention
in Ref.~\cite{martyn2021grand}. 

Another convention is the Reflection convention, which uses a reflection gate $R$ instead of $W$
\begin{equation} \label{eq:reflection_operator}
  R(x) \equiv 
  \begin{bmatrix}
    x & \sqrt{1-x^2} \\
    \sqrt{1-x^2} & -x \\
  \end{bmatrix}.
\end{equation}
The relationship between $W$ and $R$ is given by
\begin{equation}
  W(x) = ie^{-i\frac{\pi}{4}Z}R(x)e^{-i\frac{\pi}{4}Z}.
\end{equation}
Therefore, Eq.~\eqref{eq:QSP_sequence_Wx_convention} is rewritten as
\begin{align}
  e^{i\phi'_0 Z}\prod_{k=1}^{d}W(x)e^{i\phi'_k Z} 
    &= e^{i\phi_0 Z}\prod_{k=1}^{d}R(x)e^{i\phi_k Z} \notag \\
    &\equiv R_{\Phi},
\end{align}
where 
\begin{equation} \label{eq:convert_Wx_convention_to_Reflection_convention}
  \begin{cases}
    \phi_0 = \phi'_{0} + (2d-1)\frac{\pi}{4} \\
    \phi_k = \phi'_{k} - \frac{\pi}{2}\quad (k=1,2,\ldots,d-1) \\
    \phi_d = \phi'_{d} - \frac{\pi}{4}.
  \end{cases}
\end{equation}

The phases $\Phi, \Phi'\in\mathbb{R}^{d+1}$ exist, and the gate sequence
constructs the $(1, 1, 0)$-block-encoding of a polynomial function $P\in\mathbb{C}$
\begin{equation} 
  \bra{0}W_{\Phi'}\ket{0} = \bra{0}R_{\Phi}\ket{0} = P(x), 
\end{equation}
if and only if the conditions (i)-(iv) in Sec.~\ref{section:HS_using_QSVT} hold.
If $P_{\mathbb{\Re}}$ satisfies the conditions (v) and (vi) in Sec.~\ref{section:HS_using_QSVT},
then there exists $P\in\mathbb{C}$ that satisfies $\mathrm{Re}(P)=P_{\mathbb{\Re}}$ and the above conditions 
(i)--(iv).

\begin{figure}[tbp]
  \[\Qcircuit @C=1.5em @R=1.5em {
    \lstick{\ket{0}} & \gate{H} & \ctrlo{1}       & \ctrl{1}         & \gate{H} & \qw \\
    \lstick{\ket{0}} & \qw      & \gate{R_{\Phi}} & \gate{R_{-\Phi}} & \qw      & \qw
  }\]
  \caption{Quantum circuit $U_{\mathrm{P_{\mathbb{\Re}}}}$ that constructs a $(1, 1, 0)$-block-encoding of $P_{\mathbb{\Re}}$.}
  \label{fig:circuit_of_real_QSP}
\end{figure}
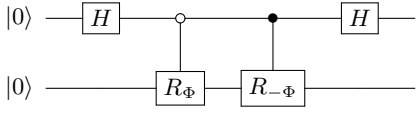

Since $R^{*}(x)=R(x)$, if the complex conjugate of $R_{\Phi}$ is taken, we can get
\begin{equation} \label{eq:QSP_reflection_convention_star}
  R_{\Phi}^{*}
  = e^{-i\phi_0 Z}\prod_{k=1}^{d}R(x)e^{-i\phi_k Z} = \begin{bmatrix}
    P^{*}(x) & \cdot \\
    \cdot    & \cdot
  \end{bmatrix},
\end{equation}
and $R_{\Phi}^{*}$ can be denoted as $R_{-\Phi}$. The quantum circuit
in Fig.~\ref{fig:circuit_of_real_QSP} constructs the $(1, 2, 0)$-block-encoding of $P_{\mathbb{\Re}}$:
\begin{align}
  &\bra{0}\bra{0} U_{\mathrm{P_{\mathbb{\Re}}}} \ket{0}\ket{0} \notag \\
    &= \bra{+}\bra{0}\left(
      \ket{0}\bra{0}\otimes R_{\Phi}+\ket{1}\bra{1}\otimes R_{-\Phi}
    \right)\ket{+}\ket{0} \notag \\
    &= \frac{P(x)+P^{*}(x)}{2} \notag \\
    &= P_{\mathbb{\Re}}(x). \label{eq:implement_real_function_by_QSP}
\end{align}

Now, we derive the result of QSVT from that of QSP.
Suppose that $U$ is a $(1, a, 0)$-block-encoding of a matrix $A$ such that
$A = \sum_{k=1}^{r}\sigma_k\ket{w_k}\bra{v_k}$. The unitaries $U$ and $e^{i\phi\Pi}$, where
$\Pi = 2\ket{0}_a\bra{0}-I$, act on the two-dimensional invariant subspaces
$\mathrm{Span}\left( \ket{0}_a\ket{v_k},\ket{\bot}\ket{v_k} \right)$ and 
$\mathrm{Span}\left( \ket{0}_a\ket{w_k},\ket{\bot}\ket{w_k} \right)$, where $\ket{\bot}$ satisfies
$\braket{\bot|0}_a=0$. The unitary $U$ acts on these invariant subspaces as follows:
\begin{equation} \label{eq:decomposition_of_U}
  \begin{split}
    U\ket{0}_a\ket{v_k}  &= \sigma_k\ket{0}_a\ket{w_k} + \sqrt{1-\sigma_{k}^{2}}\ket{\bot}\ket{w_k}, \\
    U\ket{\bot}\ket{v_k} &= \sqrt{1-\sigma_{k}^{2}}\ket{0}_a\ket{w_k} -\sigma_k\ket{\bot}\ket{w_k}.
  \end{split}
\end{equation}
Therefore, $U$ becomes
\begin{align}
  U &= \sum_{k}\begin{bmatrix}
    \sigma_k                & \sqrt{1-\sigma_{k}^{2}} \\
    \sqrt{1-\sigma_{k}^{2}} & -\sigma_k
  \end{bmatrix}\otimes \ket{w_k}\bra{v_k} \notag \\
    &= \sum_{k}R(\sigma_k)\otimes \ket{w_k}\bra{v_k},
\end{align}
and $U^{\dagger}$ becomes
\begin{align}
  U^{\dagger} &= \sum_{k}R^{\dagger}(\sigma_k)\otimes \left(\ket{w_k}\bra{v_k}\right)^{\dagger} \notag \\
              &= \sum_{k}R(\sigma_k)\otimes \ket{v_k}\bra{w_k}.
\end{align}
Moreover, $\Pi$ acts on the invariant subspaces as follows:
\begin{align}
    \Pi\otimes I_s 
      &= (\ket{0}_a\bra{0}-\ket{\bot}\bra{\bot})\otimes \sum_{k}\ket{w_k}\bra{v_k} \notag \\
      &= \sum_{k} \begin{bmatrix}
            1 & 0 \\
            0 & -1
        \end{bmatrix} \otimes \ket{w_k}\bra{v_k} \notag \\
      &= \sum_{k}Z \otimes \ket{w_k}\bra{v_k}.
\end{align}
Therefore, $e^{i\phi\Pi}$ becomes
\begin{equation} \label{eq:decomposition_of_exp}
  e^{i\phi\Pi} = \sum_{k}e^{i\phi Z} \otimes \ket{w_k}\bra{v_k}.
\end{equation}

Now, the alternating phase modulation sequence $U_{\Phi}$ defined in Eq.
~\eqref{eq:alternating_phase_modulation_sequence_for_odd_polynomial}
becomes the $(1,a,0)$-block-encoding of $P^{(\mathrm{SV})}(A)$ 
as for odd $d$:
\begin{align}
  U_{\Phi} &= \sum_{k}e^{i\phi_0 Z}\prod_{j=1}^{d}\left(
    R(\sigma_k)e^{i\phi_j Z}
  \right)\otimes \ket{w_k}\bra{v_k} \notag \\
          &= \begin{bmatrix}
              \sum_{k}P(\sigma_k)\ket{w_k}\bra{v_k} & \cdot \\
              \cdot & \cdot
          \end{bmatrix} \notag \\
          &= \begin{bmatrix}
            P^{(\mathrm{SV})}(A) & \cdot \\
            \cdot & \cdot
        \end{bmatrix}, 
\end{align}
and for even $d$ we can similarly derive.
We can construct the $(1,a+1,0)$-block-encoding of the real polynomial function $P^{(\mathrm{SV})}_{\Re}$ like QSP.
Since $U_{-\Phi}$ is the $(1,a,0)$-block-encoding of $P^{*(\mathrm{SV})}$, we can construct a quantum circuit
$U_{\mathrm{P^{(\mathrm{SV})}_{\mathbb{\Re}}}}$:
\begin{align}
  & \bra{0}_b\bra{0}_a U_{\mathrm{P^{(\mathrm{SV})}_{\mathbb{\Re}}}} \ket{0}_b\ket{0}_a \notag \\
  &= (\bra{0}_b H)\bra{0}_a\left(
    \ket{0}_b\bra{0}\otimes U_{\Phi} \right. \notag \\
  & \qquad \left. +\ket{1}_b\bra{1}\otimes U_{-\Phi}
  \right)(H\ket{0}_b)\ket{0}_a \notag \\
  &= \frac{\bra{0}_a (U_{\Phi}+U_{-\Phi})\ket{0}_a}{2} \notag \\
  &= \frac{P^{(\mathrm{SV})}(A)+P^{*(\mathrm{SV})}(A)}{2} \notag \\
  &= P_{\Re}^{(\mathrm{SV})}(A).
\end{align}

\bibliography{main}

\begin{thebibliography}{48}%
\makeatletter
\providecommand \@ifxundefined [1]{%
 \@ifx{#1\undefined}
}%
\providecommand \@ifnum [1]{%
 \ifnum #1\expandafter \@firstoftwo
 \else \expandafter \@secondoftwo
 \fi
}%
\providecommand \@ifx [1]{%
 \ifx #1\expandafter \@firstoftwo
 \else \expandafter \@secondoftwo
 \fi
}%
\providecommand \natexlab [1]{#1}%
\providecommand \enquote  [1]{``#1''}%
\providecommand \bibnamefont  [1]{#1}%
\providecommand \bibfnamefont [1]{#1}%
\providecommand \citenamefont [1]{#1}%
\providecommand \href@noop [0]{\@secondoftwo}%
\providecommand \href [0]{\begingroup \@sanitize@url \@href}%
\providecommand \@href[1]{\@@startlink{#1}\@@href}%
\providecommand \@@href[1]{\endgroup#1\@@endlink}%
\providecommand \@sanitize@url [0]{\catcode `\\12\catcode `\$12\catcode `\&12\catcode `\#12\catcode `\^12\catcode `\_12\catcode `\%12\relax}%
\providecommand \@@startlink[1]{}%
\providecommand \@@endlink[0]{}%
\providecommand \url  [0]{\begingroup\@sanitize@url \@url }%
\providecommand \@url [1]{\endgroup\@href {#1}{\urlprefix }}%
\providecommand \urlprefix  [0]{URL }%
\providecommand \Eprint [0]{\href }%
\providecommand \doibase [0]{https://doi.org/}%
\providecommand \selectlanguage [0]{\@gobble}%
\providecommand \bibinfo  [0]{\@secondoftwo}%
\providecommand \bibfield  [0]{\@secondoftwo}%
\providecommand \translation [1]{[#1]}%
\providecommand \BibitemOpen [0]{}%
\providecommand \bibitemStop [0]{}%
\providecommand \bibitemNoStop [0]{.\EOS\space}%
\providecommand \EOS [0]{\spacefactor3000\relax}%
\providecommand \BibitemShut  [1]{\csname bibitem#1\endcsname}%
\let\auto@bib@innerbib\@empty
\bibitem [{\citenamefont {Grover}(1996)}]{grover1996fast}%
  \BibitemOpen
  \bibfield  {author} {\bibinfo {author} {\bibfnamefont {L.~K.}\ \bibnamefont {Grover}},\ }\bibfield  {title} {\bibinfo {title} {A fast quantum mechanical algorithm for database search},\ }in\ \href@noop {} {\emph {\bibinfo {booktitle} {Proceedings of the twenty-eighth annual ACM symposium on Theory of computing}}}\ (\bibinfo  {publisher} {ACM Press},\ \bibinfo {address} {New York},\ \bibinfo {year} {1996})\ pp.\ \bibinfo {pages} {212--219}\BibitemShut {NoStop}%
\bibitem [{\citenamefont {Shor}(1994)}]{shor1994algorithms}%
  \BibitemOpen
  \bibfield  {author} {\bibinfo {author} {\bibfnamefont {P.~W.}\ \bibnamefont {Shor}},\ }\bibfield  {title} {\bibinfo {title} {Algorithms for quantum computation: discrete logarithms and factoring},\ }in\ \href@noop {} {\emph {\bibinfo {booktitle} {Proceedings 35th Annual Symposium on Foundations of Computer Science}}}\ (\bibinfo  {publisher} {IEEE Computer Society},\ \bibinfo {address} {USA},\ \bibinfo {year} {1994})\ pp.\ \bibinfo {pages} {124--134}\BibitemShut {NoStop}%
\bibitem [{\citenamefont {Harrow}\ \emph {et~al.}(2009)\citenamefont {Harrow}, \citenamefont {Hassidim},\ and\ \citenamefont {Lloyd}}]{harrow2009quantum}%
  \BibitemOpen
  \bibfield  {author} {\bibinfo {author} {\bibfnamefont {A.~W.}\ \bibnamefont {Harrow}}, \bibinfo {author} {\bibfnamefont {A.}~\bibnamefont {Hassidim}},\ and\ \bibinfo {author} {\bibfnamefont {S.}~\bibnamefont {Lloyd}},\ }\bibfield  {title} {\bibinfo {title} {Quantum algorithm for linear systems of equations},\ }\href@noop {} {\bibfield  {journal} {\bibinfo  {journal} {Phys. Rev. Lett.}\ }\textbf {\bibinfo {volume} {103}},\ \bibinfo {pages} {150502} (\bibinfo {year} {2009})}\BibitemShut {NoStop}%
\bibitem [{\citenamefont {Childs}\ \emph {et~al.}(2017)\citenamefont {Childs}, \citenamefont {Kothari},\ and\ \citenamefont {Somma}}]{childs2017quantum}%
  \BibitemOpen
  \bibfield  {author} {\bibinfo {author} {\bibfnamefont {A.~M.}\ \bibnamefont {Childs}}, \bibinfo {author} {\bibfnamefont {R.}~\bibnamefont {Kothari}},\ and\ \bibinfo {author} {\bibfnamefont {R.~D.}\ \bibnamefont {Somma}},\ }\bibfield  {title} {\bibinfo {title} {Quantum algorithm for systems of linear equations with exponentially improved dependence on precision},\ }\href@noop {} {\bibfield  {journal} {\bibinfo  {journal} {SIAM J. Comput.}\ }\textbf {\bibinfo {volume} {46}},\ \bibinfo {pages} {1920} (\bibinfo {year} {2017})}\BibitemShut {NoStop}%
\bibitem [{\citenamefont {Feynman}(1982)}]{feynman1982simulating}%
  \BibitemOpen
  \bibfield  {author} {\bibinfo {author} {\bibfnamefont {R.~P.}\ \bibnamefont {Feynman}},\ }\bibfield  {title} {\bibinfo {title} {Simulating physics with computers},\ }\href@noop {} {\bibfield  {journal} {\bibinfo  {journal} {Int. J. Theor. Phys.}\ }\textbf {\bibinfo {volume} {21}},\ \bibinfo {pages} {467} (\bibinfo {year} {1982})}\BibitemShut {NoStop}%
\bibitem [{\citenamefont {Lloyd}(1996)}]{lloyd1996universal}%
  \BibitemOpen
  \bibfield  {author} {\bibinfo {author} {\bibfnamefont {S.}~\bibnamefont {Lloyd}},\ }\bibfield  {title} {\bibinfo {title} {Universal quantum simulators},\ }\href@noop {} {\bibfield  {journal} {\bibinfo  {journal} {Science}\ }\textbf {\bibinfo {volume} {273}},\ \bibinfo {pages} {1073} (\bibinfo {year} {1996})}\BibitemShut {NoStop}%
\bibitem [{\citenamefont {Gaitan}(2020)}]{gaitan2020finding}%
  \BibitemOpen
  \bibfield  {author} {\bibinfo {author} {\bibfnamefont {F.}~\bibnamefont {Gaitan}},\ }\bibfield  {title} {\bibinfo {title} {Finding flows of a navier--stokes fluid through quantum computing},\ }\href@noop {} {\bibfield  {journal} {\bibinfo  {journal} {Npj Quantum Inf.}\ }\textbf {\bibinfo {volume} {6}},\ \bibinfo {pages} {1} (\bibinfo {year} {2020})}\BibitemShut {NoStop}%
\bibitem [{\citenamefont {Gaitan}(2021)}]{gaitan2021finding}%
  \BibitemOpen
  \bibfield  {author} {\bibinfo {author} {\bibfnamefont {F.}~\bibnamefont {Gaitan}},\ }\bibfield  {title} {\bibinfo {title} {Finding solutions of the navier-stokes equations through quantum computing--recent progress, a generalization, and next steps forward},\ }\href@noop {} {\bibfield  {journal} {\bibinfo  {journal} {Adv. Quantum Technol.}\ }\textbf {\bibinfo {volume} {4}},\ \bibinfo {pages} {2100055} (\bibinfo {year} {2021})}\BibitemShut {NoStop}%
\bibitem [{\citenamefont {Engel}\ \emph {et~al.}(2019)\citenamefont {Engel}, \citenamefont {Smith},\ and\ \citenamefont {Parker}}]{engel2019quantum}%
  \BibitemOpen
  \bibfield  {author} {\bibinfo {author} {\bibfnamefont {A.}~\bibnamefont {Engel}}, \bibinfo {author} {\bibfnamefont {G.}~\bibnamefont {Smith}},\ and\ \bibinfo {author} {\bibfnamefont {S.~E.}\ \bibnamefont {Parker}},\ }\bibfield  {title} {\bibinfo {title} {Quantum algorithm for the vlasov equation},\ }\href@noop {} {\bibfield  {journal} {\bibinfo  {journal} {Phys. Rev. A}\ }\textbf {\bibinfo {volume} {100}},\ \bibinfo {pages} {062315} (\bibinfo {year} {2019})}\BibitemShut {NoStop}%
\bibitem [{\citenamefont {Dodin}\ and\ \citenamefont {Startsev}(2021)}]{dodin2021quantum}%
  \BibitemOpen
  \bibfield  {author} {\bibinfo {author} {\bibfnamefont {I.~Y.}\ \bibnamefont {Dodin}}\ and\ \bibinfo {author} {\bibfnamefont {E.~A.}\ \bibnamefont {Startsev}},\ }\bibfield  {title} {\bibinfo {title} {On applications of quantum computing to plasma simulations},\ }\href@noop {} {\bibfield  {journal} {\bibinfo  {journal} {Phys. Plasmas}\ }\textbf {\bibinfo {volume} {28}},\ \bibinfo {pages} {092101} (\bibinfo {year} {2021})}\BibitemShut {NoStop}%
\bibitem [{\citenamefont {Novikau}\ \emph {et~al.}(2022)\citenamefont {Novikau}, \citenamefont {Startsev},\ and\ \citenamefont {Dodin}}]{novikau2022quantum}%
  \BibitemOpen
  \bibfield  {author} {\bibinfo {author} {\bibfnamefont {I.}~\bibnamefont {Novikau}}, \bibinfo {author} {\bibfnamefont {E.~A.}\ \bibnamefont {Startsev}},\ and\ \bibinfo {author} {\bibfnamefont {I.~Y.}\ \bibnamefont {Dodin}},\ }\bibfield  {title} {\bibinfo {title} {Quantum signal processing for simulating cold plasma waves},\ }\href@noop {} {\bibfield  {journal} {\bibinfo  {journal} {Phys. Rev. A}\ }\textbf {\bibinfo {volume} {105}},\ \bibinfo {pages} {062444} (\bibinfo {year} {2022})}\BibitemShut {NoStop}%
\bibitem [{\citenamefont {Ameri}\ \emph {et~al.}()\citenamefont {Ameri}, \citenamefont {Cappellaro}, \citenamefont {Krovi}, \citenamefont {Loureiro},\ and\ \citenamefont {Ye}}]{ameri2023quantum}%
  \BibitemOpen
  \bibfield  {author} {\bibinfo {author} {\bibfnamefont {A.}~\bibnamefont {Ameri}}, \bibinfo {author} {\bibfnamefont {P.}~\bibnamefont {Cappellaro}}, \bibinfo {author} {\bibfnamefont {H.}~\bibnamefont {Krovi}}, \bibinfo {author} {\bibfnamefont {N.~F.}\ \bibnamefont {Loureiro}},\ and\ \bibinfo {author} {\bibfnamefont {E.}~\bibnamefont {Ye}},\ }\href@noop {} {\bibinfo {title} {A quantum algorithm for the linear vlasov equation with collisions}},\ \bibinfo {note} {arXiv:2303.03450 (2023)}\BibitemShut {NoStop}%
\bibitem [{\citenamefont {Cao}\ \emph {et~al.}(2013)\citenamefont {Cao}, \citenamefont {Papageorgiou}, \citenamefont {Petras}, \citenamefont {Traub},\ and\ \citenamefont {Kais}}]{cao2013quantum}%
  \BibitemOpen
  \bibfield  {author} {\bibinfo {author} {\bibfnamefont {Y.}~\bibnamefont {Cao}}, \bibinfo {author} {\bibfnamefont {A.}~\bibnamefont {Papageorgiou}}, \bibinfo {author} {\bibfnamefont {I.}~\bibnamefont {Petras}}, \bibinfo {author} {\bibfnamefont {J.}~\bibnamefont {Traub}},\ and\ \bibinfo {author} {\bibfnamefont {S.}~\bibnamefont {Kais}},\ }\bibfield  {title} {\bibinfo {title} {Quantum algorithm and circuit design solving the poisson equation},\ }\href@noop {} {\bibfield  {journal} {\bibinfo  {journal} {New J. Phys.}\ }\textbf {\bibinfo {volume} {15}},\ \bibinfo {pages} {013021} (\bibinfo {year} {2013})}\BibitemShut {NoStop}%
\bibitem [{\citenamefont {Wang}\ \emph {et~al.}(2020)\citenamefont {Wang}, \citenamefont {Wang}, \citenamefont {Li}, \citenamefont {Fan}, \citenamefont {Wei},\ and\ \citenamefont {Gu}}]{wang2020quantum}%
  \BibitemOpen
  \bibfield  {author} {\bibinfo {author} {\bibfnamefont {S.}~\bibnamefont {Wang}}, \bibinfo {author} {\bibfnamefont {Z.}~\bibnamefont {Wang}}, \bibinfo {author} {\bibfnamefont {W.}~\bibnamefont {Li}}, \bibinfo {author} {\bibfnamefont {L.}~\bibnamefont {Fan}}, \bibinfo {author} {\bibfnamefont {Z.}~\bibnamefont {Wei}},\ and\ \bibinfo {author} {\bibfnamefont {Y.}~\bibnamefont {Gu}},\ }\bibfield  {title} {\bibinfo {title} {Quantum fast poisson solver: the algorithm and complete and modular circuit design},\ }\href@noop {} {\bibfield  {journal} {\bibinfo  {journal} {Quantum Inf. Process.}\ }\textbf {\bibinfo {volume} {19}},\ \bibinfo {pages} {170} (\bibinfo {year} {2020})}\BibitemShut {NoStop}%
\bibitem [{\citenamefont {Costa}\ \emph {et~al.}(2019)\citenamefont {Costa}, \citenamefont {Jordan},\ and\ \citenamefont {Ostrander}}]{costa2019quantum}%
  \BibitemOpen
  \bibfield  {author} {\bibinfo {author} {\bibfnamefont {P.~C.~S.}\ \bibnamefont {Costa}}, \bibinfo {author} {\bibfnamefont {S.}~\bibnamefont {Jordan}},\ and\ \bibinfo {author} {\bibfnamefont {A.}~\bibnamefont {Ostrander}},\ }\bibfield  {title} {\bibinfo {title} {Quantum algorithm for simulating the wave equation},\ }\href@noop {} {\bibfield  {journal} {\bibinfo  {journal} {Phys. Rev. A}\ }\textbf {\bibinfo {volume} {99}},\ \bibinfo {pages} {012323} (\bibinfo {year} {2019})}\BibitemShut {NoStop}%
\bibitem [{\citenamefont {Suau}\ \emph {et~al.}(2021)\citenamefont {Suau}, \citenamefont {Staffelbach},\ and\ \citenamefont {Calandra}}]{suau2021practical}%
  \BibitemOpen
  \bibfield  {author} {\bibinfo {author} {\bibfnamefont {A.}~\bibnamefont {Suau}}, \bibinfo {author} {\bibfnamefont {G.}~\bibnamefont {Staffelbach}},\ and\ \bibinfo {author} {\bibfnamefont {H.}~\bibnamefont {Calandra}},\ }\bibfield  {title} {\bibinfo {title} {Practical quantum computing: Solving the wave equation using a quantum approach},\ }\href@noop {} {\bibfield  {journal} {\bibinfo  {journal} {ACM Trans. Quantum Comput.}\ }\textbf {\bibinfo {volume} {2}},\ \bibinfo {pages} {1} (\bibinfo {year} {2021})}\BibitemShut {NoStop}%
\bibitem [{\citenamefont {Berry}\ \emph {et~al.}(2007)\citenamefont {Berry}, \citenamefont {Ahokas}, \citenamefont {Cleve},\ and\ \citenamefont {Sanders}}]{berry2007efficient}%
  \BibitemOpen
  \bibfield  {author} {\bibinfo {author} {\bibfnamefont {D.~W.}\ \bibnamefont {Berry}}, \bibinfo {author} {\bibfnamefont {G.}~\bibnamefont {Ahokas}}, \bibinfo {author} {\bibfnamefont {R.}~\bibnamefont {Cleve}},\ and\ \bibinfo {author} {\bibfnamefont {B.~C.}\ \bibnamefont {Sanders}},\ }\bibfield  {title} {\bibinfo {title} {Efficient quantum algorithms for simulating sparse hamiltonians},\ }\href@noop {} {\bibfield  {journal} {\bibinfo  {journal} {Commun. Math. Phys.}\ }\textbf {\bibinfo {volume} {270}},\ \bibinfo {pages} {359} (\bibinfo {year} {2007})}\BibitemShut {NoStop}%
\bibitem [{\citenamefont {Childs}\ and\ \citenamefont {Kothari}(2010)}]{childs2010limitations}%
  \BibitemOpen
  \bibfield  {author} {\bibinfo {author} {\bibfnamefont {A.~M.}\ \bibnamefont {Childs}}\ and\ \bibinfo {author} {\bibfnamefont {R.}~\bibnamefont {Kothari}},\ }\bibfield  {title} {\bibinfo {title} {Limitations on the simulation of non-sparse hamiltonians},\ }\href@noop {} {\bibfield  {journal} {\bibinfo  {journal} {Quantum Comput. Inf.}\ }\textbf {\bibinfo {volume} {10}},\ \bibinfo {pages} {669} (\bibinfo {year} {2010})}\BibitemShut {NoStop}%
\bibitem [{\citenamefont {Berry}\ and\ \citenamefont {Childs}(2012)}]{berry2012black}%
  \BibitemOpen
  \bibfield  {author} {\bibinfo {author} {\bibfnamefont {D.~W.}\ \bibnamefont {Berry}}\ and\ \bibinfo {author} {\bibfnamefont {A.~M.}\ \bibnamefont {Childs}},\ }\bibfield  {title} {\bibinfo {title} {Black-box hamiltonian simulation and unitary implementation},\ }\href@noop {} {\bibfield  {journal} {\bibinfo  {journal} {Quantum Inf. Comput.}\ }\textbf {\bibinfo {volume} {16}},\ \bibinfo {pages} {29} (\bibinfo {year} {2012})}\BibitemShut {NoStop}%
\bibitem [{\citenamefont {Childs}\ and\ \citenamefont {Wiebe}(2012)}]{childs2012hamiltonian}%
  \BibitemOpen
  \bibfield  {author} {\bibinfo {author} {\bibfnamefont {A.~M.}\ \bibnamefont {Childs}}\ and\ \bibinfo {author} {\bibfnamefont {N.}~\bibnamefont {Wiebe}},\ }\bibfield  {title} {\bibinfo {title} {Hamiltonian simulation using linear combinations of unitary operations},\ }\href@noop {} {\bibfield  {journal} {\bibinfo  {journal} {Quantum Inf. Comput.}\ }\textbf {\bibinfo {volume} {12}},\ \bibinfo {pages} {901} (\bibinfo {year} {2012})}\BibitemShut {NoStop}%
\bibitem [{\citenamefont {Berry}\ \emph {et~al.}(2014)\citenamefont {Berry}, \citenamefont {Childs}, \citenamefont {Cleve}, \citenamefont {Kothari},\ and\ \citenamefont {Somma}}]{berry2014exponential}%
  \BibitemOpen
  \bibfield  {author} {\bibinfo {author} {\bibfnamefont {D.~W.}\ \bibnamefont {Berry}}, \bibinfo {author} {\bibfnamefont {A.~M.}\ \bibnamefont {Childs}}, \bibinfo {author} {\bibfnamefont {R.}~\bibnamefont {Cleve}}, \bibinfo {author} {\bibfnamefont {R.}~\bibnamefont {Kothari}},\ and\ \bibinfo {author} {\bibfnamefont {R.~D.}\ \bibnamefont {Somma}},\ }\bibfield  {title} {\bibinfo {title} {Exponential improvement in precision for simulating sparse hamiltonians},\ }in\ \href@noop {} {\emph {\bibinfo {booktitle} {Proceedings of the forty-sixth annual ACM symposium on Theory of computing}}}\ (\bibinfo  {publisher} {ACM Press},\ \bibinfo {address} {New York},\ \bibinfo {year} {2014})\ pp.\ \bibinfo {pages} {283--292}\BibitemShut {NoStop}%
\bibitem [{\citenamefont {Berry}\ \emph {et~al.}(2015{\natexlab{a}})\citenamefont {Berry}, \citenamefont {Childs},\ and\ \citenamefont {Kothari}}]{berry2015hamiltonian}%
  \BibitemOpen
  \bibfield  {author} {\bibinfo {author} {\bibfnamefont {D.~W.}\ \bibnamefont {Berry}}, \bibinfo {author} {\bibfnamefont {A.~M.}\ \bibnamefont {Childs}},\ and\ \bibinfo {author} {\bibfnamefont {R.}~\bibnamefont {Kothari}},\ }\bibfield  {title} {\bibinfo {title} {Hamiltonian simulation with nearly optimal dependence on all parameters},\ }in\ \href@noop {} {\emph {\bibinfo {booktitle} {2015 IEEE 56th annual symposium on foundations of computer science}}}\ (\bibinfo  {publisher} {IEEE Computer Society},\ \bibinfo {address} {USA},\ \bibinfo {year} {2015})\ pp.\ \bibinfo {pages} {792--809}\BibitemShut {NoStop}%
\bibitem [{\citenamefont {Berry}\ \emph {et~al.}(2015{\natexlab{b}})\citenamefont {Berry}, \citenamefont {Childs}, \citenamefont {Cleve}, \citenamefont {Kothari},\ and\ \citenamefont {Somma}}]{berry2015simulating}%
  \BibitemOpen
  \bibfield  {author} {\bibinfo {author} {\bibfnamefont {D.~W.}\ \bibnamefont {Berry}}, \bibinfo {author} {\bibfnamefont {A.~M.}\ \bibnamefont {Childs}}, \bibinfo {author} {\bibfnamefont {R.}~\bibnamefont {Cleve}}, \bibinfo {author} {\bibfnamefont {R.}~\bibnamefont {Kothari}},\ and\ \bibinfo {author} {\bibfnamefont {R.~D.}\ \bibnamefont {Somma}},\ }\bibfield  {title} {\bibinfo {title} {Simulating hamiltonian dynamics with a truncated taylor series},\ }\href@noop {} {\bibfield  {journal} {\bibinfo  {journal} {Phys. Rev. Lett.}\ }\textbf {\bibinfo {volume} {114}},\ \bibinfo {pages} {090502} (\bibinfo {year} {2015}{\natexlab{b}})}\BibitemShut {NoStop}%
\bibitem [{\citenamefont {Berry}\ and\ \citenamefont {Novo}(2016)}]{berry2016corrected}%
  \BibitemOpen
  \bibfield  {author} {\bibinfo {author} {\bibfnamefont {D.~W.}\ \bibnamefont {Berry}}\ and\ \bibinfo {author} {\bibfnamefont {L.}~\bibnamefont {Novo}},\ }\bibfield  {title} {\bibinfo {title} {Corrected quantum walk for optimal hamiltonian simulation},\ }\href@noop {} {\bibfield  {journal} {\bibinfo  {journal} {Quantum Inf. Comput.}\ }\textbf {\bibinfo {volume} {16}},\ \bibinfo {pages} {1295} (\bibinfo {year} {2016})}\BibitemShut {NoStop}%
\bibitem [{\citenamefont {Novo}\ and\ \citenamefont {Berry}(2017)}]{novo2017improved}%
  \BibitemOpen
  \bibfield  {author} {\bibinfo {author} {\bibfnamefont {L.}~\bibnamefont {Novo}}\ and\ \bibinfo {author} {\bibfnamefont {D.~W.}\ \bibnamefont {Berry}},\ }\bibfield  {title} {\bibinfo {title} {Improved hamiltonian simulation via a truncated taylor series and corrections},\ }\href@noop {} {\bibfield  {journal} {\bibinfo  {journal} {Quantum Inf. Comput.}\ }\textbf {\bibinfo {volume} {17}},\ \bibinfo {pages} {623} (\bibinfo {year} {2017})}\BibitemShut {NoStop}%
\bibitem [{\citenamefont {Childs}\ \emph {et~al.}(2018)\citenamefont {Childs}, \citenamefont {Maslov}, \citenamefont {Nam}, \citenamefont {Ross},\ and\ \citenamefont {Su}}]{childs2018toward}%
  \BibitemOpen
  \bibfield  {author} {\bibinfo {author} {\bibfnamefont {A.~M.}\ \bibnamefont {Childs}}, \bibinfo {author} {\bibfnamefont {D.}~\bibnamefont {Maslov}}, \bibinfo {author} {\bibfnamefont {Y.}~\bibnamefont {Nam}}, \bibinfo {author} {\bibfnamefont {N.~J.}\ \bibnamefont {Ross}},\ and\ \bibinfo {author} {\bibfnamefont {Y.}~\bibnamefont {Su}},\ }\bibfield  {title} {\bibinfo {title} {Toward the first quantum simulation with quantum speedup},\ }\href@noop {} {\bibfield  {journal} {\bibinfo  {journal} {Proc. Natl. Acad. Sci. U.S.A.}\ }\textbf {\bibinfo {volume} {115}},\ \bibinfo {pages} {9456} (\bibinfo {year} {2018})}\BibitemShut {NoStop}%
\bibitem [{\citenamefont {Low}\ and\ \citenamefont {Chuang}(2017)}]{low2017optimal}%
  \BibitemOpen
  \bibfield  {author} {\bibinfo {author} {\bibfnamefont {G.~H.}\ \bibnamefont {Low}}\ and\ \bibinfo {author} {\bibfnamefont {I.~L.}\ \bibnamefont {Chuang}},\ }\bibfield  {title} {\bibinfo {title} {Optimal hamiltonian simulation by quantum signal processing},\ }\href@noop {} {\bibfield  {journal} {\bibinfo  {journal} {Phys. Rev. Lett.}\ }\textbf {\bibinfo {volume} {118}},\ \bibinfo {pages} {010501} (\bibinfo {year} {2017})}\BibitemShut {NoStop}%
\bibitem [{\citenamefont {Low}\ and\ \citenamefont {Chuang}(2019)}]{low2019hamiltonian}%
  \BibitemOpen
  \bibfield  {author} {\bibinfo {author} {\bibfnamefont {G.~H.}\ \bibnamefont {Low}}\ and\ \bibinfo {author} {\bibfnamefont {I.~L.}\ \bibnamefont {Chuang}},\ }\bibfield  {title} {\bibinfo {title} {Hamiltonian simulation by qubitization},\ }\href@noop {} {\bibfield  {journal} {\bibinfo  {journal} {Quantum}\ }\textbf {\bibinfo {volume} {3}},\ \bibinfo {pages} {163} (\bibinfo {year} {2019})}\BibitemShut {NoStop}%
\bibitem [{\citenamefont {Gily{\'e}n}\ \emph {et~al.}(2019)\citenamefont {Gily{\'e}n}, \citenamefont {Su}, \citenamefont {Low},\ and\ \citenamefont {Wiebe}}]{gilyen2019quantum}%
  \BibitemOpen
  \bibfield  {author} {\bibinfo {author} {\bibfnamefont {A.}~\bibnamefont {Gily{\'e}n}}, \bibinfo {author} {\bibfnamefont {Y.}~\bibnamefont {Su}}, \bibinfo {author} {\bibfnamefont {G.~H.}\ \bibnamefont {Low}},\ and\ \bibinfo {author} {\bibfnamefont {N.}~\bibnamefont {Wiebe}},\ }\bibfield  {title} {\bibinfo {title} {Quantum singular value transformation and beyond: exponential improvements for quantum matrix arithmetics},\ }in\ \href@noop {} {\emph {\bibinfo {booktitle} {Proceedings of the 51st Annual ACM SIGACT Symposium on Theory of Computing}}}\ (\bibinfo  {publisher} {ACM Press},\ \bibinfo {address} {New York},\ \bibinfo {year} {2019})\ pp.\ \bibinfo {pages} {193--204}\BibitemShut {NoStop}%
\bibitem [{\citenamefont {Nielsen}\ and\ \citenamefont {Chuang}(2010)}]{nielsen2010quantum}%
  \BibitemOpen
  \bibfield  {author} {\bibinfo {author} {\bibfnamefont {M.~A.}\ \bibnamefont {Nielsen}}\ and\ \bibinfo {author} {\bibfnamefont {I.~L.}\ \bibnamefont {Chuang}},\ }\href@noop {} {\emph {\bibinfo {title} {Quantum Computation and Quantum Information}}}\ (\bibinfo  {publisher} {American Association of Physics Teachers},\ \bibinfo {year} {2010})\BibitemShut {NoStop}%
\bibitem [{\citenamefont {Szegedy}(2004)}]{szegedy2004quantum}%
  \BibitemOpen
  \bibfield  {author} {\bibinfo {author} {\bibfnamefont {M.}~\bibnamefont {Szegedy}},\ }\bibfield  {title} {\bibinfo {title} {Quantum speed-up of markov chain based algorithms},\ }in\ \href@noop {} {\emph {\bibinfo {booktitle} {45th Annual IEEE symposium on foundations of computer science}}}\ (\bibinfo  {publisher} {IEEE Computer Society},\ \bibinfo {address} {USA},\ \bibinfo {year} {2004})\ pp.\ \bibinfo {pages} {32--41}\BibitemShut {NoStop}%
\bibitem [{\citenamefont {Martyn}\ \emph {et~al.}(2021)\citenamefont {Martyn}, \citenamefont {Rossi}, \citenamefont {Tan},\ and\ \citenamefont {Chuang}}]{martyn2021grand}%
  \BibitemOpen
  \bibfield  {author} {\bibinfo {author} {\bibfnamefont {J.~M.}\ \bibnamefont {Martyn}}, \bibinfo {author} {\bibfnamefont {Z.~M.}\ \bibnamefont {Rossi}}, \bibinfo {author} {\bibfnamefont {A.~K.}\ \bibnamefont {Tan}},\ and\ \bibinfo {author} {\bibfnamefont {I.~L.}\ \bibnamefont {Chuang}},\ }\bibfield  {title} {\bibinfo {title} {Grand unification of quantum algorithms},\ }\href@noop {} {\bibfield  {journal} {\bibinfo  {journal} {PRX Quantum}\ }\textbf {\bibinfo {volume} {2}},\ \bibinfo {pages} {040203} (\bibinfo {year} {2021})}\BibitemShut {NoStop}%
\bibitem [{\citenamefont {Paetznick}\ and\ \citenamefont {Svore}(2014)}]{paetznick2014repeat}%
  \BibitemOpen
  \bibfield  {author} {\bibinfo {author} {\bibfnamefont {A.}~\bibnamefont {Paetznick}}\ and\ \bibinfo {author} {\bibfnamefont {K.~M.}\ \bibnamefont {Svore}},\ }\bibfield  {title} {\bibinfo {title} {Repeat-until-success: Non-deterministic decomposition of single-qubit unitaries},\ }\href@noop {} {\bibfield  {journal} {\bibinfo  {journal} {Quantum Inf. Comput.}\ }\textbf {\bibinfo {volume} {14}},\ \bibinfo {pages} {1277} (\bibinfo {year} {2014})}\BibitemShut {NoStop}%
\bibitem [{\citenamefont {Daskin}\ and\ \citenamefont {Kais}(2017)}]{daskin2017ancilla}%
  \BibitemOpen
  \bibfield  {author} {\bibinfo {author} {\bibfnamefont {A.}~\bibnamefont {Daskin}}\ and\ \bibinfo {author} {\bibfnamefont {S.}~\bibnamefont {Kais}},\ }\bibfield  {title} {\bibinfo {title} {An ancilla-based quantum simulation framework for non-unitary matrices},\ }\href@noop {} {\bibfield  {journal} {\bibinfo  {journal} {Quantum Inf. Process.}\ }\textbf {\bibinfo {volume} {16}},\ \bibinfo {pages} {1} (\bibinfo {year} {2017})}\BibitemShut {NoStop}%
\bibitem [{\citenamefont {Grover}(2005)}]{grover2005quantum}%
  \BibitemOpen
  \bibfield  {author} {\bibinfo {author} {\bibfnamefont {L.~K.}\ \bibnamefont {Grover}},\ }\bibfield  {title} {\bibinfo {title} {Fixed-point quantum search},\ }\href@noop {} {\bibfield  {journal} {\bibinfo  {journal} {Phys. Rev. Lett.}\ }\textbf {\bibinfo {volume} {95}},\ \bibinfo {pages} {150501} (\bibinfo {year} {2005})}\BibitemShut {NoStop}%
\bibitem [{\citenamefont {Tulsi}\ \emph {et~al.}(2006)\citenamefont {Tulsi}, \citenamefont {Grover},\ and\ \citenamefont {Patel}}]{tulsi2006new}%
  \BibitemOpen
  \bibfield  {author} {\bibinfo {author} {\bibfnamefont {T.}~\bibnamefont {Tulsi}}, \bibinfo {author} {\bibfnamefont {L.~K.}\ \bibnamefont {Grover}},\ and\ \bibinfo {author} {\bibfnamefont {A.}~\bibnamefont {Patel}},\ }\bibfield  {title} {\bibinfo {title} {A new algorithm for fixed point quantum search},\ }\href@noop {} {\bibfield  {journal} {\bibinfo  {journal} {Quantum Inf. Comput.}\ }\textbf {\bibinfo {volume} {6}},\ \bibinfo {pages} {483} (\bibinfo {year} {2006})}\BibitemShut {NoStop}%
\bibitem [{\citenamefont {Yoder}\ \emph {et~al.}(2014)\citenamefont {Yoder}, \citenamefont {Low},\ and\ \citenamefont {Chuang}}]{yoder2014fixed}%
  \BibitemOpen
  \bibfield  {author} {\bibinfo {author} {\bibfnamefont {T.~J.}\ \bibnamefont {Yoder}}, \bibinfo {author} {\bibfnamefont {G.~H.}\ \bibnamefont {Low}},\ and\ \bibinfo {author} {\bibfnamefont {I.~L.}\ \bibnamefont {Chuang}},\ }\bibfield  {title} {\bibinfo {title} {Fixed-point quantum search with an optimal number of queries},\ }\href@noop {} {\bibfield  {journal} {\bibinfo  {journal} {Phys. Rev. Lett.}\ }\textbf {\bibinfo {volume} {113}},\ \bibinfo {pages} {210501} (\bibinfo {year} {2014})}\BibitemShut {NoStop}%
\bibitem [{\citenamefont {Brassard}\ \emph {et~al.}(2002)\citenamefont {Brassard}, \citenamefont {H{\o}yer}, \citenamefont {Mosca},\ and\ \citenamefont {Tapp}}]{Brassard2002quantum}%
  \BibitemOpen
  \bibfield  {author} {\bibinfo {author} {\bibfnamefont {G.}~\bibnamefont {Brassard}}, \bibinfo {author} {\bibfnamefont {P.}~\bibnamefont {H{\o}yer}}, \bibinfo {author} {\bibfnamefont {M.}~\bibnamefont {Mosca}},\ and\ \bibinfo {author} {\bibfnamefont {A.}~\bibnamefont {Tapp}},\ }\bibfield  {title} {\bibinfo {title} {Quantum amplitude amplification and estimation},\ }\href@noop {} {\bibfield  {journal} {\bibinfo  {journal} {Quantum Comput. Inf.}\ }\textbf {\bibinfo {volume} {305}},\ \bibinfo {pages} {53} (\bibinfo {year} {2002})}\BibitemShut {NoStop}%
\bibitem [{\citenamefont {Low}\ \emph {et~al.}(2016)\citenamefont {Low}, \citenamefont {Yoder},\ and\ \citenamefont {Chuang}}]{low2016methodology}%
  \BibitemOpen
  \bibfield  {author} {\bibinfo {author} {\bibfnamefont {G.~H.}\ \bibnamefont {Low}}, \bibinfo {author} {\bibfnamefont {T.~J.}\ \bibnamefont {Yoder}},\ and\ \bibinfo {author} {\bibfnamefont {I.~L.}\ \bibnamefont {Chuang}},\ }\bibfield  {title} {\bibinfo {title} {Methodology of resonant equiangular composite quantum gates},\ }\href@noop {} {\bibfield  {journal} {\bibinfo  {journal} {Phys. Rev. X}\ }\textbf {\bibinfo {volume} {6}},\ \bibinfo {pages} {041067} (\bibinfo {year} {2016})}\BibitemShut {NoStop}%
\bibitem [{\citenamefont {Low}\ and\ \citenamefont {Chuang}()}]{low2017hamiltonian}%
  \BibitemOpen
  \bibfield  {author} {\bibinfo {author} {\bibfnamefont {G.~H.}\ \bibnamefont {Low}}\ and\ \bibinfo {author} {\bibfnamefont {I.~L.}\ \bibnamefont {Chuang}},\ }\href@noop {} {\bibinfo {title} {Hamiltonian simulation by uniform spectral amplification}},\ \bibinfo {note} {arXiv:1707.05391 (2020)}\BibitemShut {NoStop}%
\bibitem [{\citenamefont {Chao}\ \emph {et~al.}()\citenamefont {Chao}, \citenamefont {Ding}, \citenamefont {Gily{\'e}n}, \citenamefont {Huang},\ and\ \citenamefont {Szegedy}}]{chao2020finding}%
  \BibitemOpen
  \bibfield  {author} {\bibinfo {author} {\bibfnamefont {R.}~\bibnamefont {Chao}}, \bibinfo {author} {\bibfnamefont {D.}~\bibnamefont {Ding}}, \bibinfo {author} {\bibfnamefont {A.}~\bibnamefont {Gily{\'e}n}}, \bibinfo {author} {\bibfnamefont {C.}~\bibnamefont {Huang}},\ and\ \bibinfo {author} {\bibfnamefont {M.}~\bibnamefont {Szegedy}},\ }\href@noop {} {\bibinfo {title} {Finding angles for quantum signal processing with machine precision}},\ \bibinfo {note} {arXiv:2003.02831 (2020)}\BibitemShut {NoStop}%
\bibitem [{\citenamefont {Chao}\ \emph {et~al.}(2021)\citenamefont {Chao}, \citenamefont {Ding}, \citenamefont {Gily{\'e}n}, \citenamefont {Huang},\ and\ \citenamefont {Szegedy}}]{chao2021finding}%
  \BibitemOpen
  \bibfield  {author} {\bibinfo {author} {\bibfnamefont {R.}~\bibnamefont {Chao}}, \bibinfo {author} {\bibfnamefont {D.}~\bibnamefont {Ding}}, \bibinfo {author} {\bibfnamefont {A.}~\bibnamefont {Gily{\'e}n}}, \bibinfo {author} {\bibfnamefont {C.}~\bibnamefont {Huang}},\ and\ \bibinfo {author} {\bibfnamefont {M.}~\bibnamefont {Szegedy}},\ }\href@noop {} {\bibinfo {title} {Finding angles for quantum signal processing with machine precision}} (\bibinfo {year} {2021}),\ \bibinfo {note} {\url{https://github.com/ichuang/pyqsp}, accessed: 07/2022}\BibitemShut {NoStop}%
\bibitem [{\citenamefont {Low}(2017)}]{low2017quantum}%
  \BibitemOpen
  \bibfield  {author} {\bibinfo {author} {\bibfnamefont {G.~H.}\ \bibnamefont {Low}},\ }\emph {\bibinfo {title} {Quantum signal processing by single-qubit dynamics}},\ \href@noop {} {\bibinfo {type} {Ph.d. thesis}},\ \bibinfo  {school} {Massachusetts Institute of Technology} (\bibinfo {year} {2017})\BibitemShut {NoStop}%
\bibitem [{\citenamefont {Mitarai}\ and\ \citenamefont {Mizukami}()}]{mitarai2022quantum}%
  \BibitemOpen
  \bibfield  {author} {\bibinfo {author} {\bibfnamefont {K.}~\bibnamefont {Mitarai}}\ and\ \bibinfo {author} {\bibfnamefont {W.}~\bibnamefont {Mizukami}},\ }\href@noop {} {\bibinfo {title} {Perturbation theory with quantum signal processing}},\ \bibinfo {note} {arXiv:2210.00718 (2022)}\BibitemShut {NoStop}%
\bibitem [{\citenamefont {Prakash}(2014)}]{prakash2014quantum}%
  \BibitemOpen
  \bibfield  {author} {\bibinfo {author} {\bibfnamefont {A.}~\bibnamefont {Prakash}},\ }\href@noop {} {\emph {\bibinfo {title} {Quantum algorithms for linear algebra and machine learning}}}\ (\bibinfo  {publisher} {University of California, Berkeley},\ \bibinfo {year} {2014})\BibitemShut {NoStop}%
\bibitem [{\citenamefont {Giovannetti}\ \emph {et~al.}(2008)\citenamefont {Giovannetti}, \citenamefont {Lloyd},\ and\ \citenamefont {Maccone}}]{giovannetti2008architectures}%
  \BibitemOpen
  \bibfield  {author} {\bibinfo {author} {\bibfnamefont {V.}~\bibnamefont {Giovannetti}}, \bibinfo {author} {\bibfnamefont {S.}~\bibnamefont {Lloyd}},\ and\ \bibinfo {author} {\bibfnamefont {L.}~\bibnamefont {Maccone}},\ }\bibfield  {title} {\bibinfo {title} {Architectures for a quantum random access memory},\ }\href@noop {} {\bibfield  {journal} {\bibinfo  {journal} {Phys. Rev. A}\ }\textbf {\bibinfo {volume} {78}},\ \bibinfo {pages} {052310} (\bibinfo {year} {2008})}\BibitemShut {NoStop}%
\bibitem [{\citenamefont {\textit{et al.}}(2021)}]{ANIS2021Qiskit}%
  \BibitemOpen
  \bibfield  {author} {\bibinfo {author} {\bibfnamefont {M.~S.~A.}\ \bibnamefont {\textit{et al.}}},\ }\href@noop {} {\bibinfo {title} {Qiskit: An open-source framework for quantum computing}} (\bibinfo {year} {2021})\BibitemShut {NoStop}%
\bibitem [{\citenamefont {Chen}(1984)}]{chen1984introduction}%
  \BibitemOpen
  \bibfield  {author} {\bibinfo {author} {\bibfnamefont {F.~F.}\ \bibnamefont {Chen}},\ }\href@noop {} {\emph {\bibinfo {title} {Introduction to plasma physics and controlled fusion}}}\ (\bibinfo  {publisher} {Springer},\ \bibinfo {address} {New York},\ \bibinfo {year} {1984})\ pp.\ \bibinfo {pages} {224--232}\BibitemShut {NoStop}%
\end{thebibliography}%
\end{document}